\documentclass[a4paper,fleqn,usenatbib]{mnras}

\usepackage{mathptmx}
\usepackage[figuresright]{rotating}
\usepackage[graphicx]{realboxes}
\usepackage{adjustbox}
\usepackage{epstopdf}
\usepackage[T1]{fontenc}
\usepackage{ae,aecompl}

\usepackage{graphicx}	% Including Figure files
\usepackage{amsmath}	% Advanced maths commands
\usepackage{amssymb}	% Extra maths symbols
\usepackage{pdflscape}
\title[Investigation on Young radio AGNs based on SDSS spectroscopy]{Investigation on Young radio AGNs based on SDSS spectroscopy}

\author[Mai Liao et al.]{Mai Liao,$^{1,2}$\thanks{E-mail: liaomai@shao.ac.cn}
Minfeng Gu,$^{1}$\thanks{E-mail: gumf@shao.ac.cn}
\\
$^{1}$Shanghai Astronomical Observatory, Chinese Academy of Sciences, 80 Nandan Road, Shanghai 200030, China \\
$^{2}$University of Chinese Academy of Sciences, 19A Yuquanlu, Beijing 100049, China \\
}

\date{Accepted XXX. Received YYY; in original form ZZZ}

\pubyear{2019}

\begin{document}
\label{firstpage}
\pagerange{\pageref{firstpage}--\pageref{lastpage}}
\maketitle

\begin{abstract}
The Gigahertz Peaked Spectrum (GPS) sources, Compact Steep Spectrum (CSS) radio sources, and High Frequency Peakers (HFP) radio sources are thought to be young radio AGNs, at the early stage of AGN evolution. We investigated the optical properties of the largest sample of 126 young radio AGNs based on the spectra in SDSS DR12. We find the black hole masses $M_{\rm BH}$ range from $10^{7.32}$ to $10^{9.84}$ $\rm M_{\sun}$, and the Eddington ratios $R_{\rm edd}$ vary from $10^{-4.93}$ to $10^{0.37}$, suggesting that young radio AGNs have various accretion activities and not all are accreting at high accretion rate. Our young radio sources generally follow the evolutionary trend towards large-scale radio galaxies with increasing linear size and decreasing accretion rate in the radio power - linear size diagram. The radio properties of low luminosity young radio AGNs with low $R_{\rm edd}$ are discussed. The line width of [O III] $\lambda5007$ core ($\sigma_{\rm [O III]}$) is found to be a good surrogate of stellar velocity dispersion $\sigma_{\ast}$. The radio luminosity $L_{\rm 5GHz}$ correlates strongly with [O III] core luminosity $L_{\rm [O III]}$, suggesting that radio activity and accretion are closely related in young radio sources. We find one object that can be defined as narrow-line Seyfert 1 galaxies (NLS1s), representing a population of young AGNs both with young jet and early accretion activity. The optical variabilities of 15 quasars with multi-epoch spectroscopy were investigated. Our results show that the optical variability in young AGN quasars presents low variations ($\leqslant60\%$) similar to the normal radio-quiet quasars.
%We also present the optical spectra of two GPS quasars observed with the Lijiang 2.4m telescope in 2012.
%when incorporating the Eddington ratio into the radio power - linear size distribution
\end{abstract}


\section{Introduction}

Young radio active galactic nuclei (AGNs) are a subclass of AGNs characterized by an intrinsically compact radio morphology and a peaked radio spectrum, including Gigahertz Peaked Spectrum (GPS) radio sources, Compact Steep Spectrum (CSS) radio sources and High Frequency Peaker (HFP) radio sources. The peak frequency in radio spectrum are around 1 GHz, 100 MHz, and above 5 GHz for GPS, CSS and HFP, respectively. Both GPS and HFP sources are with linear size (LS) below 1kpc, while CSS sources are well confined wihtin 20kpc \citep{od98}. These radio AGNs are identified in both galaxies and quasars, with galaxies being at lower redshift, less variable than quasars and have symmetric radio morphologies, while quasars are detected at both low and high redshifts and show core-jet structures \citep{od98}. It is usually believed that these radio sources will perhaps eventually evolve into large-scale radio sources, known as Fanaroff-Riley (FR) I and FR II sources based on the youth scenario \citep{od98,2003PASA...20...69P,2011MNRAS.416.1135R}, which is strongly supported by the measurements of dynamical and spectral age, about $10^{3}-10^{5}$ years \citep{1998A&A...336L..37O,2003PASA...20...19M,2003PASA...20...69P,2009AN....330..193G,2012ApJS..198....5A}. Therefore, they are important to study the triggering of radio/accretion activity and evolution of AGN as well as the relationship between the accretion and jet. 
%\textbf{GPS/CSS/HFP sources are usually believed to be young radio objects and be at the early stage of AGN evolution. They will perhaps eventually evolve into large scale radio sources, known as Fanaroff-Riley (FR) I and FR II sources \citep{od98,2003PASA...20...69P,2011MNRAS.416.1135R}}

While many works have focused on radio investigations of young radio AGNs, their optical properties have not been sufficiently studied and there are only a few studies on various samples of young radio AGNs. \cite{2009MNRAS.398.1905W} investigated the optical properties of a sample of 65 young radio galaxies (both GPS and CSS sources) based on the data collected from the literature. They found that most young radio AGNs have relatively small black hole (BH) masses and high Eddington ratios which are similar to those of Narrow-line Seyfert 1 galaxies (NLS1s). \cite{2012ApJ...757..140S} compiled a sample of 34 low-redshift young radio galaxies including high-excitation galaxies (HEGs) and low-excitation galaxies (LEGs) to study their accretion properties. The BH mass distribution was found to be similar to large-scale radio galaxies, while accretion activity presented a diversity. While these two works give us clues on the accretion process in young radio AGNs, the source number in their sample is rather limited. A larger sample is needed to further systematically study the optical properties of young radio AGNs, especially the combination of the radio and optical properties will give clues on the AGN evolution and jet-disk relation.

NLS1s are a special class of AGNs. It is well known that they have smaller black hole masses and higher accretion rates compared to typical broad-line AGNs, suggesting that NLS1s are candidates of young AGNs and may be in the early stage of AGN accretion activity \citep{2008RMxAC..32...86K}. Recent works show that some radio-loud NLS1s are CSS-like sources and they share same radio properties with young radio AGNs \citep{2006komossa1,2010AJ....139.2612G,2014MNRAS.441..172C,Gu2015}. These objects may represent an AGN population with both jet and accretion activity at early stage of evolution. Searching for NLS1s in the sample of young radio AGNs enables us to find more targets, and then to investigate the jet and accretion, their relation, and AGN evolution in general.

In this work, our goal is to systematically study accretion properties, evolution in young radio AGNs and to search for NLS1s candidates by analyzing the SDSS spectra of a larger sample. In addition we will study stellar velocity dispersion versus gas velocity dispersion, jet-disk relation and optical variability. Section 2 presents the details of sample selection. The spectral analysis is shown in Section 3, in which the measurements of emission lines and stellar velocity dispersions are given. In Section 4, we show the main results. Section 5 gives discussions and the main results are summarized in Section 6. Throughout this paper. the cosmological parameters $H_0 = 70\, \mathrm{km\,s^{-1}\, Mpc^{-1}}$, $\Omega_\mathrm{m} = 0.3$, and $\Omega_{\lambda} = 0.7$ is adopted. The spectral index $\alpha_{\lambda}$ is defined as $f_{\lambda}$ $\propto$ $\lambda^{{\alpha}_{\lambda}}$ with $f_{\lambda}$ being the flux density at wavelength $\lambda$.

\section{Sample selection}

To systematically study the optical properties of the largest possible sample of young radio AGNs, we firstly collected all available radio-selected samples of young radio AGNs in the literature. Besides GPS, CSS and HFP sources, the Compact Symmetric Objects (CSO) are also included in our sample, which are also believed to be young AGNs, and share similar properties as GPS radio galaxies \citep{od98}. The source classification (i.e., CSS/GPS/HFP/CSO) is directly taken from the literature based on radio properties (e.g., radio size, spectral shape and peak frequency, and/or radio morphology).

The CSS sources were collected from \cite{Spencer1989}, \cite{Fanti1990} (3C and PW catalogue combined), \cite{Fanti2001} (B3/VLA sample), \cite{Kunert2002} (the FIRST sample), \cite{E04} (ATCA sample), \cite{Snellen2004} (the CORALZ sample) and \cite{2010MNRAS.408.2261K}. For GPS sources, the radio-selected samples of   \cite{Stanghellini1998} (1-Jansky catalogues), \cite{Snellen98} (faint WENSS sample), \cite{Snellen2002} (Parkes 0.5 Jy sample), \cite{E04} (ATCA sample) and \cite{Snellen2004} (the CORALZ sample) were included. We collected HFP sources from \cite{Dallacasa2000} and \cite{S09}, and CSO sources from \cite{Peck2000} (COINS sample), \cite{Snellen2004} (the CORALZ sample), and \cite{AB12}. This selection produces a sample of 250 CSS, 148 GPS, 116 HFP, and 80 CSO radio sources. After removing the overlap sources in the literature, the sample consists of 545 radio sources, which is the parent sample of our study.  %including A CSS, B GPS, C HFP, D CSO.

Then, we searched for the SDSS spectroscopic counterparts of the young radio AGNs within 2 arcsec from the NED position\footnote{http://ned.ipac.caltech.edu} for all the objects in parent sample. This results in our optical sample, consisting of 147 sources.
%, of which there are 90 quasars and 57 galaxies according to SDSS classification. 
%The detailed information of the sample are presented in Table 1, including 82 CSS, 20 GPS, 39 HFP and 6 CSO. We retained all spectra for those objects with multiple spectroscopic observations in order to analyse their optical variability, but only the spectrum with highest median signal-to-noise ratio (S/N) was used in sample statistical analysis, which is listed in Table 1. The linear size and radio luminosity at 5 GHz in Table 1 were collected from the literature.
\subsection{Excluding blazar-type objects }
It is generally believed that young radio AGNs have the least variability in the total flux density when compared with flat-spectrum radio sources \citep{od98}. Moreover, GPS and CSS quasars are not as strongly beamed as the core-dominated quasars because the optical polarization of GPS/CSS radio sources are very low in comparison to core-dominated quasars \citep{od98}. However, it should be noticed that a large portion of these sources studied in the literature have limited radio data and they may be contaminated by blazars, which can be temporarily inverted in their radio spectrum. Simultaneous long-term radio observations at several frequencies are needed to search genuine GPS and HFP radio sources.

The source nature of HFP/GPS sources has been explored in various works. \cite{2007A&A...475..813O} and \cite{2008A&A...479..409O} carried out simultaneous multi-frequency VLA observations to investigate the variability and polarization properties of bright HFPs in \cite{Dallacasa2000}. \cite{mo10} and \cite{mo12} studied the variability and morphology of faint HFPs in \cite{S09}, while systematic multi-frequency studies of GPS radio sources were shown in \cite{2012A&A...544A..25M}. We checked the GPS and HFP objects of our parent and optical sample with these works. We find 77 (14 $\%$) objects out of the parent sample and 21 (14 $\%$) sources in optical sample show significant flux density and spectral variability, or do not preserve the convex spectrum all the time with flat radio spectrum at various epochs. These sources are most likely blazar-type sources and should be excluded when studying genuine young radio AGNs properties.

\subsection{Final sample}
After removing the blazar-type sources from the parent and optical sample, it results in 468 and 126 sources in final parent and optical sample, respectively.

The detailed information of the final optical sample are presented in Table 1, including 82 CSS, 19 GPS, 19 HFP and 6 CSO for 71 QSOs and 55 galaxies according to SDSS classification. We retained all spectra for those objects with multiple spectroscopic observations in order to analyse their optical variability, but only the spectrum with highest median signal-to-noise ratio (S/N) was used in sample statistical analysis, which is listed in Table 1.
%The LS and radio luminosity at $5 GHz$ in Table 1 were collected from literature.
%The linear size and radio luminosity at 5 GHz in Table 1 were collected from the literature

The redshift distribution of our sample is shown in Figure 1, from which the redshift of galaxies range from 0.025 to 0.730, and quasars cover broad range of 0.077 $-$ 3.594.

\begin{figure}
\centering
%\plotone{redshiftnew.eps}
\includegraphics[width=3.4in]{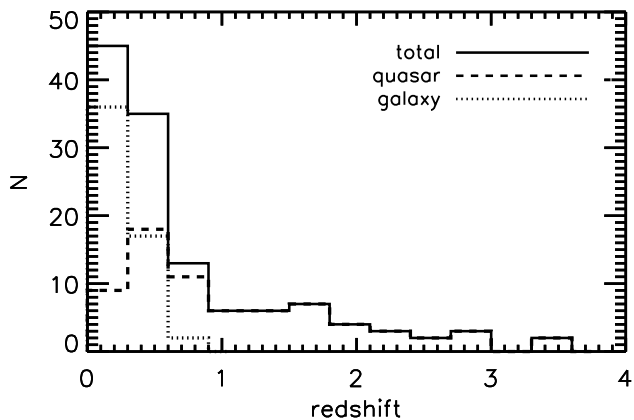}
\vspace*{-0.2 cm}\caption{The redshift distribution of our sample. The dotted, dashed and solid lines represent galaxies, quasars and whole sample, respectively.}
\vskip-10pt
\end{figure}

\subsection{Two additional GPS sources}
  
Many young radio AGNs do not have optical spectral data in SDSS or other archives. In order to increase available spectrum, we observed two GPS sources HB89 1127$-$145 and CGRaBS J1424+2256 on May 11 and 12, 2012, respectively, by using the Yunnan Faint Object Spectrograph Camera (YFOSC) installed on Lijiang 2.4m telescope at Yunnan Observatory \citep{2010gfss.conf...63B}, Chinese Academy of Sciences. Two objects were observed by using G8 grating with a dispersion of 1.5 \AA\ $ \rm pixel^{-1}$, and a wavelength coverage of 4970 - 9830 \AA, and the exposure time is 1800 seconds for each target. The spectra were reduced in IRAF following standard process, including bias subtraction, flat field correction and cosmic ray removal. Neon and helium lamps were used for wavelength calibration. The spectra were flux calibrated using the standard stars observed on the same night. The spectra of two GPS sources are shown in Figure 2, from which Mg II, C IV and Ly $\alpha$ lines are clearly seen.

For this work, these two sources were not used in following statistic analysis to avoid systematic bias between Lijiang telescope and SDSS. We just presented the spectra information for HB89 1127$-$145 and CGRaBS J1424+2256, including the emission lines measurements (see details in Table 2), black hole mass and Eddingtion ratio estimation (see details in Table 1).
  
%Following the method in Section 3 and 4.1, we measured emission lines, and then estimated the black hole mass and Eddington ratio for two sources. In HB89 $1127-145$ $(z=1.184)$, the broad Mg II can be well modelled with one Gaussian, with FWHM = 3676.62 $\rm km\ s^{-1}$ and its luminosity is 1.2$\times$ 10$^{44}$ $\rm erg\ s^{-1}$. These result in a black hole mass of log $M_{\rm BH}$ = 8.78 $\rm M_{\sun}$, and bolometric luminosity of 3.04$\times$ 10$^{46}$ $\rm erg\ s^{-1}$, indicating Eddington ratio of log $R_{\rm edd} =  -0.41$. In CGRaBS J1424+2256 ($z$=3.620), the FWHM is 5461.37 $\rm km\ s^{-1}$ and the luminosity is 1.07$\times$ 10$^{46}$ $\rm erg\ s^{-1}$ for broad C IV line. Based on these measurements, we estimated black hole mass as log $M_{\rm BH}$ = 9.55 $\rm M_{\sun}$, and the bolometric luminosity is 9.4$\times$ 10$^{47}$ $\rm erg\ s^{-1}$ resulting log $R_{\rm edd}$ = 0.30. CGRaBS J1424+2256, thus is comparable to the sources at highest redshift, and with the most massive black hole and largest Eddington ratio in our SDSS sample.

   \begin{figure*}
   \centering
   \centering
   \includegraphics[width=3.4in]{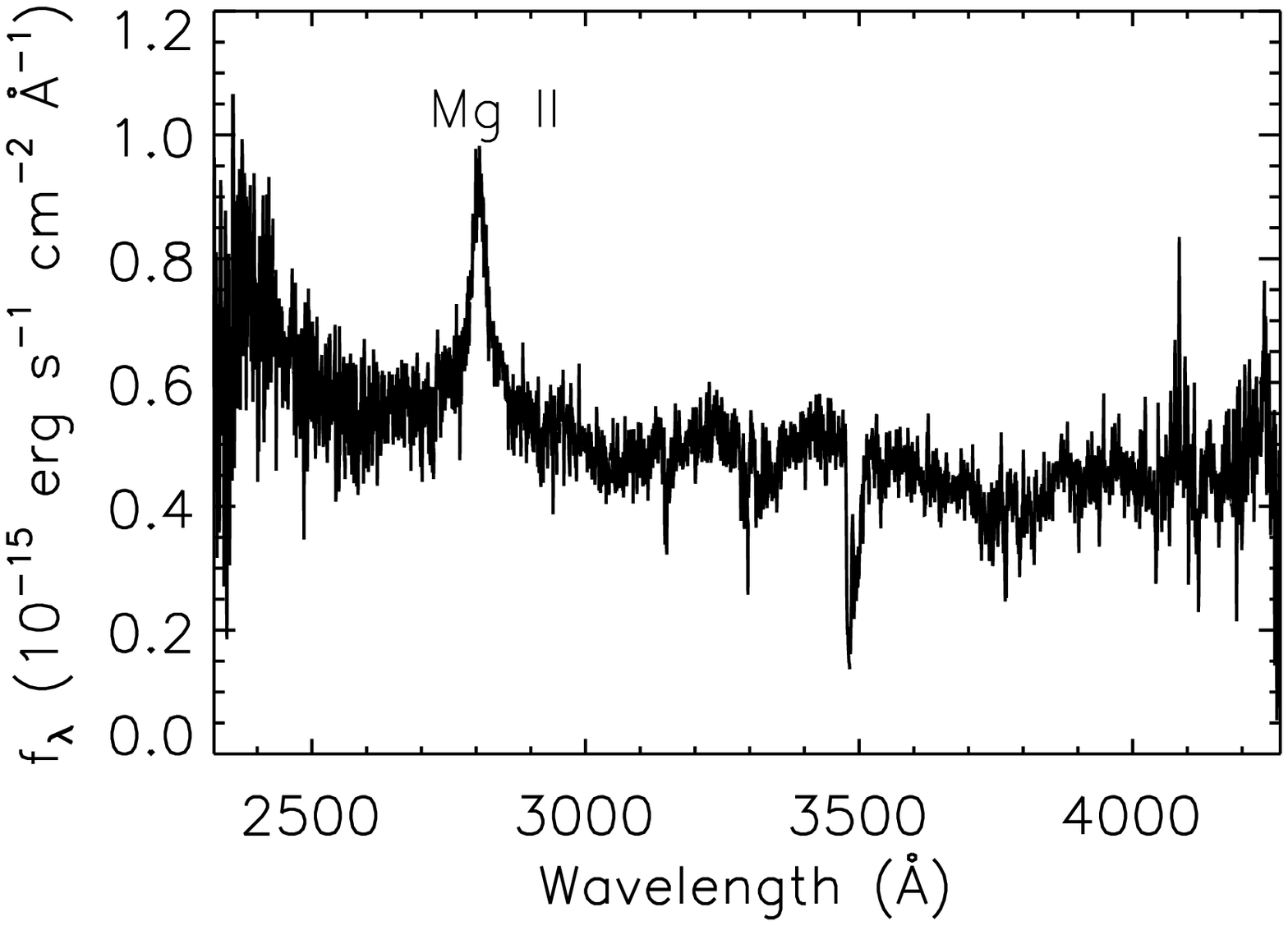}
   \includegraphics[width=3.4in]{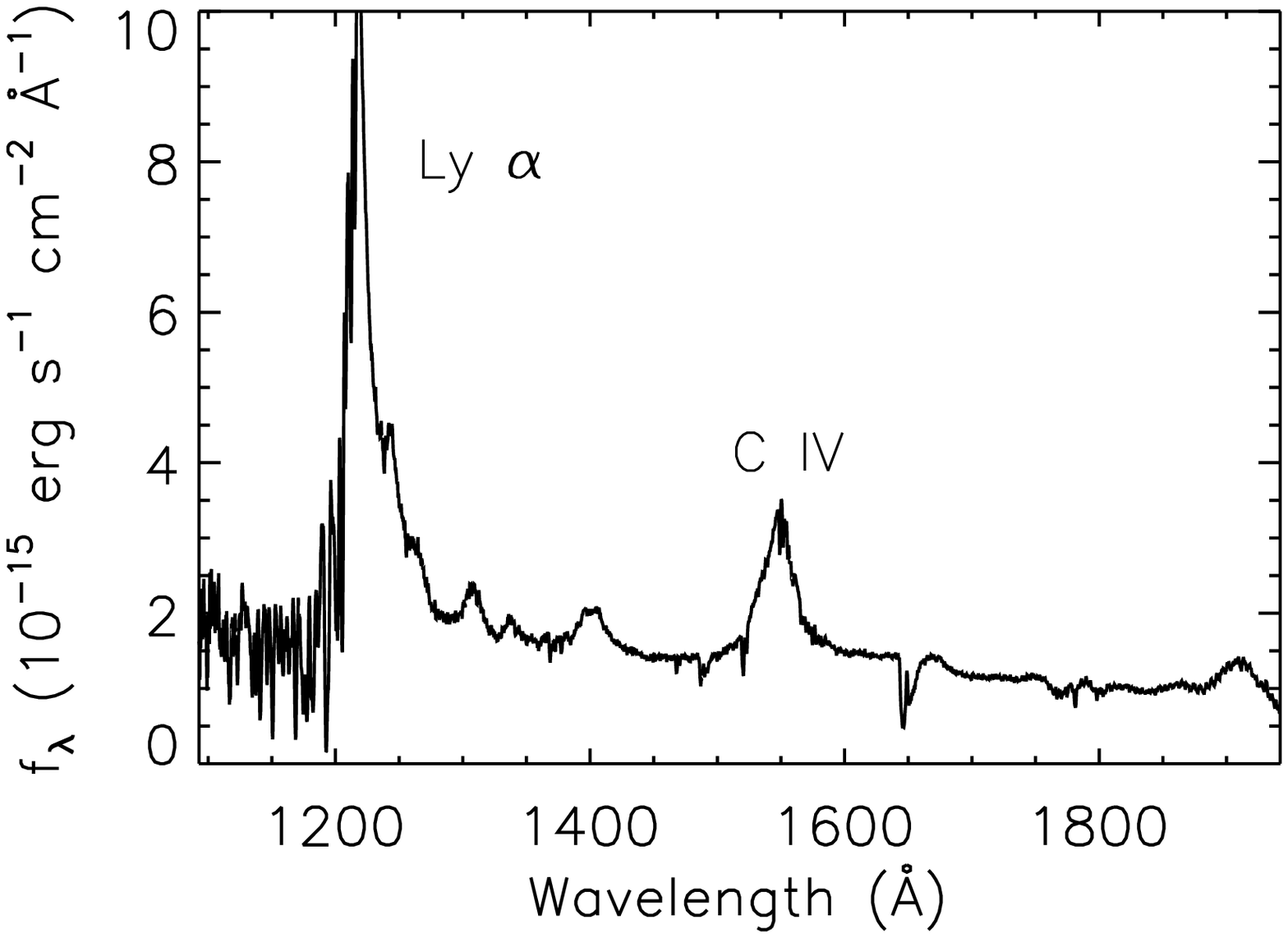}
   %\plotone{BH.eps}
   
   %\gridline{\fig{BH.eps}{{0.5\textwidth}}{(a)}
            % \fig{Edd_ratio.eps}{0.5\textwidth}{(b)}}
   \vspace*{-0.2 cm}\caption{The spectra of HB89 1127-145 ($left$) and CGRaBS J1424+2256 ($right$) observed using Lijiang 2.4m Telescope. Two spectra were corrected for Galactic extinction and shifted to source rest-frame, with main emission lines indicated. \label{fig11}}
   \vskip-10pt
   \end{figure*}

\section{Spectroscopic analysis} \label{sec:floats}
The SDSS spectra were firstly corrected for Galactic extinction with the reddening map of \cite{1998ApJ...500..525S}, and then were shifted to the rest-frame wavelength by using the redshift from the header of SDSS spectra. The spectra of galaxies and low-redshift quasars in our sample may be contaminated by the host galaxy starlight, which should be removed in order to study AGN emission. Firstly, 
we used the 4000 \AA\ break index $D_{\rm 4000}$ (defined as the flux ratio in 4000$-$4100 \AA\ to 3850$-$3950 \AA) introduced by \cite{1999ApJ...527...54B} where $D_{\rm4000}$ = $\int^{4100}_{4000}f_{\lambda}d\lambda/\int^{3950}_{3850}f_{\lambda}d\lambda$, to evaluate the significance of stellar component. For those sources with $D_{\rm 4000}$ $\le$ 1, the contribution of host galaxy can be ignored and the spectra is dominated by AGN nuclei \citep{2005ApJ...620..629D,wang09}. 
In 55 galaxies, the $D_{\rm 4000}$ of all but two sources are greater than 1, while it is the case in 16 out of 71 quasars. 

In this work, we mainly focus on the measurements of emission lines, i.e., H$\alpha$, H$\beta$, Mg II and C IV broad lines and [O III], [O II], [S II] narrow lines. To ensure the reliability of line measurements, we used the ratio of the integral line flux for the whole profile to the rms deviation around these lines to determine if the spectra is valuable to be analyzed for the lines, as some spectra have rather low S/N and the emission lines are not significant from visually inspection. The rms deviation were calculated by using the pixel-to-pixel deviation in the continuum-subtracted spectrum (see details in Section 3.1) in the spectral region surrounding the relevant lines. We found a significant correlation between the flux ratio and S/N for all our spectra. The median value of flux ratio is about 70 \AA for the sources with S/N $<$ 4. The sources above 70 \AA thus correspond to those with S/N $\ge4$ in general. Tentatively in this work, the spectra with the flux ratio of all the focused lines to the rms deviation of the continumm below 70 \AA were not further analyzed. %for the spectrum fitting\footnote{The limited value of 70 seems a little arbitrary but it's practical for our spectrum fitting}. 
There are 16 sources meeting the criterion\footnote{ We retained the spectra of SDSS J082504.56+315957.0 and SDSS J120902.79+411559.2 due to the $D_{\rm4000}$ $>$ 1 and SN $>$ 10. We can measure the $\sigma_{\ast}$ for the spectra.}, of which all have low-quality spectra with median S/N $<$ 10 except for one objects.

According to S/N and $D_{\rm 4000}$, the remaining 110 young radio AGNs were divided into three groups: 

(A) 30 galaxies and 7 quasars, the spectra with S/N $\ge$ 10 and $D_{\rm 4000}>1$;

(B) 10 galaxies and 9 quasars, the spectra with S/N $<$ 10 and $D_{\rm 4000}$ $>$ 1;

(C) 53 quasars and 1 galaxy, the spectra with $D_{\rm 4000}$ $\le$ 1.

\subsection{Continuum subtraction}
The continuum was fitted differently depending on the significance of host galaxy and S/N.
Due to the significant host contribution in the spectra of group A sources, we used the direct-pixel-fitting code Penalized Pixel-Fitting (PPXF) of \cite{Cappellari2004} and \cite{Cappellari2017} to fit the continuum and measure the stellar velocity dispersions $\sigma_{\ast}$. We focused on the spectral range of 3540 $-$ 6900 \AA, which covers absorption lines Ca H+K, and Mg b triplet. The emission lines were masked out before the fitting. A linear combination of 156 single stellar population (SSP) templates \citep{vazdekis2010} from MILES library \citep{sancetal06lib} with a broadening by line-of-sight velocity distribution (LOSVD) was applied by performing a $\chi^2$  minimization. %Because mainly concerning the line measurement in this work, except using combined SSP templates, we also utilize function of additive Legendre polynomial in the code to correct the template continuum shape. We can get good and reasonable fitting results from that practice notwithstanding the AGN continuum is cannot completely ignore which is rather weak compared to host component.
The stellar templates give good fit to the spectra (see example in Figure 3) for all sources, when adding a low-order Legendre polynomial on the continuum \cite[e.g.,][]{2012ApJ...757..140S}. %In some cases, a power-law AGN continuum is needed, however, the AGN contribution is rather weak compared to host component.}
 
Due to low S/N in group B and its resulting large uncertainties in extracting host component, we simply fitted the continuum around individual emission lines with a single power-law.

For 54 sources in group C, the continuum is dominated by AGN emission. Therefore, a single power-law and Fe II emission, were applied to fit local continuum for H$\alpha$, H$\beta$, Mg II (see Figure 4), with the fitting windows similar to \cite{Shen et al.(2011)}. In general, Fe II lines are very weak nearby C IV, therefore, only a single power-law was employed to fit C IV, as did in \cite{Shen et al.(2011)}.

\subsection{Emission line measurements}

We fitted the emission lines using Gaussian profiles on the continuum-subtracted spectra. All the narrow line components of H$\alpha $, [N II] $\lambda\lambda6548,6584$, [S II] $\lambda\lambda6716,6731$, H$ \beta$, [O II] $\lambda 3727$ and Mg II were modelled with a single Gaussian component. The [O III] $\lambda\lambda4959,5007$ were fitted using one or double Gaussian profiles, when necessary in cases of complex shapes, i.e., asymmetric blue or red wings. The broad components of H$\alpha$, H$ \beta $, and Mg II  were fitted with one or multiple Gaussian profiles (up to three). We fitted C IV with three broad Gaussians following \cite{Shen et al.(2011)} despite some attempts also considered the narrow component in the literature \citep{2004ApJ...617..171B,2007ApJ...666..757S}. We imposed an upper (lower) limit on the line width of narrow (broad) components with FWHM $<1000\ \rm km\ s^{-1}$ ($\ge1000\ \rm km\ s^{-1}$).
 
In this work, the line region of H$\alpha$ and H$\beta $, Mg II and C IV were fitted individually. The line width and offset of narrow lines H$\alpha$, and [N II] $\lambda\lambda6548,6584$ were constrained to be same. The same treatments were applied for narrow H$\beta $ and the line core of [O III] $\lambda\lambda4959,5007$. %The width and offset of narrow H$\beta $ are tied to those of core [O III] $\lambda5007$\footnote{In few cases, we free the width and offset of narrow H$\alpha$ and H$\beta$ in order to achieve better and more reasonable fitting results for these two lines.}. 
The line width and offset of [S II] $\lambda\lambda6716,6731$ were tied to each other. The flux ratio of [N II] doublet was fixed to be 2.96 and [O III] doublet was constrained to be 3 \citep{Shen et al.(2011)}. 
% The width and offset of the narrow component of H$\alpha$ [N II] $\lambda\lambda6548,6584$ are tied to each other

The $\chi^2$-minimization was used to obtain the best fit for the continuum and emission lines. To estimate the errors in the measured quantities, we generated 100 mock spectra by adding a random Gaussian noise to original spectra using the flux density errors, and then estimated the  uncertainty as the standard deviation of measurements from those mock spectra. The examples of spectral fitting are presented in Figure 3 and Figure 4. The measured flux and line width of studied emission lines are displayed in Table 2. All the widths of the narrow lines were corrected by the instrumental resolution of SDSS using equation of $\rm FWHM_{\rm intrinsic} = \sqrt {{\rm FWHM_{\rm observed}^2} - {\rm FWHM_{\rm instrumental}^2}}$.

\begin{figure*}
	\centering
\includegraphics[width=6in]{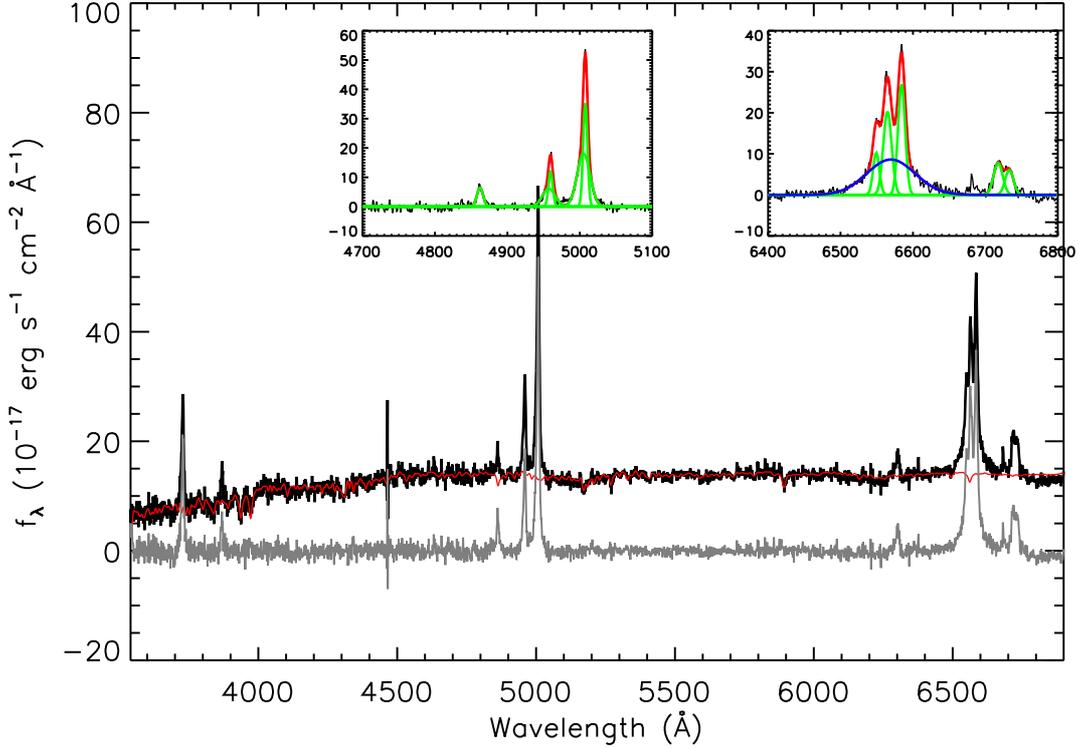}

\vspace*{-0.2 cm} \caption{The spectral fitting of SDSS J124419.96+405136.8 ($z=0.249$) as an example of source in group A, where the significant contribution of host galaxy on the continuum was fitted using PPXF. The black and grey lines are the original and the continuum-subtracted spectra, respectively. The red line is the fitted continuum. The two insets represent the spectral fitting around H$\alpha$ and H$\beta$ lines. The blue, green, and red lines are the fitted broad, narrow and total components, respectively.} %\textbf{ H$_2$CO is at the velocity center of spectra.}}
    \vskip-10pt
\end{figure*}

\begin{figure*}
	\centering
\includegraphics[width=6in]{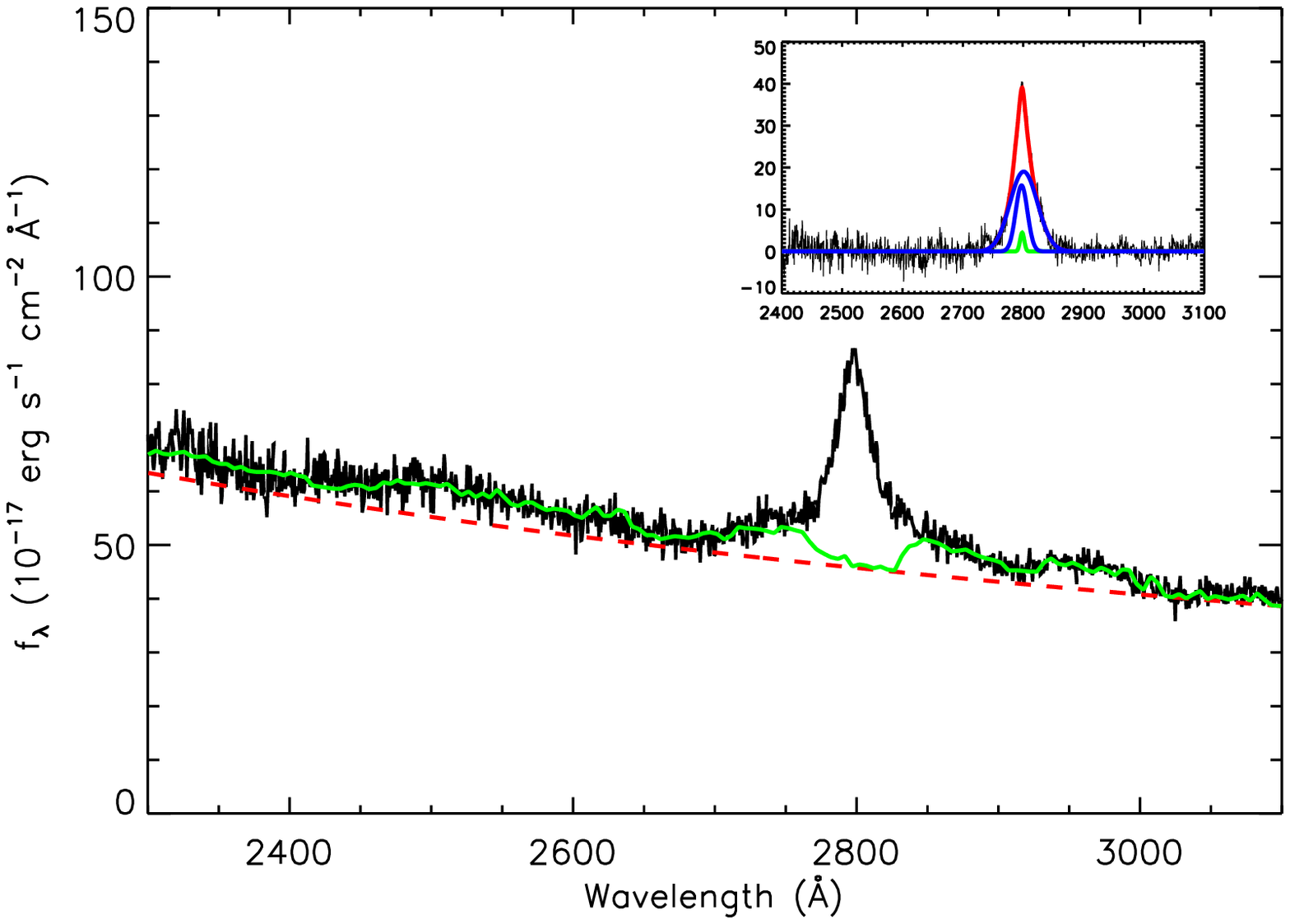}

\vspace*{-0.2 cm} \caption{The spectral fitting of SDSS J005905.51+000651.6 ($z=0.719$) as an example of source in group C, where the AGN emission is dominant and the continuum is fitted with a single power-law and Fe II emission. The black line is the original spectrum. The power-law and total fitted continuum including Fe II emission are shown with the red dashed and green lines, respectively. The inset shows the line fitting for Mg II. Lines and symbols as in Figure 3.}
    \vskip-10pt
\end{figure*}

%\subsection{Final measurements}

As mentioned earlier, the spectral fitting was carried out for 110 out of 126 young radio AGNs, consisting of 69 quasars and 41 galaxies. Our spectral analysis shows that 57 quasars have broad emission lines (FWHM $\ge$ $1000\ \rm km\ s^{-1}$)\footnote{It should be noticed that our definition of broad line is stricter than that of SDSS classification, i.e., any lines with FWHM $>500\ \rm km\ s^{-1}$},  % with four sources (SDSS J094525.90+352103.6, SDSS J124419.96+405136.8, SDSS J134733.36+121724.2, SDSS J151141.26+051809.2) have $\sigma_{\ast}$ measurements.
while only narrow emission lines were detected in the remaining 12 quasars, of which three objects have $\sigma_{\ast}$ measurements (i.e., SDSS J102844.26+384436.8, SDSS J140942.44+360415.9, and SDSS J161148.52+404020.9). In comparison, the broad emission lines were found in two galaxies (i.e., SDSS J090615.53+463619.0 and SDSS J160246.39+524358.3). %have  in group A and there are no broad line for galaxies in group B and C. 

\section{Results}
\subsection{Narrow lines versus $\sigma_{\ast}$}
  
It is well known that there is a close correlation between the BH mass and the stellar velocity dispersion ($\sigma_{\ast}$) of the bulge in nearby normal galaxies. This relation is a key to understand the formation and evolution of galaxies \citep{2000ApJ...539L...9F,2002ApJ...574..740T,KormendyHo2013,2013ApJ...764..184M}. However, usually $\sigma_{\ast}$ is difficult to measure in bright AGNs, and thus the line width of narrow line [OIII] was frequently used as a surrogate for $\sigma_{\ast}$ because the gravity of bulge dominates the global kinematics of Narrow Line Region (NLR) in AGNs \citep{nels96,boroson03,GH05,komossa07}. The $\sigma_{\rm [OIII]}$ has been commonly used as a surrogate for the $\sigma_{\ast}$ of host bulge in AGNs after removing wing component and there is a strong relationship between them. \citep{komossa07,2011ApJ...739...28X,craccoetal16}. This can be tested for our sample when both $\sigma_{\ast}$ and [O III]$\lambda5007$ line width are directly measured.
 
 %Only six objects both have broad emission line and $\sigma_{\ast}$ in our sample, so the number of the objects is not enough to allow us to directly investigate the relation of M - $\sigma_{\ast}$
 
We only considered those spectra with high S/N $\geq$ 10, and line width measurements of [O III] and [S II] at high significance ($>3 \sigma$). The relationship between the stellar velocity dispersion ($\sigma_{\ast}$) and the velocity dispersion of [O III] line core ($\sigma_{\rm [OIII]}$=FWHM$_{[\rm OIII]}$/2.35), and [S II] velocity dispersion ($\sigma_{\rm [SII]}$=FWHM$_{[\rm SII]}$/2.35) are shown in Figure 5. The correlation analyses show that the velocity dispersion of [O III] core component has a strong correlation with $\sigma_{\ast}$ for 33 sources, with a Spearman rank correlation coefficient $r_{\rm s}$ = 0.65 and the probability $P_{\rm null}$ less than  $10^{-4}$ for the null hypothesis of no correlation, while $\sigma_{\ast}$ correlates less strongly with velocity dispersion of [S II] ($r_{\rm s}$ = 0.38, $P_{\rm null}$ = 0.03) for 33 objects. This suggests that [O III] can be a better surrogate than [S II] at least in our sample of young radio AGNs, which is supported by the mean values of $-0.002\pm0.110$, and $0.035\pm0.120$ for log($\sigma_{\rm [OIII]}/\sigma_{\ast}$), and log($\sigma_{\rm [SII]}/\sigma_{\ast}$), respectively (see right panels in Figure 5). 

In our work, the strong blueshifted wing of [O III] was found in young radio AGNs and the width of [O III] tends to be broad \citep{Gelderman1994,od02,ho08,Kim2013}. Such blueshifted/broad wing may be caused by the outflow or the strong interaction between jet and ambient gas or ISM when jet expands outward \citep{labianolet}. 
%The [O III] doublet was fitted with double Gaussians separately if there appears a blueshifted wing component \citep{2011ApJ...739...28X}. 
Our results show that 15 out of 68 sources with [O III] measurements present blueshifted wing. We find that the line width of [O III] core strongly correlates with $\sigma_{\ast}$, indicating it can be well used as a surrogate of $\sigma_{\ast}$ in young radio AGNs. Therefore, we can derive the $\sigma_{\ast}$ from the width of core component of [O III] ($\sigma_{\ast}$ = $\rm FWHM_{\rm [O III]}$/2.35) if there is no $\sigma_{\ast}$ measurement for the type 2 sources in our sample, to estimate their black hole mass.

However, large scatter and deviation from the equivalent line are clearly seen at $\sigma_{\rm [OIII]}>300\rm~ km~s^{-1}$. Such broad line width could be related with jet$-$ISM interaction \cite[e.g.,][]{Kim2013}. In principle, [S II] should be a better surrogate of $\sigma_{\ast}$ than [O III] because the latter may suffer from outflow/interaction. However, we found a weaker relationship between the width of [S II] and $\sigma_{\ast}$ than $\sigma_{\rm [OIII]}$ $-$ $ \sigma_{\ast}$ relation, not as expected. This may at least partly be caused by the asymmetric [S II] profile found in many of our sources, however is not fully fit in our single-Gaussian spectral fitting. Although the details are unknown, the asymmetric profile is most likely caused by the interaction between jet and ambient medium, similarly as [O III] but less significant.

%However, large scatter and deviation from the equivalent line are clearly seen at $\sigma_{\rm [OIII]}>300\rm~ km~s^{-1}$. Such broad line width could be related with jet$-$ISM interaction \cite[e.g.,][]{Kim2013}. In principle, [S II] should be a better surrogate of $\sigma_{\ast}$ than [O III] because the latter may suffer from outflow/interaction. However, we found a weaker relationship between the width of [S II] and $\sigma_{\ast}$ than $\sigma_{\rm [OIII]}$ $-\ \sigma_{\ast}$ relation, not as expected. This may at least partly be caused by the asymmetric [S II] profile found in many of our sources, however is not fully fit in our single-Gaussian spectral fitting.
%Although the details are unknown, the asymmetric profile is most likely caused by the interaction between jet and ambient medium, similarly as [O III] but less significant.
 
 \begin{figure*}
 \centering
 \includegraphics[width=3.4in]{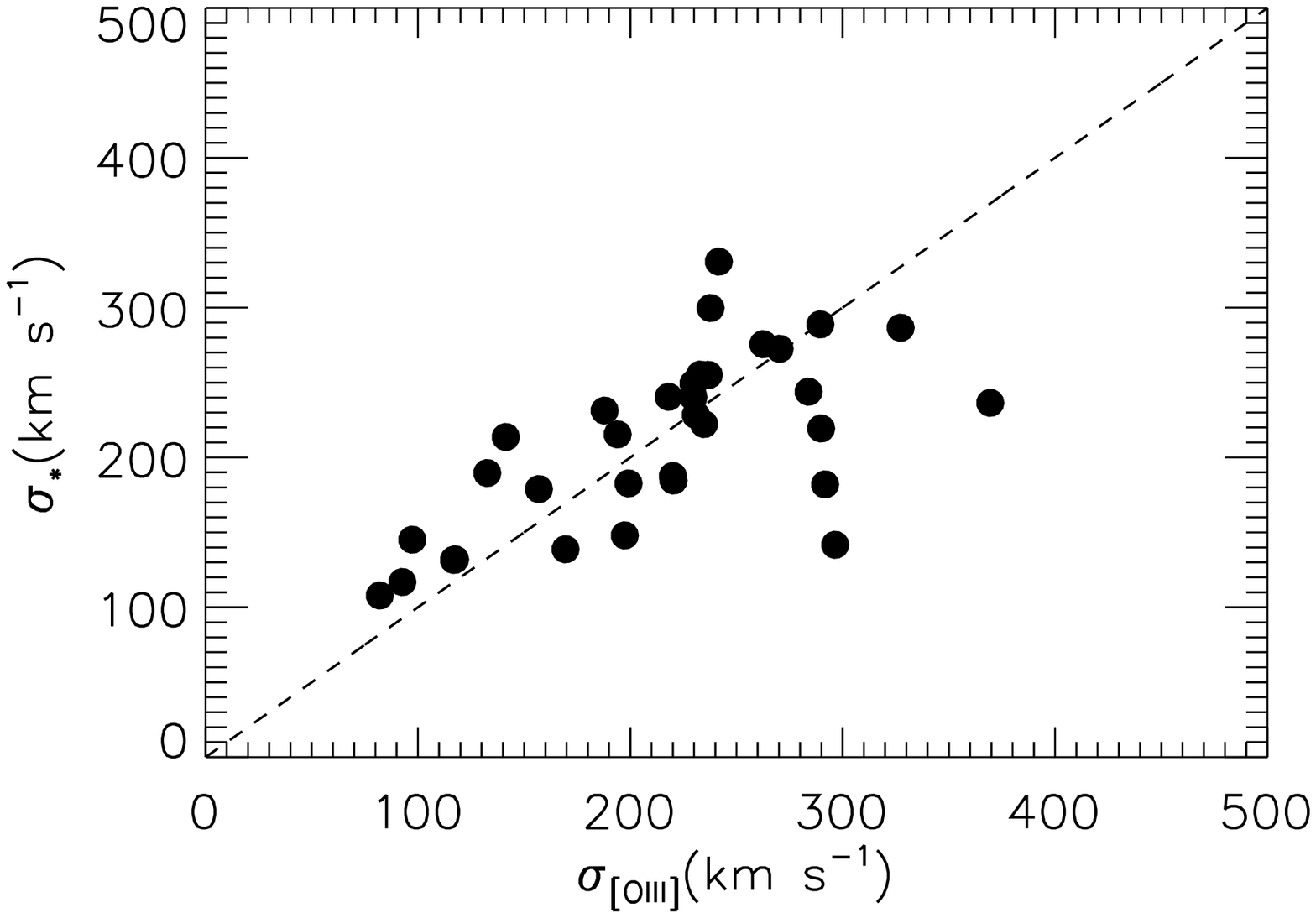}
 \includegraphics[width=3.4in]{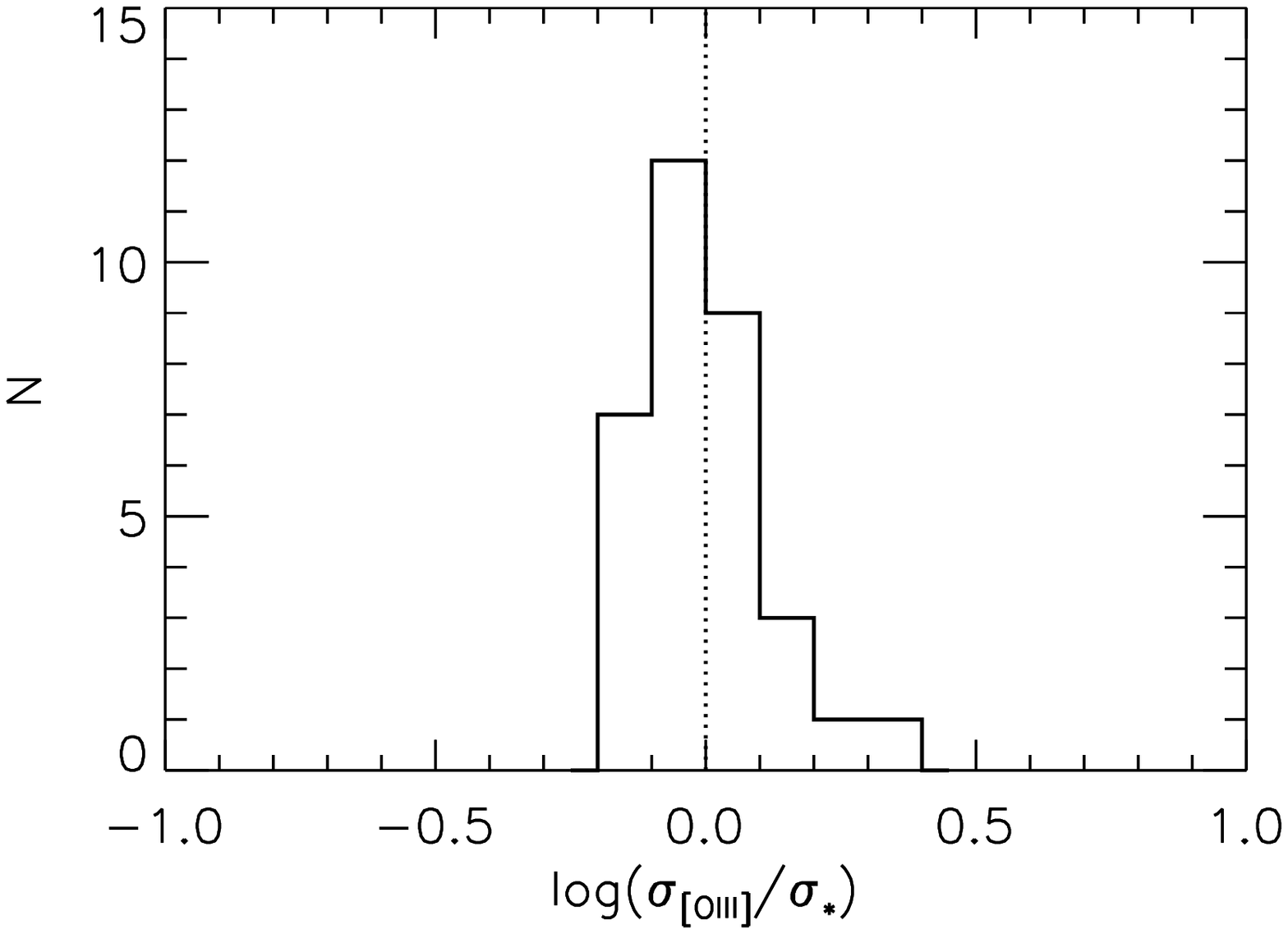}
 \includegraphics[width=3.4in]{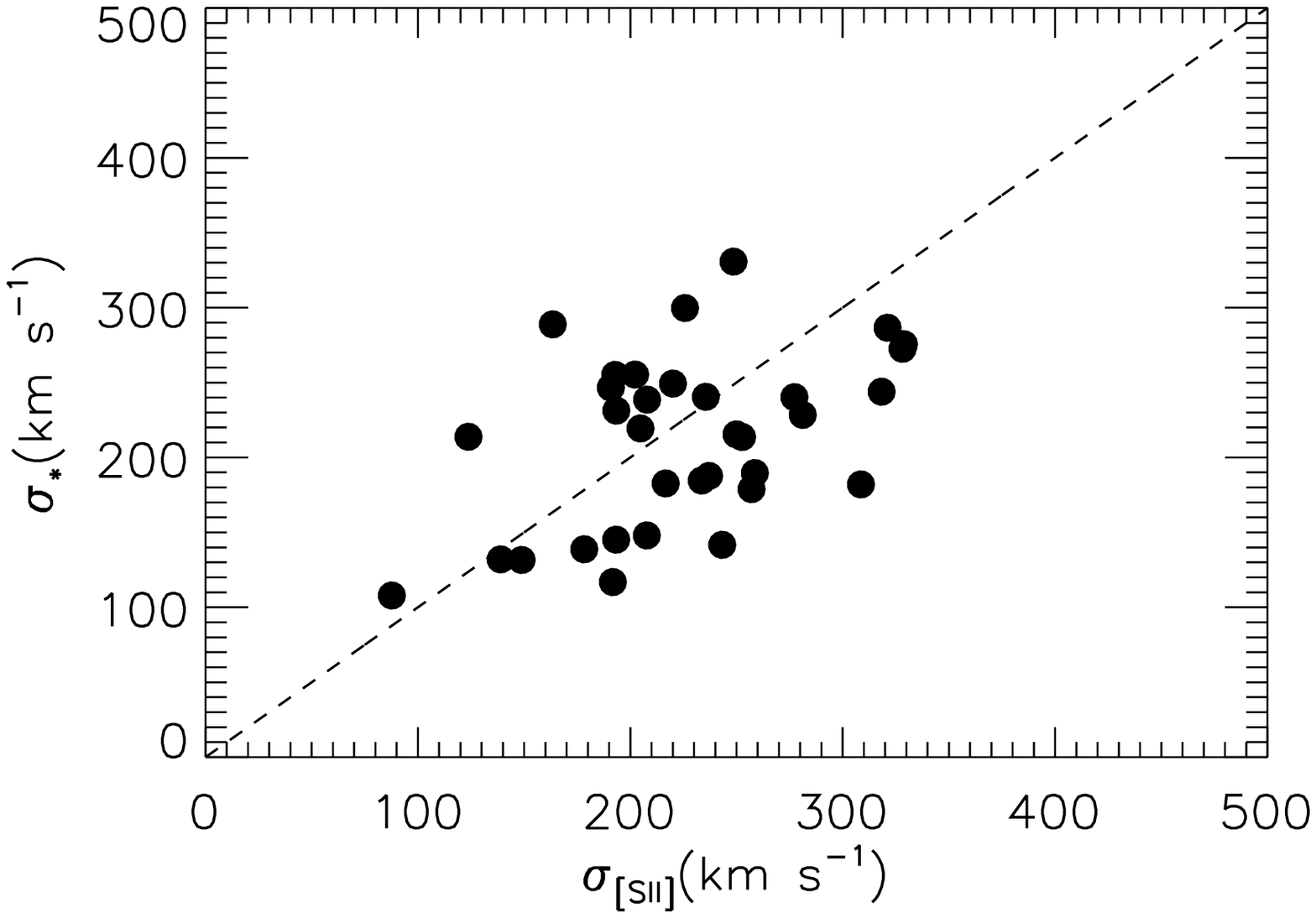}
 \includegraphics[width=3.4in]{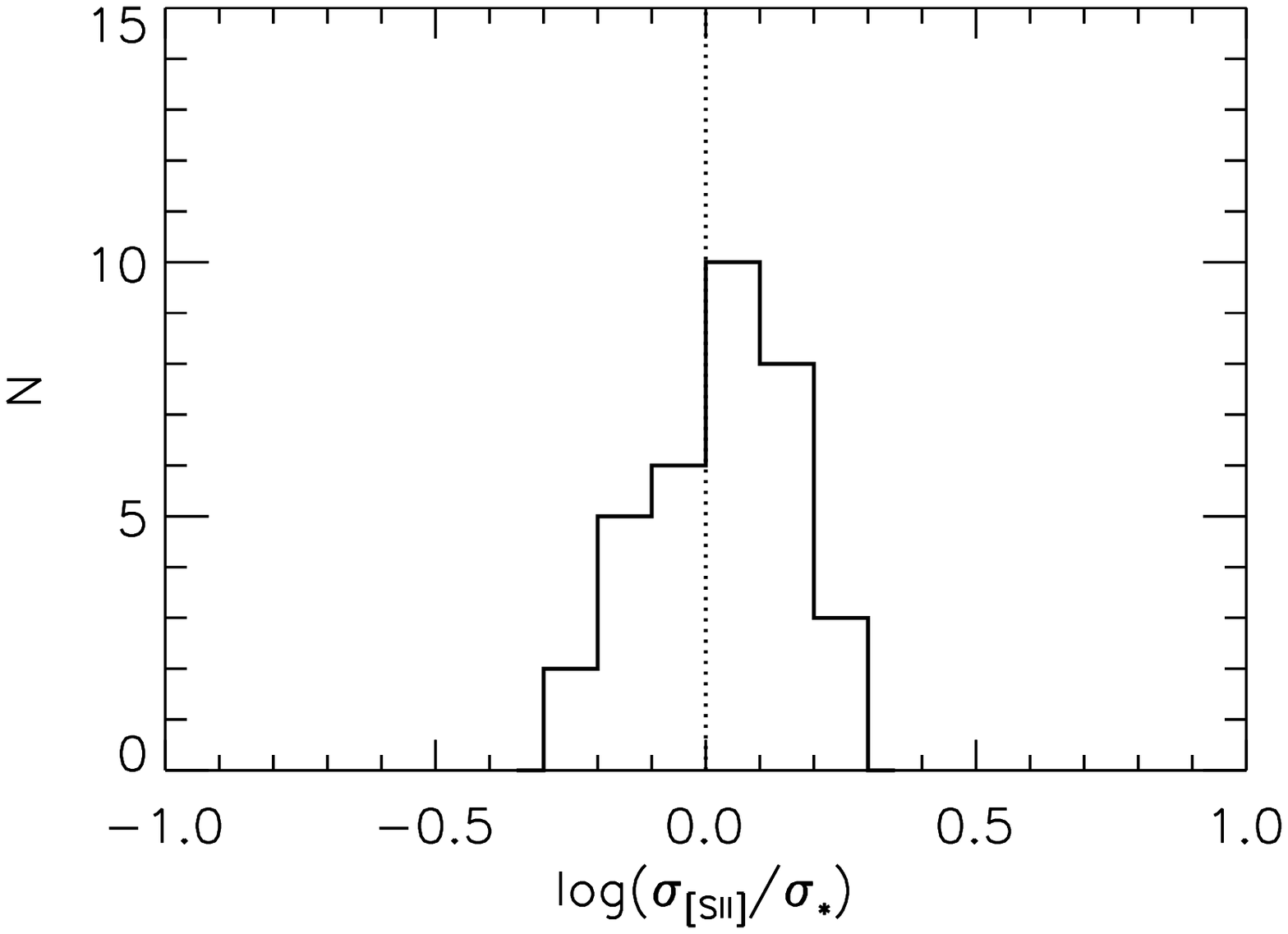}
\vspace*{-0.2 cm}\caption{$\sigma_{\ast}$ and narrow lines: $upper~ left-$ $\sigma_{\ast}$ versus $\sigma_{\rm [OIII]}$=FWHM$_{[\rm OIII]}$/2.35 of [O III] core component; $upper ~right-$ Histogram of log($\sigma_{\rm [OIII]}/ \sigma_{\ast}$); $lower~ left-$ $\sigma_{\ast}$ versus $\sigma_{\rm [SII]}$=FWHM$_{[\rm SII]}$/2.35; $lower~ right-$ Histogram of log($\sigma_{\rm [SII]}/ \sigma_{\ast}$). Dashed lines represent the 1:1 relation. The dotted vertical lines mark the mean value of zero. \label{fig6}}
\vskip-10pt
  \end{figure*}

\subsection{Black hole mass and Eddington Ratio}
The black hole mass and Eddington ratio are two 
fundamental parameters in AGN studies, which can be used to study the accretion mode (e.g., advection-dominated accretion flow or standard accretion disc), and the source position in AGN evolution (e.g., NLS1s) \citep{2006komossa1}.
\subsubsection{Black hole mass estimation}
Young radio AGNs are radio-loud sources, therefore their optical continuum emission may be more or less affected by the synchrotron emission from jets (e.g., in GPS and HFP) \citep{bh95}. This can be studied by directly comparing the relationship between the continuum luminosity and the broad emission line luminosity in our sources with that of radio-quiet AGNs, in which the relativistic jets are absent or weak. The comparisons are shown in Figure 6 for $L_{5100 \rm \AA}$ and $ L_{\rm H\beta}$, $L_{3000 \rm \AA}$ and $ L_{\rm Mg II}$, and $L_{1350 \rm \AA}$ and $ L_{\rm C IV}$, with the data of radio-quiet AGNs taken from the literature \citep{Vestergaard06,ko06}. The dashed lines in Figure 6 show the linear fit to the relation between continuum luminosity to line luminosity for radio-quiet AGNs. While radio-quiet quasars distribute tightly around the line, our young radio AGNs show a large scatter with most objects below the line. This indicates that the continuum luminosity of our sample is larger than that of radio-quiet AGNs at the fixed emission line luminosity, and the deviation can even be larger than 0.5 dex in some cases. The K-S test was also appiled to check whether the difference between the continuum and line luminosity is significant or not. The test results show that the difference is significant with P-value of 0.01, 0.001, 0.002 for the ratios of line-to-continuum luminosity in $ L_{\rm H\beta}$/$L_{5100 \rm \AA}$, $ L_{\rm Mg II}$/$L_{3000 \rm \AA}$ and $ L_{\rm C IV}$/$L_{1350 \rm \AA}$ between our and radio-quiet sources, respectively,  implying that the continuum emission is likely contaminated by the jet emission in our sample.  %The evident deviations below the dashed line can be seen in H$\beta$  Mg II and C IV (a,b,c panel in Figure 4.).

\begin{figure*}
	\centering
\includegraphics[width=3.4in]{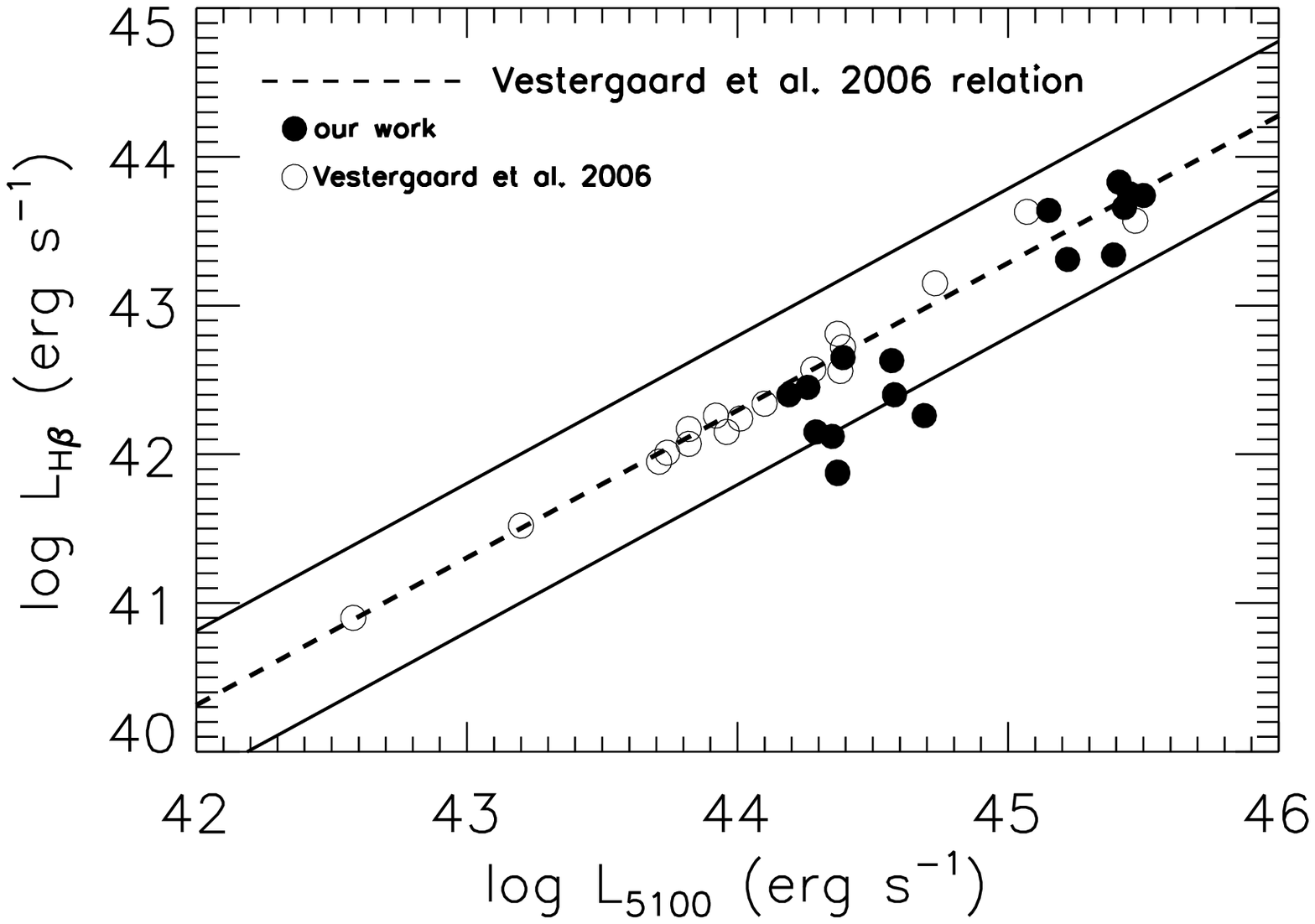}
\includegraphics[width=3.4in]{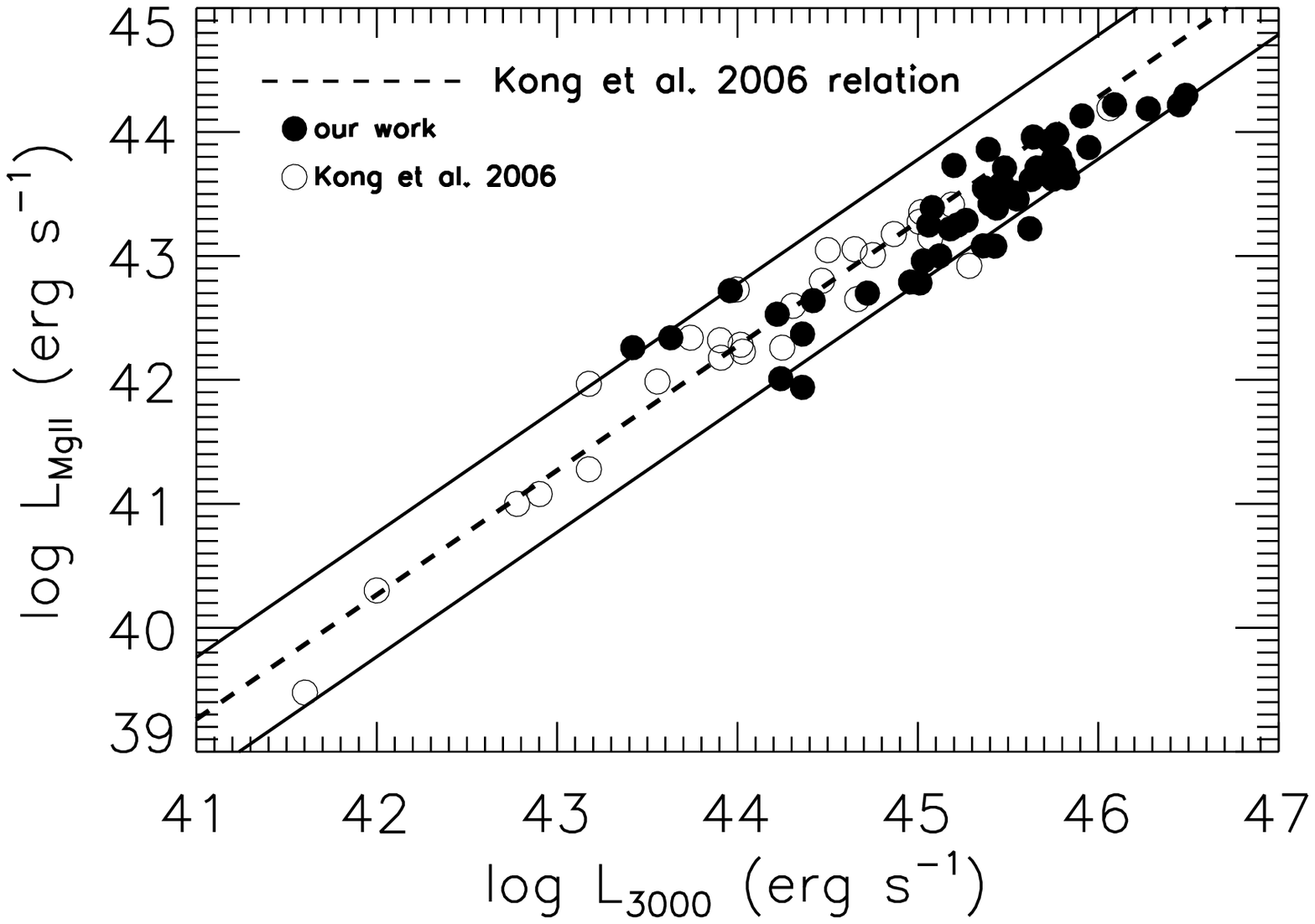}
\includegraphics[width=3.4in]{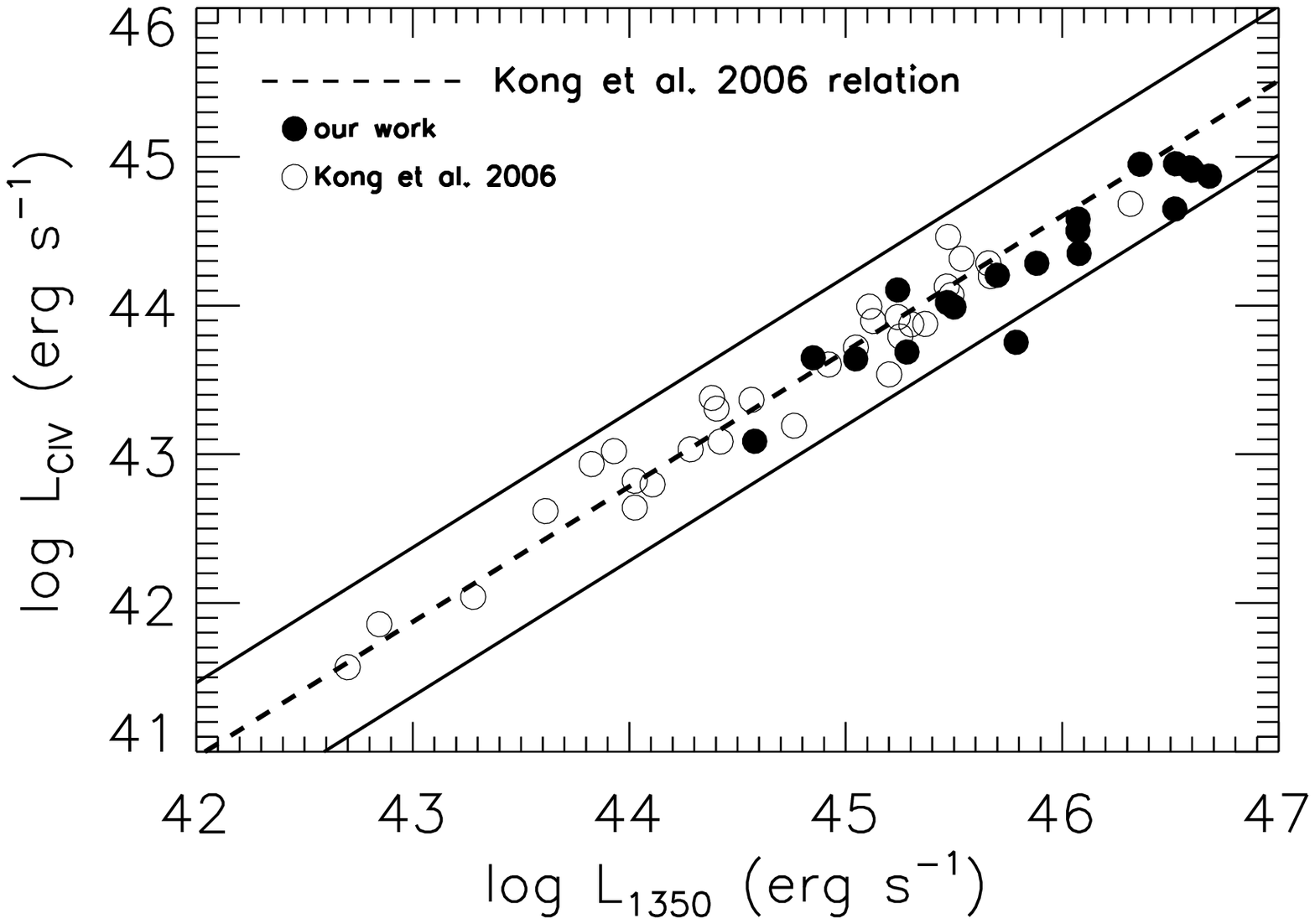}
\vspace*{-0.2 cm} \caption{The luminosities of broad H$\beta$, Mg II, and C IV versus the continuum luminosities at 5100 \AA, 3000 \AA, and 1350 \AA, respectively. The filled circles are the sources in our work, while open circles denote the radio-quiet quasars in the literature. The dashed lines represent the linear fit between the continuum and broad line luminosities for radio-quiet quasars in the literature. The solid lines show 0.5 dex deviation from the dashed lines.}
    \vskip-10pt
\end{figure*} 

\begin{figure*}
	\centering
	%\plotone{redshiftnew.eps}
	\includegraphics[width=6.0in]{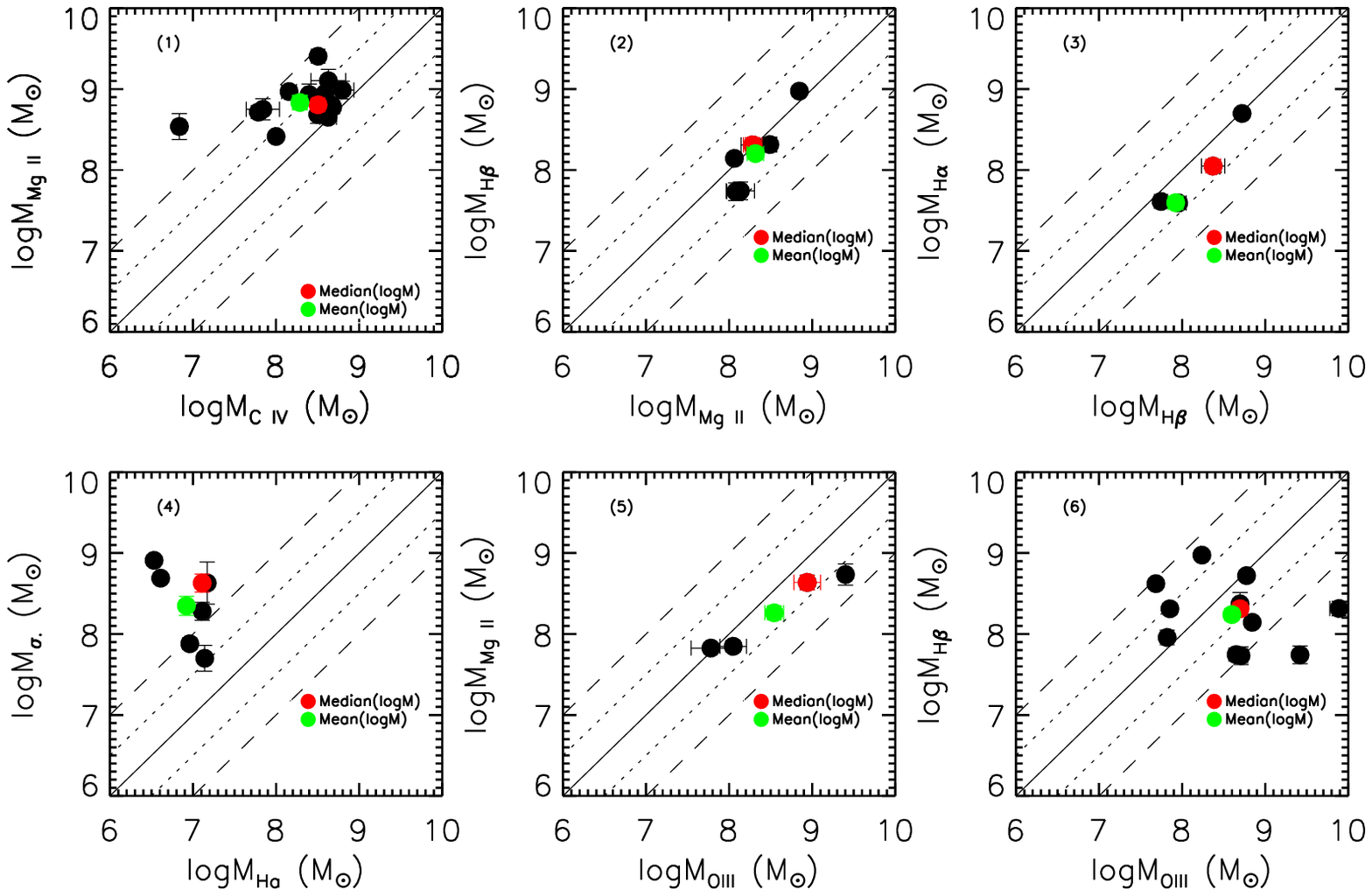}
	\vspace*{-0.2 cm}\caption{Black hole mass comparisons between various broad lines, or between broad line and $\sigma_{\ast}$, or between broad lines and [O III] lines. The red circles and the green circles are median value and mean value, respectively. Solid lines represent the 1:1 relation. The dotted lines and the dashed lines show 0.5 dex and 1 dex deviation from the solid lines, respectively.}
	\vskip-10pt
\end{figure*}

Due to the likely presence of jet emission in the continuum, we used the line width and luminosity of various broad emission lines to calculate the black hole mass instead of continuum luminosity in order to avoid overestimation. %\footnote{The priority ranking of the broad lines used for BH masses estimation is H$ \beta $ (H$\alpha$), Mg II, C IV. When H$ \beta $, H$\alpha$ both exit, we use the H$ \beta $ preferentially.}. 
The empirical relations used for H$\alpha$ and H$ \beta $ \citep{gh005,Vestergaard06}, and Mg II and C IV \citep{ko06} are as following:
\begin{equation}
 M_{\rm BH}=1.3\times10^{6} \left(\frac{L_{\rm H\alpha}}{10^{42}\
\rm ergs\ s^{-1}} \right)^{0.57} \left(\frac{\rm FWHM_{\rm
H\alpha}}{10^{3}\ \rm km\ s^{-1}} \right)^{2.06} \rm M_{\sun}
\end{equation}

\begin{equation}
M_{\rm BH}=4.7\times10^{6} \left(\frac{L_{\rm H\beta}}{10^{42}\
\rm ergs\ s^{-1}} \right)^{0.63} \left(\frac{\rm FWHM_{\rm
H\beta}}{10^{3}\ \rm km\ s^{-1}} \right)^{2} \rm M_{\sun}
\end{equation}

\begin{equation}
M_{\rm BH}=2.9\times10^{6} \left(\frac{L_{\rm Mg II}}{10^{42}\
\rm ergs\ s^{-1}} \right)^{0.57} \left(\frac{\rm FWHM_{\rm Mg
II}}{10^{3}\ \rm km\ s^{-1}} \right)^{2} \rm M_{\sun}
\end{equation}

\begin{equation}
M_{\rm BH}=4.6\times10^{5} \left(\frac{L_{\rm C IV}}{10^{42}\
\rm ergs\ s^{-1}} \right)^{0.60} \left(\frac{\rm FWHM_{\rm C
IV}}{10^{3}\ \rm km\ s^{-1}} \right)^{2} \rm M_{\sun}
\end{equation}

Due to the torus obscuration (Son 2012, Berton 2015) (see details panel (4) in Figure 7 of BH mass comparison between broad H$\alpha$ and $\sigma_{\ast}$ with more than one order of magnitude difference in median values), we preferred to use $\sigma_{\ast}$ with the relation of $M_{\rm BH}$ - $\sigma_{\ast}$ in \cite{KormendyHo2013} to get the black hole mass for the sources both with broad line and $\sigma_{\ast}$ measurements:
\begin{equation}
\log\left( \frac{M_{\mathrm{BH}}}{\rm M_{\sun}} \right)=8.49 + 4.38 \times \log \left( \frac{\sigma_{*}}{200\, \mathrm{km\, s^{-1}}} \right)
\end{equation}

For the sources without significant stellar absorption but with broad lines, we used the broad lines to estimate the BH mass (empirical relations (1)-(4)). In this situation, when more than one broad line is available in the source, we preferred a priority sequence of H$ \beta $ (H$\alpha$), Mg II, C IV to estimate the black hole mass. In the sources with only [O III] $\lambda 5007$ line core available, we estimated the BH masses using the width of [O III] line core.
%If the sources without $\sigma_{\ast}$ and broad lines, we used the width of [O III] core to derive  $\sigma_{\ast}$, to get their BH masses.}

The method used to estimate the black hole mass for each source is listed in Column 11 of Table 1.

We investigated for the sources with both broad lines and $\sigma_{\ast}$, or both broad lines and [O III] lines, or various broad lines simultaneously (C IV and Mg II, Mg II and H$\beta$, H$\beta$ and H$\alpha$) to check the consistence of black hole masses estimation. Our results in Figure 7 show that the median values of black hole mass estimation from various methods are generally consistent with each other within the typical uncertainty of 0.5 dex except for broad $\rm H\alpha$ - $\sigma_{\ast}$, supporting broadly consistent results when using different methods. 
However, we noticed that the black hole masses estimated from C IV are systematically lower those from Mg II, which is apparently different from Mg II - $\rm H\beta$ comparison. Indeed, the median values of log ($M_{\rm BH}^{\rm MgII}$/$M_{\rm BH}^{\rm C IV}$) and log ($M_{\rm BH}^{\rm H\beta}$/$M_{\rm BH}^{\rm Mg II}$) are 0.42 dex and 0.03 dex, respectively. This may be related with the non-virialized component, usually blueshifts in C IV line, as suggested in \cite{2008ApJ...680..169S}. The authors found that log ($M_{\rm BH}^{\rm MgII}$/$M_{\rm BH}^{\rm C IV}$) is correlated with the C IV - Mg II blueshift, and the C IV estimator tends to give smaller virial masses than the Mg II estimator for objects with small blueshifts ($\leqslant$1000~km/s) but larger C IV masses for larger blueshifts. 
Consistently, we found that the available C IV - Mg II blueshifts in six of our sources from \cite{2008ApJ...680..169S} are around or less than 1000 km/s.   
While this effect may cause the uncertainty in C IV-based mass, the median value of C IV - Mg II mass difference is less than typical uncertainty of 0.5 dex. Moreover, the black hole masses are only estimated in eight sources using C IV line in our sample. Therefore, our conclusion will not be altered considering the small source fraction and generally consistent C IV - Mg II mass estimations. % , supporting the broadly consistent results and no systematic influence the later conclusions.}

%Previous work (Greene & Ho 2005; Trakhtenbrot & Netzer 2012) showed that M_BH  calibrations based on low ionization lines (i.e. Hα, and Mg II ) generally show good agreement with the Hβ M_BH estimator with a typical scatter of <=0.2 dex, but the recalibration using the C IV high ionization line is more problematic and shows large scatter (0.4 - 0.5 dex), possibly driven by the C IV profiles which show large blueshifts with non virial component.
  
\subsubsection{Eddington ratio estimation}
We calculated the bolometric luminosity assuming $L_{\rm bol}=10L_{\rm BLR}$ \cite[e.g.,][]{li06} for AGNs with broad line measurements, where $L_{\rm BLR}$ is the luminosity of broad line region. $L_{\rm BLR}$ was estimated following the method of \cite{1997MNRAS.286..415C} by scaling the strong broad
emission lines H$ \alpha $, H$ \beta $, Mg II and C IV to the quasar template spectrum of \cite{Francis91}, where Ly$ \alpha $ is is used as a reference of 100. The total relative BLR flux is 555.77, and H$ \alpha$, H$ \beta$, Mg II and C IV are 77, 22, 34, and 63, respectively \citep{Francis91,1997MNRAS.286..415C}. For example, when H$\beta$ and Mg II are available, we calculated the bolometric luminosity using the relation $ L_{\rm bol} = 10 \times 555.77 \times (L_{\rm H \beta} + L_{\rm Mg II})/(22+34)$.
For those AGNs without broad line measurements, we estimated the bolometric luminosity from [O III] line by using the relationship between two parameters. Following \cite{2015A&A...578A..28B}, we constructed a self-consistent relation $L_{\rm bol}$ - $L_{\rm [OIII]}$ from the sources having both broad lines and [O III] line in our sample. The $L_{\rm bol}$ - $L_{\rm [OIII]}$ relation is shown in Figure 8 as
\begin{equation}
\log\left(\frac{L_{\mathrm{bol}}}{\textrm{erg s}^{-1}}\right) = (10.01\pm4.8) + (0.82\pm0.11)\log\left(\frac{L_{\mathrm{[OIII]}}}{\textrm{erg s}^{-1}}\right) \, .
\end{equation}
It is in good agreement with that of \cite{2015A&A...578A..28B}. Our relation is then used to estimate $L_{\rm bol}$, similarly as the relation of \cite{2015A&A...578A..28B} used for CSS sources in \cite{2016A&A...591A..98B}.

%\textbf{We find that our derived $L_{\rm bol}$ $-$ $L_{\rm [OIII]}$ relation agree well with that of \cite{2015A&A...578A..28B} of which that relation was further used to calculate the bolometric luminosity for CSS young radio sources in the work of \cite{2016A&A...591A..98B}}. 
By using the black hole mass and bolometric luminosity, the Eddington ratio $R_{\rm edd}= L_{\rm bol}/ L_{\rm edd}$ can be calculated where $L_{\rm edd} = 1.38 \times 10^{38} M_{\rm BH}/\rm M_{\sun}\ erg\ s^{-1}.$

\begin{figure}
\centering
%\plotone{redshiftnew.eps}
\includegraphics[width=3.4in]{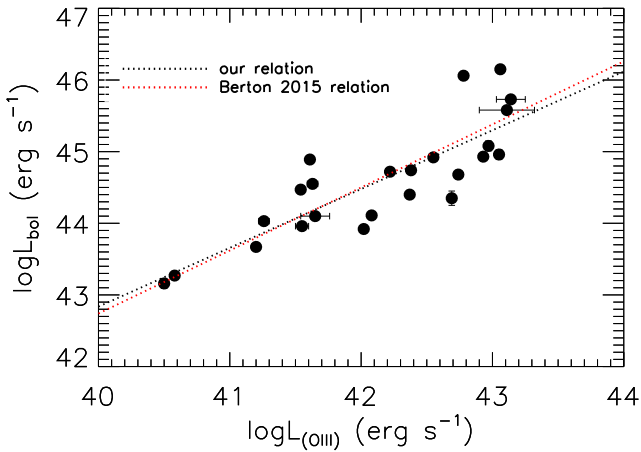}
\vspace*{-0.2 cm}\caption{The relation between the luminosity of [O III] $\lambda 5007$ and the bolometric luminosity derived from broad lines. The black dashed line is the best fit, and the red dashed line is the relation from Berton et al. (2015).}
\vskip-10pt
\end{figure}

\subsubsection{Black hole mass and Eddington ratio distributions}    
In our sample, we were able to estimate BH masses for 106 sources (69 quasars and 37 galaxies), and Eddington ratios for 69 quasars and 33 out of 37 galaxies since four galaxies have no broad line nor [O III] line measurements. The distribution of BH masses and Eddington ratios are shown in Figure 9. We find that the BH masses ($M_{\rm BH}$) have a distribution ranging from $10^{7.32}$ to $10^{9.84} \rm M_{\sun}$ with a peak around $10^9$ $\rm M_{\sun}$ and a median value of $10^{8.72} \rm M_{\sun}$. In Figure 9, it seems that galaxies have on average larger masses than quasars. The average values of log $M_{\rm BH}$ are 8.61 and 8.73 for quasars and galaxies, respectively. This difference is further checked with K-S test, showing a significant difference in $M_{\rm BH}$ distributions between galaxies and quasars with P-value of 0.039. This difference in BH mass distribution is also found between large samples of quasars in \cite{Shen et al.(2011)} and galaxies in \cite{2005MNRAS.362....9B} when using same methods to estimate the BH mass as in our work. The average values of log $M_{\rm BH}$ are 8.71 and 8.88 for the radio-loud quasars and galaxies, respectively, and the difference is confirmed with P-value of 0.036 from K-S test. The Eddington ratios $R_{\rm edd}$ have a wide distribution that range from $10^{-4.93}$ to $10^{0.37}$ with most QSOs at $R_{\rm edd}>0.01$. This is obviously that most QSOs are at $R_{\rm edd}>0.01$, while most galaxies cover the range of $R_{\rm edd}<0.01$. As quasars usually have larger luminosity than galaxies, the larger black hole mass will be expected if two populations have similar accretion rate. The slightly larger BH mass in galaxies would be related with the lower accretion rate/Eddington ratio in these sources as shown in Figure 9, although they have lower luminosities than quasars.

\begin{figure*}
	\centering
\includegraphics[width=3.4in]{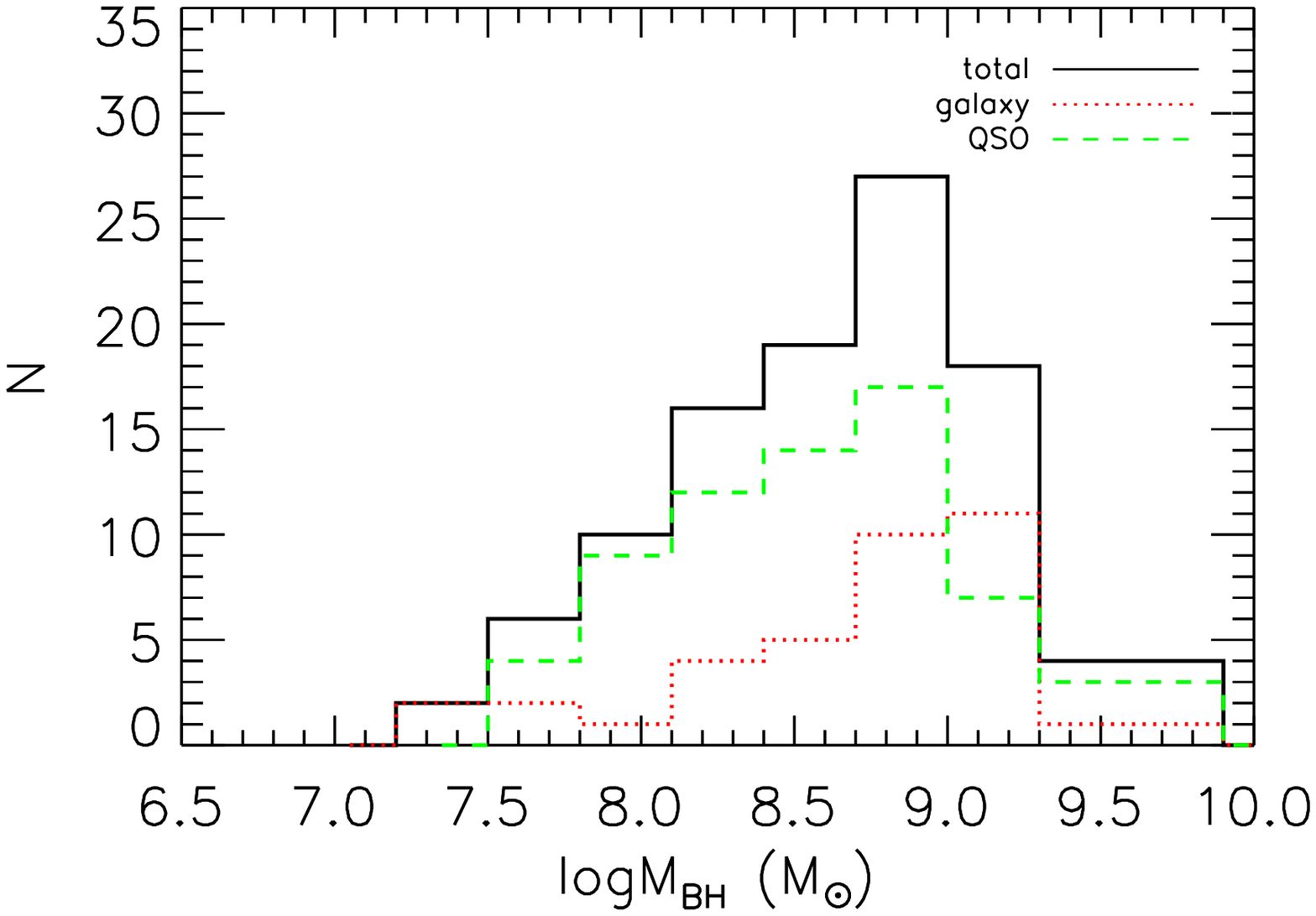}
\includegraphics[width=3.4in]{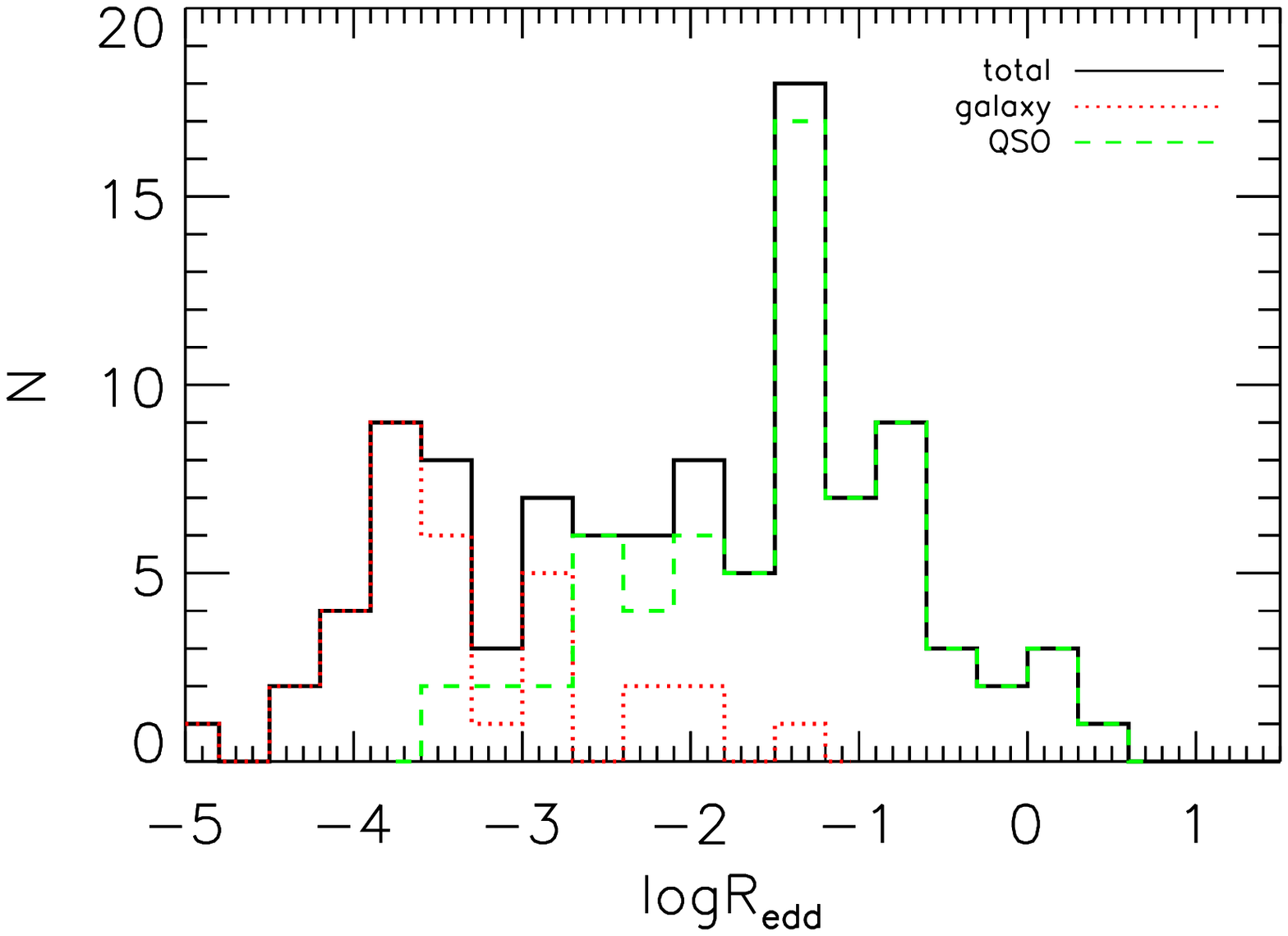}

\vspace*{-0.2 cm} \caption{Histogram of the black hole mass ($left$) and Eddington ratio ($right$) of our sample.}
    \vskip-10pt
\end{figure*}

\subsection{ $ M_{\rm BH}$, $R_{\rm edd}$ versus LS}

The radio linear size is available for 109 sources from the literature, out of which 89 have BH masses, and 85 sources have Eddingtion ratio measurements.
The distributions between linear size and BH mass, and Eddingtion ratio are shown in Figure 10. To further study this, we tentatively took 69 low-redshift large-scale radio galaxies in \cite{Hu16} as a comparison sample. The BH masses and Eddington ratios were directly collected from their work, which were originally used to study the accretion properties. The linear size of these sources were taken from the literature. The comparison in Figure 10 shows that the spread in $M_{\rm BH}$ for our young radio sources is much larger than large-scale radio galaxies. A large fraction of young radio AGNs (77 out of 89, i.e., $\sim$ 87\%) have comparable BH masses to large-scale radio galaxies ($>10^{8} \rm M_{\odot}$), however the rest objects have relatively lower black hole masses. The mean values of BH masses are $10^{8.64} \rm M_{\odot}$, and $10^{8.92} \rm M_{\odot}$ for young radio AGNs, and large-scale radio galaxies, respectively. We found a mild correlation at 98\% confidence level between LS and BH mass for the combined sample of young radio AGNs and large-scale radio galaxies. This indicates systematically lower black hole masses in young radio AGNs in comparison to large-scale radio galaxies. In contrast, we only found a trend of decreasing Eddington ratio with increasing LS in the combined sample.  The mean values of Eddingtion ratio are $10^{-2.26}$ and $10^{-3.05}$ for young radio AGNs, and large-scale radio galaxies, respectively.
%There is a obvious trend that the Eddingtion ratios of young radio AGNs are systematically higher than that of large-scale radio galaxies, \textbf{in particular when compared to quasars}.
%\textbf{However, the distribution of quasars is much different from galaxies for our young radio sources, especially below 1 kpc.} 
\begin{figure*}
	\centering
\includegraphics[width=3.4in]{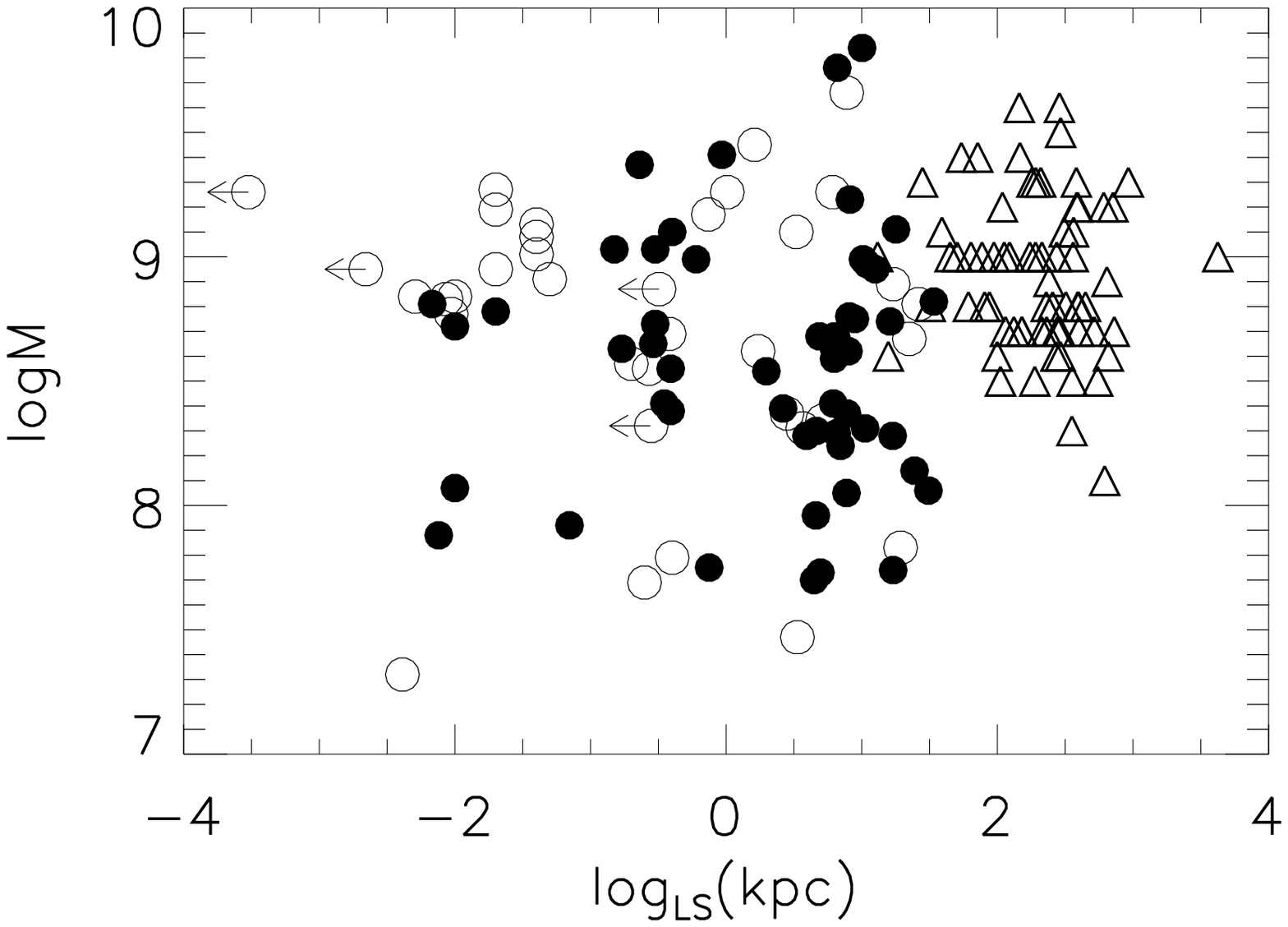}
\includegraphics[width=3.4in]{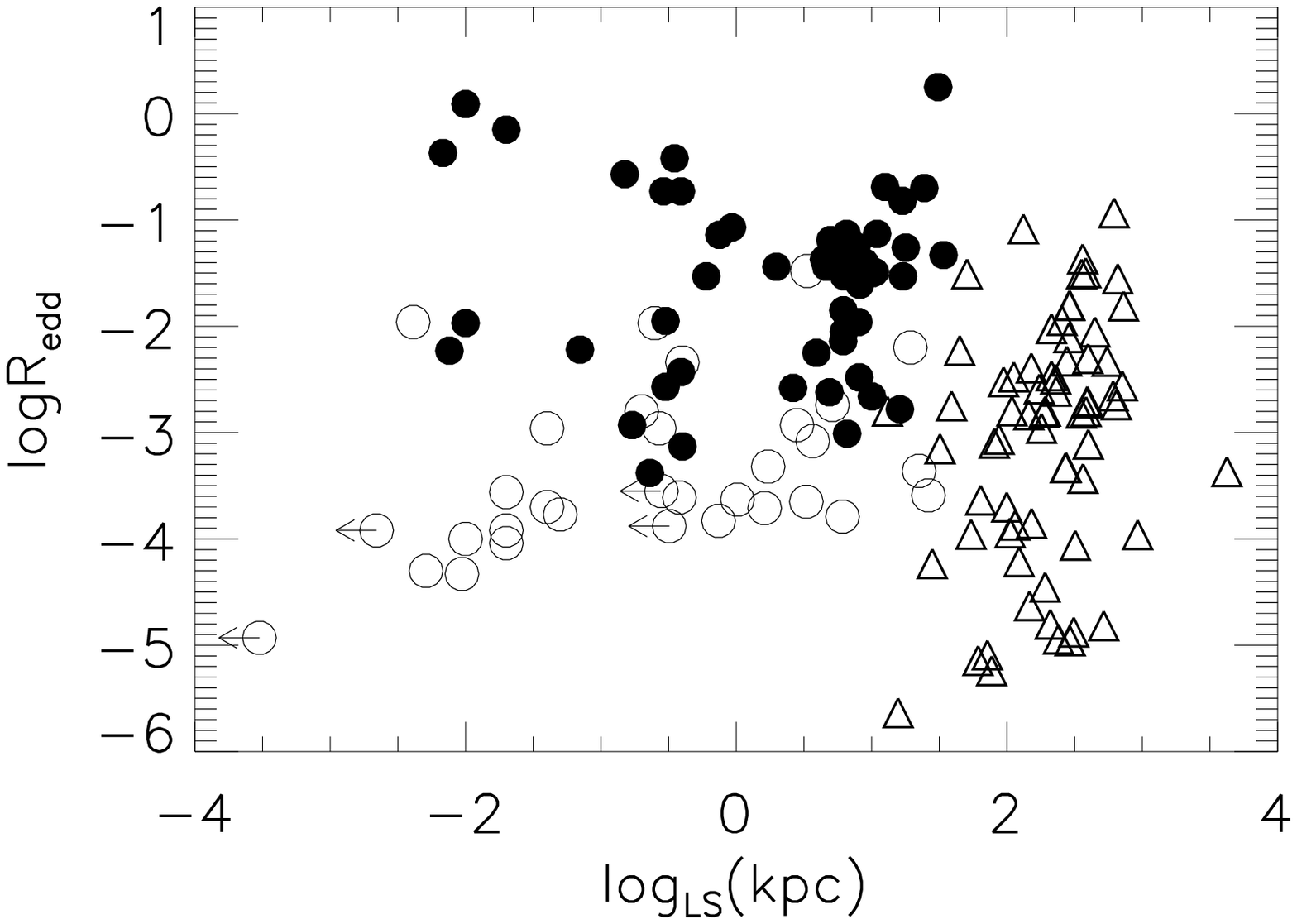}

\vspace*{-0.2 cm}\caption{BH mass versus linear size ($left$) and Eddington ratio vs. linear size ($right$). The filled and open circles are young radio AGNs in our sample classified as quasar and galaxy, respectively. The triangles are large-scale radio galaxies in Hu et al. (2016), used as comparison. The arrows represent the upper limits on the linear size.}
\vskip-10pt
\end{figure*}
%\cite{Hu16}
 
  %\begin{figure}
  %%\epsscale{.80}
  %\plotone{oiii_sii.eps}
 % \includegraphics[height=2.50in,scale=.40]{oiii_sii.eps}
 % \caption{FWHM([O III]) versus FWHM([S II]). The dashed line is same to figure 6.\label{fig8}}

 \subsection{NLS1s}
 
 %We are aim to search NLS1s in the young radio AGNs sample. 
%The criterion to classify a source as NLS1 consists of a narrower H$\beta$ line than typical broad-line AGNs with FWHM(H$\beta)_{\rm broad} $  $<2000\ \rm km\ s^{-1}$ \citep{1985ApJ...297..166O,1989ApJ...342..224G} or $<2200\ \rm km\ s^{-1}$ \citep{2006ApJS..166..128Z}, weak [O III] $\lambda5007$ and strong Fe II emission. 
According to the criteria to classify a source as NLS1 of FWHM(H$\beta)_{\rm broad} $ $<2000\ \rm km\ s^{-1}$ \citep{1985ApJ...297..166O,1989ApJ...342..224G} or $2200\ \rm km\ s^{-1}$ \citep{2006ApJS..166..128Z} , flux(total [O III] $\lambda5007$)/flux(total H$\beta$) $<$ 3 and $R_{4750}$ $>$ 0.4 \citep{2001A&A...372..730V}, which is defined as the flux ratio $R_{4750}$ of Fe II complex between rest wavelength 4434 \AA\ and 4684 \AA\ to that of total H$\beta$, we find one NLS1s in our sample (SDSS J133108.29+303032.9) within 11 sources (1 GPS, 1 HFP and 9 CSS) with measurements of broad H$\beta$, [O III] line and $R_{4750}$. Its original spectra and spectral fitting are presented in Figure 11.
 %and SDSS J150506.47+032630.8)
 
SDSS J133108.29+303032.9 ($z=0.850$, CSS source) has two spectroscopic observations in SDSS. The spectral region around H$\beta$ was analyzed on the BOSS spectrum with a median S/N of 37 and strong Fe II emission. The H$\beta$ can be well modeled with two broad Gaussians and one narrow Gaussian. Each of the [O III] $\lambda\lambda4959,5007$ lines was fitted using two Gaussians. The position and width of the core and wing component of [O III] $\lambda4959$ were separately tied to those of [O III] $\lambda5007$, and the FWHM of narrow H$\beta$ was tied to that of [O III] $\lambda5007$ core component. We found that the line width of broad H$\beta$ is FWHM=2073.65 $\rm km\ s^{-1}$, and the flux ratio of [O III] $\lambda5007$ to total H$\beta$ is 1.02. The relative Fe II emission, $R_{4750}$ is 0.88. Following the criterion in \cite{2006ApJS..166..128Z}, SDSS J133108.29+303032.9 can be classified as NLS1. Using the line width and luminosity of broad H$\beta$, the black hole mass can be estimated, log $M_{\rm BH}$ = 8.14 $\rm M_{\sun}$, from Equation (2). By combining the luminosity of broad H$\beta$ and Mg II lines, we calculated the BLR luminosity log $L_{\rm BLR}$ = 44.58 $ \rm erg\ s^{-1}$, and thus the bolometric luminosity log $L_{\rm bol}$ = 45.58 $ \rm erg\ s^{-1}$ following the method in Section 4.1. The Eddingtion ratio of the source is 0.20. SDSS J133108.29+303032.9 is very radio loud with a radio loudness of log $R = 4.82$, which defined as the ratio between the 5 GHz and 2500 \AA\ rest-frame flux densities.
 %$\rm km\ s^{-1}$, $ \rm erg\ s^{-1}$
 
We plotted the FWHM of $(\rm H{\beta})_{broad}$ and Fe II strength $R_{4750}$ in Figure 12 to look the position of SDSS J133108.29+303032.9 with other sources (1 GPS, 1 HFP and 8 CSS) in which the broad $\rm H\beta$ and Fe II emission were measured. The plot shows that SDSS J133108.29+303032.9 locates the different space with others, especially performs apparently high Fe II emission. From the Figure 12, we find that HFP and GPS sources locate the same space with non-NLS1s.
%Thus, we can exclude other 10 sources are not NLS1s.
 
% SDSS J150506.47+032630.8 ($z=0.408$) was defined as a HFP in \cite{Dallacasa2000} and identified as a NLS1 in \cite{2006ApJS..166..128Z}. It's very radio loud with a radio loudness of log $R = 3.19$, defined as the ratio between the 1.4 GHz and 4400 \AA\ rest-frame flux densities \citep{2008ApJ...685..801Y}. We used two Gaussians to fit broad H$\beta$ component after subtracting the continuum including a single power-law and Fe II emission. \textbf{Each of the [O III] $\lambda\lambda4959,5007$ lines can be well fitted with a single Gaussian}. The broad H$\beta$ has a line width of FWHM = 1715.68 $\rm km\ s^{-1}$, and the flux ratio of [O III] $\lambda5007$ to total H$\beta$ is 0.77. The $R_{4750}$ has a value of 1.27. The black hole mass is log $M_{\rm BH}$ = 7.06 $\rm M_{\sun}$ estimated by using Equation (2), and the Eddington ratio is 1.58 obtained from the BLR-based bolometric luminosity. %0.2 in log space with combining the luminosity of Mg II and H$\beta$.
 
 \begin{figure}
 \centering
 \includegraphics[height=2.5in,scale=.40]{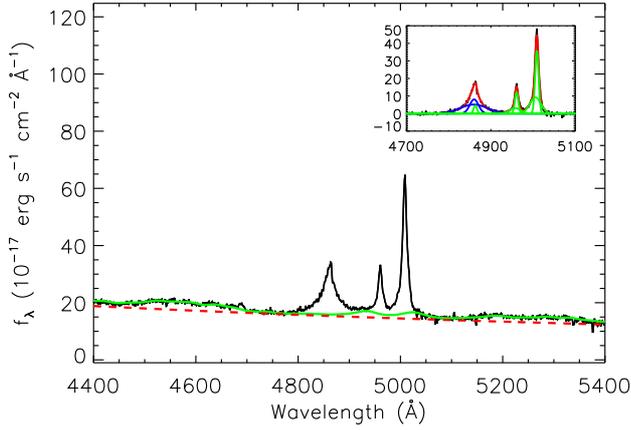}
\vspace*{-0.2 cm} \caption{The spectral fitting for two sources classified as NLS1s: SDSS J133108.29+303032.9 Lines as in Figure 3.}. %and SDSS J150506.47+032630.8 ($bottom$). \textbf{Lines as in Figure 3.}}
\vskip-10pt
 \end{figure}
 
 \begin{figure}
  \centering
  \includegraphics[height=2.5in,scale=.40]{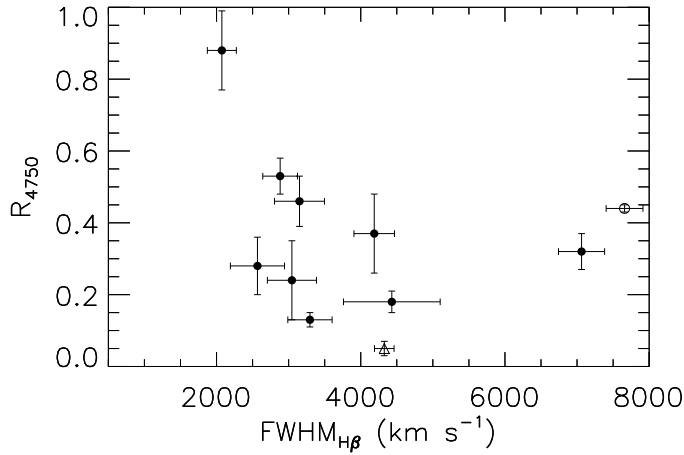}
 \vspace*{-0.2 cm} \caption{Fe II strength against the FWHM of $(\rm H{\beta})_{broad}$. The filled circles, the open circle and the open 
 triangle represent CSS, GPS and HFP sources, respectively.}. %and SDSS J150506.47+032630.8 ($bottom$). \textbf{Lines as in Figure 3.}}
 \vskip-10pt
  \end{figure}

 \subsection{$L_{\rm [OIII]}$ vs. $L_{\rm 5GHz}$}
 
We investigated the relationship between the [O III] $\lambda5007$ core and the radio (5 GHz) luminosity aiming to study the relationship between optical and radio properties. The relationship between them is shown in Figure 13 for 35 quasars and 33 galaxies where the typical uncertainty on $L_{\rm [OIII]}$ is $<$ 10\% in our spectral measurement, while it is 3\% - 10\%  for $L_{\rm 5GHz}$ from the literature. We find a significant correlation between $L_{\rm [OIII]}$ and $L_{\rm 5GHz}$ with a Spearman rank correlation coefficient $r_{\rm s}$ = 0.72, and a probability $P_{\rm null}$ less than $10^{-10}$ for the null hypothesis of no correlation. It is evident that quasars tend to be brighter in both luminosities than galaxies, which may be caused by the fact that many of quasars are at higher redshift than galaxies (see Figure 13). 
We employed the partial Kendall's $\tau$ correlation test \citep{1996MNRAS.278..919A} to assess the correlation between $L_{\rm 5GHz}$ and $L_{\rm [OIII]} $ eliminating the effect of redshift. Our partial test shows that the strong positive correlation still present with partial correlation coefficients $\tau$ = 0.2 and $P_{\rm null}$ = 0.0047.
%a partial Spearman correlation analysis was applied and the strong positive correlation is still present. 
The Ordinary Least Squares bisector (OLS bisector) method  \citep{1990ApJ...364..104I} fit to the correlation between $L_{\rm [OIII]}$ and $L_{\rm 5GHz}$ is log $L_{\rm 5GHz} = 1.38(\pm0.09) \times$ log $L_{\rm [OIII]}-22.18(\pm2.98)$.

%The relationship between the jet and accretion disk is investigated by comparing the luminosity of core [O III] $\lambda5007$ and 5 GHz radio emission. 
The strong correlation between $L_{\rm [OIII]}$ of core [O III] $\lambda5007$ and $L_{\rm 5GHz}$ for 68 objects in our sample suggests that jet activity is strongly associated with accretion process in young radio AGNs as the luminosity of core [O III] $\lambda5007$ can trace nuclei radiation \citep{2012ApJ...757..140S}. However, the dependence of $L_{\rm 5GHz}$ on $L_{\rm [OIII]}$ is steeper than that of \cite{1999AJ....118.1169X} for a sample of radio-loud AGNs with available data collected from the literature at that time. Their relation is indicated in Figure 13 as a dotted line derived from the same method of OLS bisector. We find that most of our sources lie above their relation, indicating that young radio AGNs have stronger radio emission than that of normal radio-loud AGNs at given $L_{\rm [OIII]}$. This supports the evolution scenario where young, compact AGNs are more efficient radio emitters than normal, large-scale objects, and will eventually evolve into the latter population.
%\cite[see also equation (1) in][]{1999AJ....118.1169X}
%partial correlation tau:0.20,P_null=0.0047

 \begin{figure}
  \centering
  %%\epsscale{.80}
  %\plotone{optical_radio.eps}
  \includegraphics[height=2.50in,scale=.40]{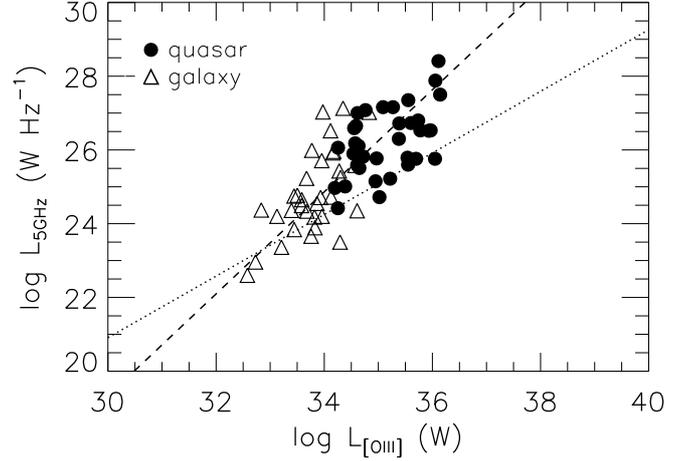}
  \vspace*{-0.2 cm} \caption{The radio luminosity $L_{\rm 5GHz}$ versus the luminosity of $\rm [OIII] \lambda 5007$ line core. The filled circles and open triangles stand for quasars and galaxies, respectively. The dashed line is the OLS bisector fit for our sources, while the dotted line denotes the relation of normal radio-loud AGNs derived from OLS bisector fit in Xu et al. (1999).\label{fig10}}
  \vskip-10pt
  \end{figure}

\subsection{The optical variability of quasars}
  
We studied the optical variability of quasars based on SDSS multi-epoch spectroscopy. To study the variability of continuum emission, we selected those quasars with $D_{\rm 4000}\leqslant1$ to minimize the contribution of host galaxy, and we fitted the continuum on global spectra. There are in total 15 quasars (i.e., 8 CSS, 5 HFP, 1 GPS and 1 CSO) with multiple spectroscopic observations, of which 12 were observed twice, 3 more than two times. The source list and the measurements of continuum and emission lines are given in Table 3. The redshift of these sources range from 0.230 to 3.593, with most objects at $z>$ 0.8 (see Table 3). 
  
We used the integrated flux of overall spectrum where the  wavelength range (in the rest-frame)  correspond to the overlaps of SDSS and BOSS spectra from 3800 to 9200 \AA\ at observed frame to study the source variability. The source variability was determined by comparing the variation between the brightest and faintest spectra as $\Delta f$ = $(f_{\rm int,b}-f_{\rm int,f})/f_{\rm int,f}$, where $f_{\rm int,b}$ and $f_{\rm int,f}$ are integrated flux at the brightest and faintest epochs, respectively) \cite[e.g.,][]{G16}. We also corrected the systematic difference between BOSS and SDSS DR7 spectra by dividing the BOSS spectrum by the systematic correction spectrum of \cite{G16} before spectra fitting when BOSS spectrum is involved. 
%To ensure data quality, only those spectra with S/N $>$ 5 were considered. For this reason, one object, SDSS J101603.13+051302.3 was excluded due to very low S/N (=2.1) of the faint spectrum. This left 21 young radio AGNs in the optical variability study.

The distribution of the optical variabilities $\Delta f$ of 15 objects is shown in Figure 14. All CSS/GPS/HFP/CSO sources distribute at low variability $\leqslant60\%$.
  
  %The variability range for CSS sources is in 0.46 - 41 $\%$, while 0.1 - 116 $\%$ for HFP sources. GPS and CSO have 16 $\%$ and 47 $\%$ variability, respectively.
  
The color variation for every object was evaluated from the difference in the spectral index between the bright and faint epochs, $\Delta \alpha$ = $\alpha_{\rm b}-\alpha_{\rm f}$, where $\alpha_{\rm b}$ and $\alpha_{\rm f}$ are the spectral indices at brightest and faintest epochs, respectively. It was further checked by the power-law fit on the difference spectrum between two epochs \cite[see details in][]{G16}. Considering $\sim4\%$ spectrophotometry uncertainty in SDSS spectra \citep{ade08}, we conservatively studied the color variation only in 7 out of 15 sources, which have reliable variations $\Delta f>20\%$. The consistent color variations between the spectral index difference and the power-law fit on the difference spectrum are found in 6 out of these 7 sources, which all exhibit bluer-when-brighter (BWB) trend (see Table 3). %While 6 sources in total show color variation above 3$\sigma$ significance.
  
We failed to find any significant variations in the line width of broad H$\beta$, Mg II and C IV lines for all sources studied. The luminosities of broad emission lines also do not change significantly in all sources, except for SDSS J104406.33+295900.9, of which C IV emission at bright epoch is larger than the faint one by about a factor of 2.5 at $>3\sigma$ significance (see Table 3).

\cite{G16} investigated the optical variability for a sample of 2169 quasars with a large redshift coverage, in which most sources are radio-quiet. Their results showed that the variability of most objects is below $60\%$. In our sample, we find that all CSS/GPS/HFP/CSO sources have variability below 60$\%$, similar to radio-quiet quasars of \cite{G16}, implying that the variability mechanism in these sources could be similar to radio-quiet quasars and the jet does not contribute much in variability mainly due to large jet viewing angle. 

\begin{figure}
  \centering
   %%\epsscale{.80}
   %\plotone{optical_radio.eps}
   \includegraphics[height=2.50in,scale=.40]{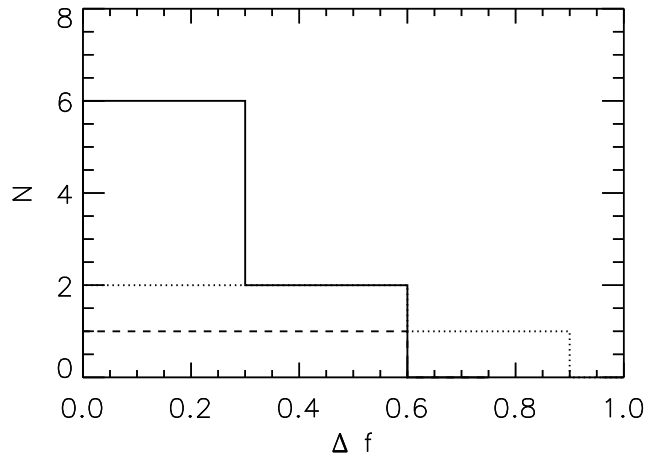}
   \vspace*{-0.2 cm}\caption{Distribution of the optical variabilities of 15 quasars with multi-epoch SDSS spectroscopy. CSS sources are shown as solid line, HFP sources as dotted line and the dashed line is for GPS/CSO sources.}
   \vskip-10pt
   \end{figure} 
  
  %% Note that the \setcounter and \renewcommand are needed here because
  %% this example is using a mix of deluxetable and tabular.  Here the
  %% deluxetable counters are set with \tablenum but the situation is a bit
  %% more complex for tabular.  Use the first command to set the Table number
  %% to ONE LESS than it should be.  The next command will auto increment it
  %% to the desired number.
\section{Discussions}
In this work, we investigated the optical properties for a sample of young radio AGNs constructed from SDSS DR12, which includes GPS, CSS, HFP and CSO radio sources.
\subsection{Black hole mass and accretion mode}
While the empirical relation between the radius of broad line region and continuum luminosity was commonly used to estimate the black hole mass in large sample of AGNs \cite[e.g.,][]{Shen et al.(2011)}, its applicability was evaluated and then the continuum luminosity was replaced by BLR luminosity for radio-loud AGNs to avoid the overestimation of the black hole mass due to the jet contribution on the continuum emission \cite[e.g.,][]{ko06}.
Previous studies have shown that at least some GPS and extreme GPS (i.e., HFP) sources may present beaming effect \citep{od98}, which could be less significant in CSS sources due to large jet viewing angles. As shown in Section 4.2.1, the deviation from radio-quiet AGNs strongly favor the presence of jet emission. It thus is more reasonable to calculate black hole mass by using BLR luminosity. The black hole masses estimated in this way in our sample cover a broad range, consistent with \cite{2009MNRAS.398.1905W} and \cite{2012ApJ...757..140S}.  

% \cite{bh95} showed that the optical composite spectrum of GPS sources resemble that of flat-spectrum radio quasars (FSRQs), suggesting that the optical continuum emission in GPS and HFP sources may be affected by jet emission.

%We test this by comparing the continuum luminosity (at 5100 $\AA$, 3000 $\AA$, 1350 $\AA$) with the line luminosity (H$\beta$, Mg II, C IV) for the quasars in our sample.

Compared to $M_{\rm BH}$ distribution, the Eddington ratios cover even a broader range from $10^{-4.93}$ to $10^{0.37}$, suggesting diversities of accretion activities as found in \cite{2012ApJ...757..140S}. Particularly, quasars tend to have higher Eddington ratios than galaxies in our sample, mostly at $>0.01$. Adopting a rough dividing value of Eddington ratio, 0.01, between the optically thin advection-dominated accretion flow and optically thick standard accretion disc \citep{Narayan95}, there exists two accretion modes in our sample.
%\textbf{Most sources have $R_{\rm edd} > 0.01$, a range explored in \cite{2009MNRAS.398.1905W}}. 
%This may be caused by the heterogeneous samples which were selected for different purpose in the literature \citep{2012ApJ...757..140S}. 

We find that the Eddington ratios of about 46\% of the sources are below 0.01, of which most are narrow-line objects. Most of the low radio luminosity sources ($L_{\rm5GHz} < 10^{26}$ $ \rm W\ Hz^{-1}$) \citep{Snellen2004,2010MNRAS.408.2261K,S09} locate at the range of $R_{\rm edd} < 0.01$ (see Figure 15).
In the study of \cite{2010MNRAS.408.2261K}, it showed that low-luminosity CSS radio sources can be classified as HEGs and LEGs, where they are usually associated with accreting in a radiatively efficient manner at high Eddington ratio ($R_{\rm edd} > 0.01$) and exhibiting radiatively inefficient accretion related to low Eddington ratio ($R_{\rm edd} < 0.01$), respectively \citep{2017A&ARv..25....2P}. Thus, our sources with low luminosity and low Eddington ratio below 0.01 may be the LEGs counterparts. We summaried the information of radio classification (CSS/GPS/HFP/CSO) and optical type (type1/type2 and HEG/LEG) in Table 4 with the estimated Eddington ratio, in which the HEGs and LEGs classification were based on EI value larger and smaller than 0.95 estimated using equation (7) from \cite{2010A&A...509A...6B} for the sources of which the measures were possible for related lines, respectively. We find that for the classified HEG and LEG sources, most of them separately correspond the $R_{\rm edd}$ $>$ 0.01 and log $R_{\rm edd}$ $<$ 0.01.

\begin{eqnarray}
\scriptsize
{\rm EI} = \log {\rm [O~III] \over H\beta}  - {1 \over 3} \left(\log {\rm [N~II] \over H\alpha} + \log {\rm [S~II] \over H\alpha } + \log { \rm [O~I] \over H\alpha} \right)
\end{eqnarray}

The strong positive correlation between the Eddington ratio and the radio luminosity at 5 GHz in our sample shown in Figure 15, implies that the accretion process is correlated with radio emission in young radio AGNs with stronger radio emission when higher accretion ratio.

%The relatively less emission in radio band in these sources compared to more powerful radio sources may be caused by the presence of less powerful jets \citep{2005ApJ...621..123U}. In turn, it associates the accretion process which can be radiatively inefficient in low-luminosity radio AGNs \citep{2012MNRAS.421.1569B}. 
%This phenomenon is supported by the strong positive correlation between Eddington ratio and radio luminosity at 5 GHz in our sample shown in Figure 12.
\begin{figure}
 %%\epsscale{.80}
 %\plotone{optical_radio.eps}
 \includegraphics[height=2.50in,scale=.40]{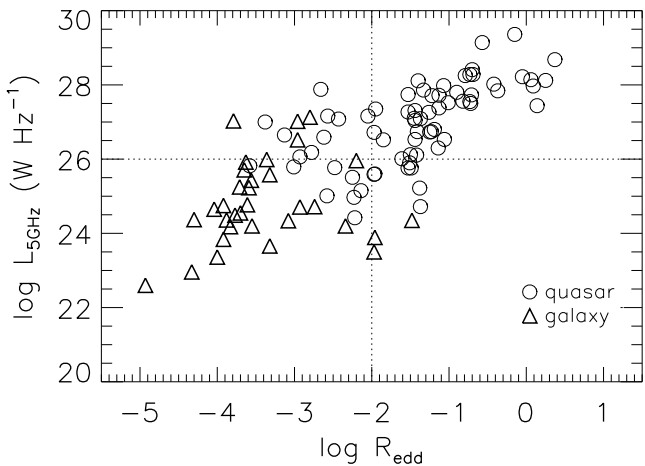} 
 \caption{The radio luminosity $L_{\rm5GHz}$ versus the Eddington ratio $R_{\rm edd}$. The circles and triangles stands for QSOs and galaxies, respectively. The vertical and horizontal dotted lines are indicated as the  separations of high and low luminosity ($10^{26}$), and high and low Eddington ratio (0.01), respectively.}
 \end{figure}
 
The mixed accretion modes as shown in young radio AGNs have implication that young jet activity could happen in various accretion systems, in others words, the sources at early stage of jet evolution are not necessarily only associated with the system with high accretion rate. 

% Nevertheless, most of our sources in the right panel of Figure 6 have higher Eddington ratios than large-scale sources, implying that systematically these young radio AGNs are at the early stage of accretion process.
%So it seems that low-luminosity type II radio sources match the LEG sources ascribing the accretion properties in nuclear. And it leads to a lower log$(R_{\rm edd})$ value in low power type II AGNs which use use luminosity of [O III] $\lambda5007$ trace the nuclear luminosity than other more powerful young radio AGNs.
%The study of \cite{2010MNRAS.408.2261K} found that low-luminosity sources have mixed accretion modes, and they can be classified as high and low excitation galaxies (HEGs and LEGs) according to optical emission line ratios. \cite{2010A&A...509A...6B} discussed that LEGs are powered by hot gas while HEGs are supplied with cold gas material. Also \cite{2010MNRAS.408.2261K} shows that the [O III] luminosity in HEGs are about 10 times brighter than LEGs and the [O III] luminosity in low-luminosity sources are lower than more powerful ones.
%.
\subsection{Evolution}
The evolution of radio sources generally can be studied either using radio luminosity vs LS as done in \cite{AB12} and \cite{2010MNRAS.408.2261K}, or using $L_{\rm [OIII]}$ versus LS \citep{2010MNRAS.408.2279K}. The long-term evolutionary scenarios were proposed as the smallest radio sources CSO/GPS/HFP (LS $<$ 1kpc) evolve into medium-sized objects CSS (LS $<$ 20 kpc) and  eventually evolve into large-scale FRII and FRI sources (LS: tens to several thousand kpc) for high luminosity and low luminosity radio sources in \cite{AB12}, respectively.
While from the optical perspective, \cite{2010MNRAS.408.2279K} suggested that HEGs and LEGs are usually associated with high accretion rate and low accretion rate, respectively, and they follow different evolution. While HEGs objects will follow evolutionary track of GPS$_{\rm HEG}$ $-$ CSS$_{\rm HEG}$ $-$ FR$_{\rm HEG}$,  the smallest-scale LEGs sources will evolve to CSS LEG radio sources, finally evolve to large-scale FR LEGs.

%and for low-luminosity radio sources with CSO/GPS/HFP $-$ MSO/CSS $-$ LSO (FRI) by  While from the optical 
%perspective, GPS$_{\rm HEG}$ $-$ CSS$_{\rm HEG}$ $-$ FR$_{\rm HEG}$ and GPS$_{\rm LEG}$ $-$ CSS$_{\rm LEG}$ $-$ FR$_{\rm LEG}$ were proposed in \cite{2010MNRAS.408.2279K}}.

In our work,  the evolution of young radio AGNs is studied by incorporating the Eddington ratio into the radio power - linear size panel shown in Figure 16.
From the radio perspective, our sources generally follow the expected evolutionary tracks towards large-scale sources as the smallest high (low) luminosity CSO/GPS/HFP objects to medium-sized high (low) luminosity CSSs to large-scale high (low) luminosity FRII/FRI sources \citep{AB12}, where the evolutionary tracks are based on parametric modeling for high luminosity and low luminosity objects, respectively. In addition to the evolution manifested from LS $-$ $L_{\rm 5GHz}$ relation, there is an evident trend of Eddington ratio decreasing with increasing LS for high luminosity radio sources ($L_{\rm 5GHz}$ $> 10^{26}\ \rm W/Hz$), in which median (log $R_{\rm edd}$) is -1.44 and -2.52 for high luminosity radio young sources and large-scale FR II radio galaxies, respectively. Similarly, this trend is also seen in low luminosity radio sources with the median (log $R_{\rm edd}$) for low luminosity radio young sources and large-scale FRI radio galaxies of -3.01 and -4.23, respectively. These results strongly indicate that the young radio AGNs will eventually evolve into large-scale radio sources in general. 

\begin{figure*}
\centering
 %%\epsscale{.80}
 %\plotone{optical_radio.eps}
  \includegraphics{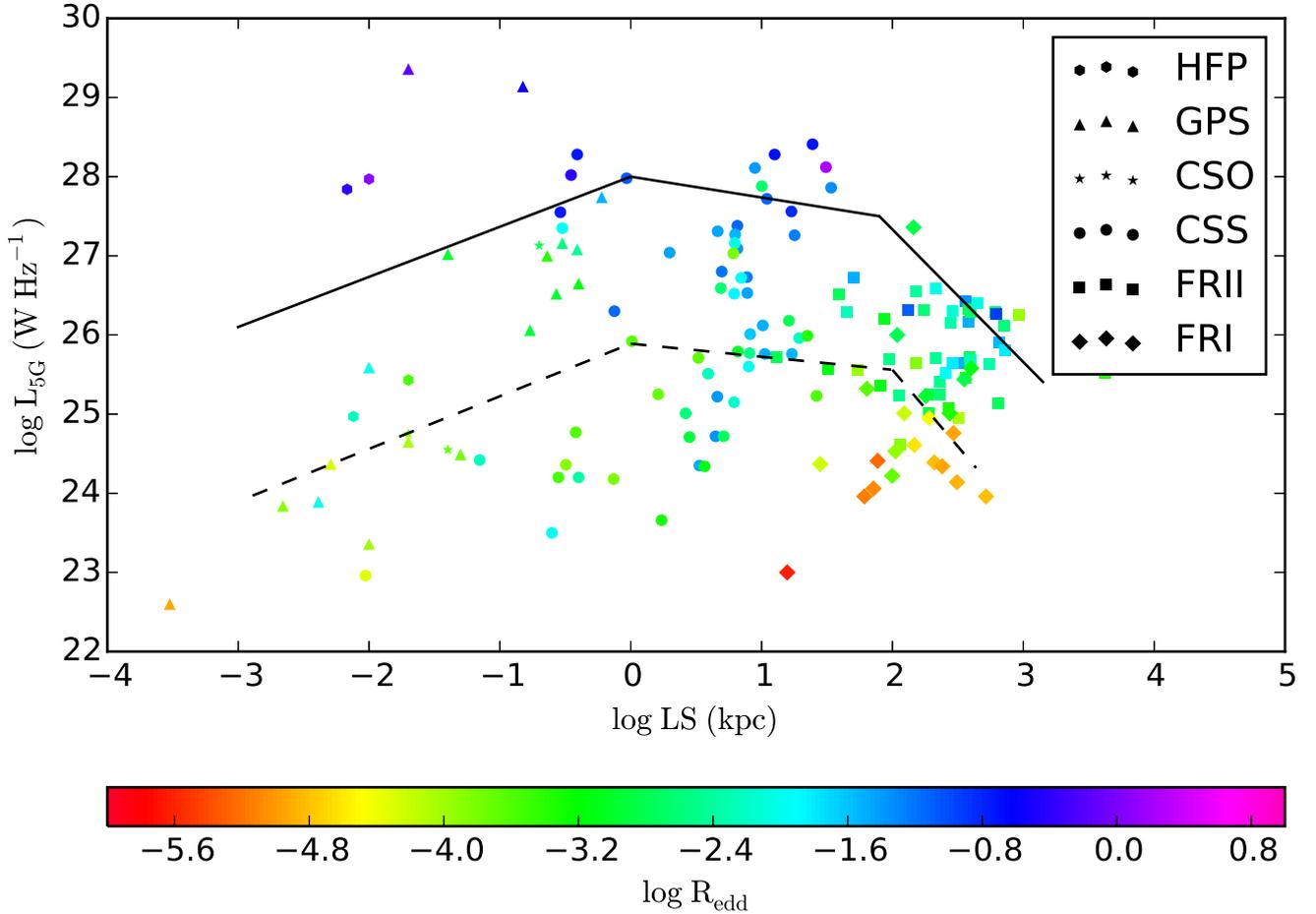}
 \vspace*{-0.2 cm}\caption{Radio power at 5 GHz vs. linear size with Eddington ratio. The HFP, GPS, CSO, CSS sources are shown as hexagons, triangles, and asterisks and circles, respectively. The squares and diamonds are for FR II and FR I sources, respectively. The value of Eddington ratio is shown with the color bar at bottom. The black solid and dashed lines are the expected evolutionary tracks based on parametric modeling for the high-luminosity and low-luminosity radio sources in An \& Baan (2012). \label{fig13}}
  \vskip-10pt
 \end{figure*}

\subsubsection{Sources with low luminosity and low accretion rate}
The Eddington ratio of low luminosity radio sources in our sample is almost less than 0.01 (see Figure 15). Although the Eddington ratios of these sources overlap with large-scale objects, they are indeed systematically larger than large-scale ones. Therefore, these low-luminosity radio sources with low $R_{\rm edd}$ could still be the ancestors of low-luminosity large-scale FR I sources (see Figure 16). Alternatively, the growing number of observations show that short-lived/dying radio sources may be associated with low-luminosity young radio AGNs \citep{2010MNRAS.408.2261K,AB12}. Their accretion phase(s) last shorter than $10^{4}-10^{5}$ years and will not evolve into large-scale object which is probably caused by the low accretion rate \citep{2017FrASS...4...38W}. This will probably result in compact radio morphology with weak radio emission or faders which show disrupted lobes without hotspots (e.g., breaking-up structures shown in \cite{2010MNRAS.408.2261K}).

The radio morphology of sources with low luminosity and low accretion rate will be useful to study the source evolution, i.e., well-confined lobes and hotspots structure, or faders, or relics. We try to check the radio morphology for our objects with low luminosity and low Eddington ratio from the previous work \cite[i.e																																																										.,][]{D09,2010MNRAS.408.2261K}. However we didn't find significant evidence of radio relics (e.g., diffused emission without radio core), thus all these sources are less likely at inactive phase. 
%Also the obvious morphology with fading was not found in these sources.
The the obvious morphology with fading was found in SDSS J155927.67+533054.4 (log $R_{\rm edd}$ = -2.74, LS = 5.13 pc, log $L_{\rm 5GHz}$ = 24.72 $\rm W\ Hz^{-1}$) from the 1.5 GHz MERLIN observation showing breaking up structure with a weak radio core in \cite{2010MNRAS.408.2261K}. The low luminosity sources in our sample SDSS J084856.57+013647.8, SDSS J100955.51+140154.2,
SDSS J154349.50+385601.3 and SDSS J154525.48+462244.3 may be also faders and undergo disturbed evolution since they show brightness and polarization asymmetry associated with interaction with surrounding medium in the radio observation in \cite{2010MNRAS.408.2261K}.
In SDSS J083139.79+460800.8 (log $R_{\rm edd}$ = -4.04, LS = 21 pc, log $L_{\rm 5GHz}$ = 24.65 $\rm W\ Hz^{-1}$), a compact double structure is clearly seen from VLBI observation in \cite{D09}. The dynamical age determined by VLBI observations is about 245$\pm$55 years in \cite{2010A&A...521A...2D} and indicates that it is indeed young and may persistently grow up. The similar case is in SDSS J151141.26+051809.2 (log $R_{\rm edd}$ = -2.23, LS = 7.6 pc, log $L_{\rm 5GHz}$ = 24.97 $\rm W\ Hz^{-1}$) in which the high-resolution VLBI images at 
8.4 and 15.4 GHz reveal it owns well confined double structure with a radio core in \cite{2012ApJS..198....5A}. The kinematic age based on the proper motion of lobes is 300 $\pm$ 140 years \citep{2012ApJS..198....5A}.
Interestingly, SDSS J124733.31+672316.4 (log $R_{\rm edd}$ = -3.92, LS = 145 pc, log $L_{\rm 5GHz}$ = 24.75 $\rm W\ Hz^{-1}$) shows double-double morphology based on VLBI observations \citep{2003PASA...20...16M}, suggesting restarted accretion and radio activities \citep{AB12}. Moreover, several sources (e.g., J093609.36+331308.3, J105731.17+405646.1, J140051.58+521606.5 and J143521.67+505122.9) are unresolved in VLBI images \citep{D09}. As shown in \cite{2017FrASS...4...38W}, these sources may be the transient radio sources (associate with short-lived radio sources) in temporary low accretion state with weak radio emission and compact morphology. Higher resolution radio observations will be needed to unveil their detailed morphology, especially for SDSS J105731.17+405646.1, the source with the lowest Eddington ratio in our sample (log $R_{\rm edd}$ = -4.93). In summary, the young radio sources with low luminosity and low accretion rate show various radio properties and activities, including short-lived/dying/transient sources, restarted sources and normal evolved sources.

%In SDSS J083139.79+460800.8 (log $R_{\rm edd}$ = -3.17, LS = 21 pc, log $L_{\rm 5GHz}$ = 24.65 $\rm W\ Hz^{-1}$), a compact double structure is clearly seen from VLBI observation with two strong, unresolved components in \cite{D09}.
% Based on above analysis, low-luminosity radio sources in low accretion state show various radio morphogy like double, probably interaction system or recurrence activity or transient sources.
%implying that some of them are expected to evolve into larger sources
%SDSS J132513.37+395553.2

% the evidence of possible fizzling out

\subsubsection{Implication from $M_{\rm BH}$ and $R_{\rm edd}$}
Naively expected, the sources with smaller LS may generally have lower BH mass and higher accretion rate. Therefore, the BH mass (accretion rate) is expected to positively (negatively) correlate with the LS in young radio AGNs. However these two trends are not evident in our sample sources alone (see Figure 10), perhaps due to the intrinsic diversities of accretion in young radio AGNs.
%, which could be related with unknown triggering mechanism of jet and accretion process.
%The complex evolution process might make young radio AGNs present variety in black hole mass and Eddington ratio (in Figure 6).

The diversity in BH and Eddington ratio is further supported by our results in Figure 17 with the distribution of black hole masses along with the Eddington ratios, where young radio AGNs in our sample can be divided into three parts using the dividing values of black hole mass $10^8 \rm M_{\sun}$ and Eddington ratio 0.01. Our sources in part three occupy the region of lower black hole masses and higher Eddington ratios compared to large-scale objects. They are similar to NLS1s, thus could be candidates of both accretion and jet at early evolutionary stage. However, some young radio AGNs overlapping with large-scale sources are also found (parts one and two in Figure 17). The sources in part one have comparable black hole masses and Eddington ratios with large-scale radio galaxies, corresponding to the sources with low radio luminosity and Eddington ratio mentioned in Section 5.2.1. Some of them may be restarted sources since they already have comparable black hole masses with large-scale objects. This is strongly supported by the double-double morphology as shown in SDSS J124733.31+672316.4 (log $M_{\rm BH}=8.95$, also see Section 5.2.1). The second box in Figure 17 consists of objects with similar black hole masses however higher Eddington ratios than large-scale ones. They may also be restarted sources however at higher accretion state than part one objects. Interestingly, there are few sources at the lower-left corner in Figure 17. This is probably because these sources are very weak, thus missed in various surveys.

 \begin{figure}
 \centering
  %%\epsscale{.80}
  %\plotone{optical_radio.eps}
  \includegraphics[height=2.50in,scale=.40]{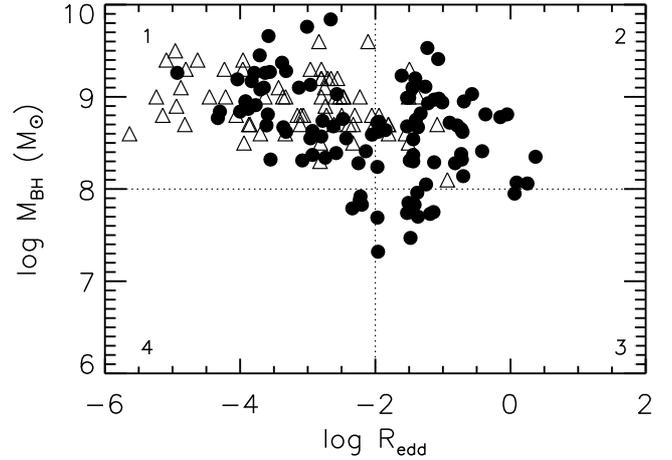}
 \vspace*{-0.2 cm}\caption{The BH mass vs. Eddington ratio for our young radio AGNs donated by filled circles and large-scale FR I/FR II radio galaxies of Hu et al. (2016) shown as triangles.\label{fig14}}
  \vskip-10pt
  \end{figure}

\subsection{NLS1s $\&$ young radio AGNs}

Several studies have focused on the association between young radio AGNs and NLS1s \citep{2008ApJ...685..801Y, 2014MNRAS.441..172C, Gu2015, 2015A&A...578A..28B, Caccianiga17}. Some of these two populations show similar radio morphology \citep{Gu2015,Caccianiga17} and are in the same range of black hole masses and Eddington accretion rates \citep{2009MNRAS.398.1905W,2016A&A...591A..98B}. The hypothesis of CSS-like sources as the parent population of flat-spectrum RLNLS1s has been proposed by \cite{2015A&A...578A..28B}. Within 11 sources (1 GPS, 1 HFP and 9 CSS) with measurements of broad H$\beta$ and [O III], we find one NLS1s (CSS source: SDSS J133108.29+303032.9), i.e., about 11$\%$ detection rate in CSS sources. %The fraction of NLS1s from our young radio AGNs sample is quite low. Most of galaxies in our sample are type II AGNs. 45 $\%$ sources with the redshift greater than 1 which do not cover the wave region of [O III] and H$\beta$. Actually we are only able to focus 19 candidates in our sample of which all are quasars and have both H$\beta$ and [O III] line. The fraction is about 11$\%$. 
In contrast, there are seven CSS-like sources in 14 RLNLS1s of \cite{Gu2015}, and six CSS-like sources out of 19 RLNLS1s in Gu et al. (2019, in preparations), resulting in a fraction of $\sim 39\%$ CSS-like sources in RLNLS1s. The discrepancy of source fraction indicates that a high fraction (~39$\%$) of radio-loud NLS1s are CSS sources but that only 11$\%$ of CSS are RL NLS1s. %So this may reflect that the fraction of similarity association between young radio source and NLS1s from the radio aspects is higher than optical band, implying that radio young does not always show youth in accretion. That may because young radio AGNs are identified from the radio band original. 
It is likely due to the observational biases, i.e., the missed large-scale extended radio emission in previous studies, as demonstrated in \cite{ric15} that such extended emission may be common, at least among the brightest radio-loud NLS1s, based on the high-sensitivity JVLA observations.

%Interestingly, two NLS1s in our sample differ from others in many aspects, especially in $\gamma-$ray emission. SDSS J150506.47+032630.8 has been classified as a blazar-type radio source with significant radio polarization \citep{2008A&A...479..409O,2009A&A...495..691M} and was included in the Fermi-LAT sources catalogues \citep{acero15}. It shows a compact core-jet structure on pc-scale \citep{dammando13}. It is included in MOJAVE sample as one of few $\gamma-$ray emitted NLS1s with the measured proper motion of $1.14\pm0.38$c \citep{2016AJ....152...12L}. %This source have significant radio polarization in the HFP quasars \citep{2008A&A...479..409O} and strong $\gamma$-ray emission resembling those of flat spectrum radio quasar \citep{dammando16}. The FWHM of this sources is a little larger than the \cite{2006ApJS..166..128Z} but consistent with the study of \cite{Shen et al.(2011)}.

Interestingly, the one NLS1 in our sample differ from others in many aspects, especially in $\gamma-$ray emission. SDSS J133108.29+303032.9, well known as 3C 286, has a core-jet structure on pc-scale with an inclination angle of 48$\degr$ between the line of sight and jet \citep{2017MNRAS.466..952A}. It has been identified in the third Fermi-LAT source catalogue \citep{acero15}. However, the origin of $\gamma$-ray emission is unclear, which may be due to the interaction between jet and interstellar medium (ISM) \citep{mig14}. The BOSS spectrum shows obvious blueshifted [O III] wing, which may be induced by outflow or jet$-$ISM interaction \citep{Gelderman1994,Kim2013}. The obtained broad $\rm H\beta$ line width of the source is consistent with \cite{2017berton1} within error.%, though the different fitting model is applied for H$\beta$ narrow component.

\subsection{Selection effect}

It should be noticed that our collection of young radio sources (starting samples) cannot be certainly considered as statistically complete and representative of the entire population due to the radio-selected samples we used have been selected using different techniques (e.g., spectral shape and peak frequency, and/or radio morphology). While it is difficult to evaluate the completeness of parent sample, we tried to study the representative of final optical sample in respect to parent sample as we requested SDSS spectrum. The distributions of radio luminosity at 5 GHz and redshift are compared in two samples (see Figure 18). The median values of redshift in optical and parent samples are 0.452 and 0.656, respectively, implying the redshift is systematically lower in optical sample. For $L_{\rm 5GHz}$, the optical sample is systematically less than parent sample, with the median values of 26.11 and 27.00 $\rm W\ Hz^{-1}$, respectively.
The K-S test results show that $L_{\rm 5GHz}$ and redshift distribution of two samples are indeed different. It seems that our optical sample is biased towards low-redshift and low-luminosity compared with parent sample.
%From the starting samples (parent sample), we selected the optical spectra from SDSS to build our optical sample. We agree with referee that our optical sample just a subset objects (126 out of 348) compared with parent sample and may not be full representation due to the selection effect. 

As shown in Figure 18, the source fraction of our optical sample to parent sample is less than 40$\%$ at $L_{\rm 5GHz}<$ 23.7 $\rm W\ Hz^{-1}$ and $L_{\rm 5GHz}>$ 26.1 $\rm W\ Hz^{-1}$, implying under-representative in these luminosity ranges. We evaluated the effect of the missed sources in the two luminosity ranges by estimating the bolometric luminosity from available R magnitude in the automated Million Optical Radio/X-ray Associations (MORX) catalog (Flesch 2016), assuming $L_{\rm bol}$ = 9.26 $\lambda$$L_{\rm 5100\AA}$ \citep{Shen et al.(2011)}, in which luminosity at 5100 \AA ~$L_{\rm 5100\AA}$ was calculated from R magnitude assuming a spectral index of 0.5. The possible location of each missed sources is shown as a dotted line in Figure 19 because the black hole mass is unknown. Although two targets is a bit far from our sample sources of which they are possibly detected in part 4 depending on black hole mass, We found that most missed sources have Eddington ratio larger than $10^{-3}$ at the black hole mass range covered by our sample sources. Therefore, the missed sources will not affect our results significantly in Figure 16.
%\textbf{While most missed sources are in the regions covered by majority of our objects, two targets is a bit far from our sample sources, and depending black hole mass are possible in part 4 with low black hole mass and Eddington ratio.}

\begin{figure*}
	\centering
	\includegraphics[width=3.4in]{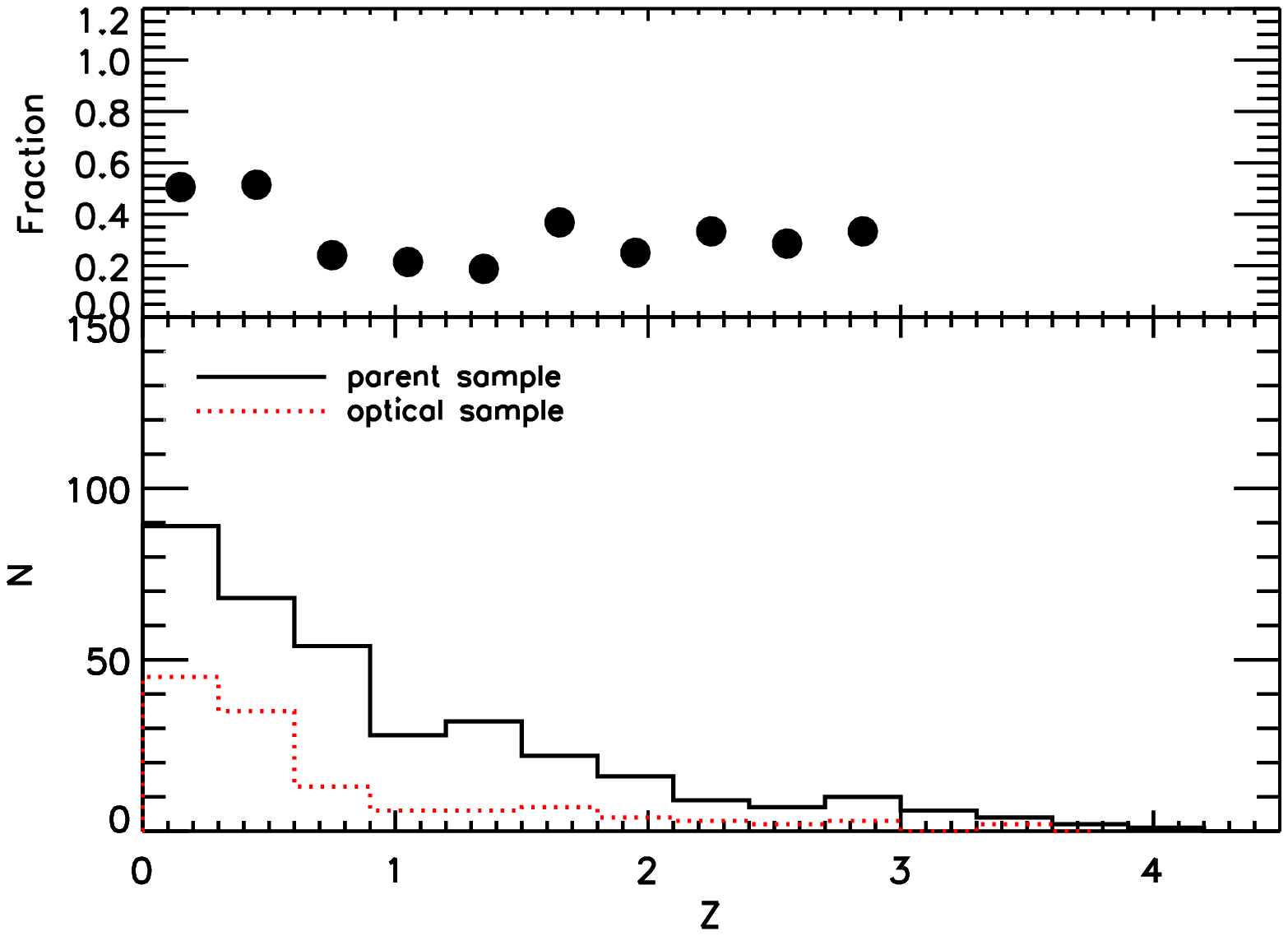}
	\includegraphics[width=3.4in]{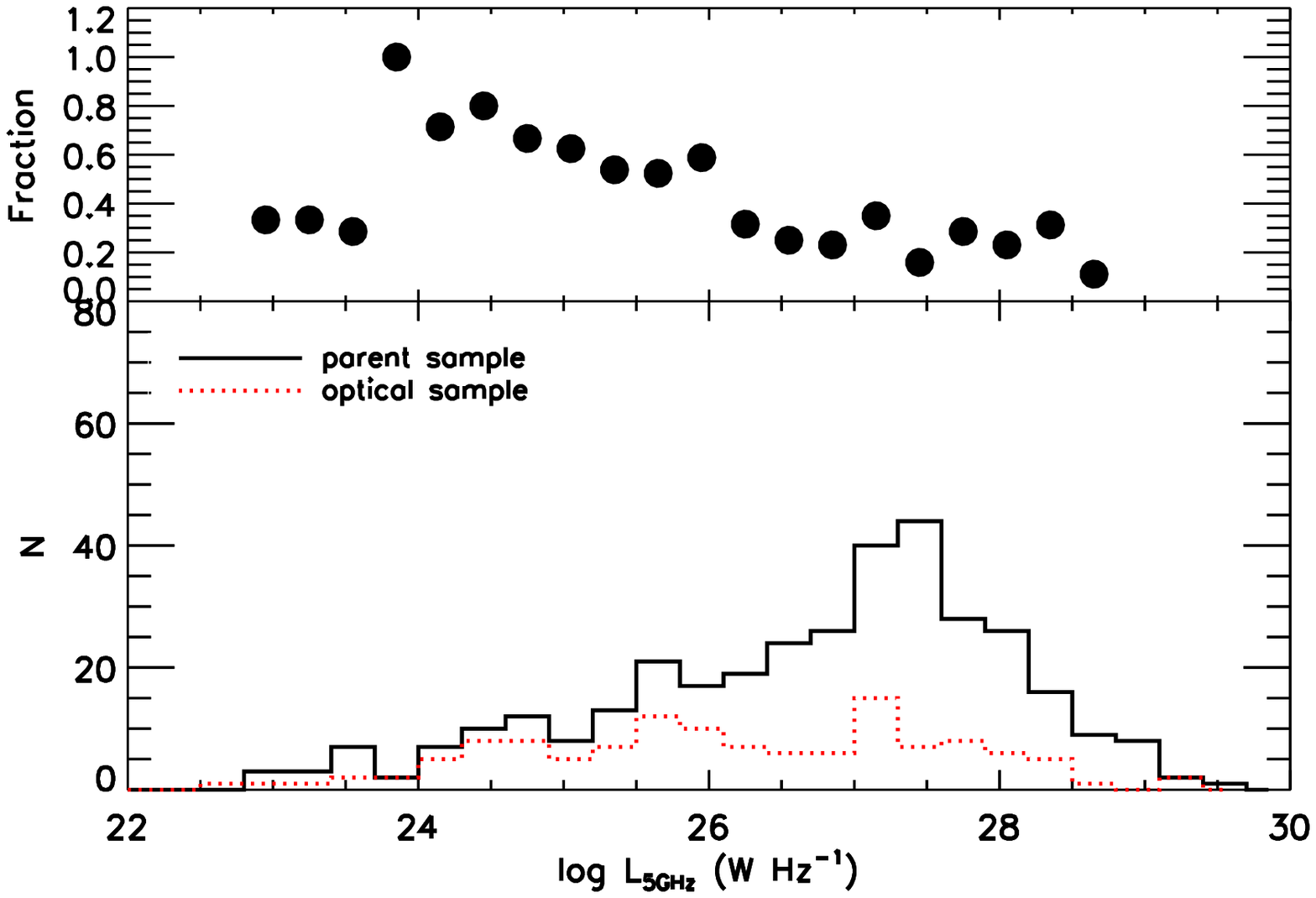}
	
	\vspace*{-0.2 cm}\caption{Redshift distribution ($left$) and $L_{\rm 5GHz}$ distribution ($right$) between parent and optical sample. The upper plots are the source fraction for redshift and $L_{\rm 5GHz}$ of optical sample to parent sample, respectively. The red dotted line and black solid line stands for optical sample and parent sample, respectively.}
	\vskip-10pt
\end{figure*}

%\begin{figure}
	%\centering
	%%\epsscale{.80}
	%\plotone{optical_radio.eps}
%\includegraphics[height=2.50in,scale=.40]{Rmag.pdf}
	%\vspace*{-0.2 cm} \caption{Optical magnitude in R band distribution. Lines as in Figure 17.\label{fig10}}
%	\vskip-10pt
%\end{figure}

\begin{figure}
	\centering
	%%\epsscale{.80}
	%\plotone{optical_radio.eps}
	\includegraphics[height=2.50in,scale=.40]{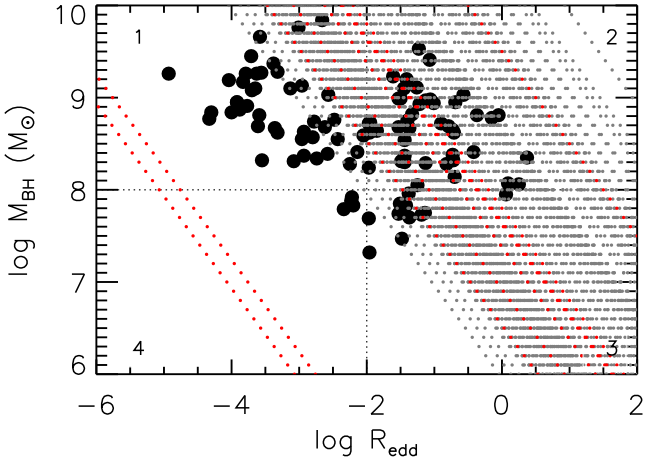}
	\vspace*{-0.2 cm} \caption{The BH mass vs. Eddington ratio for young radio AGNs donated by filled circles and missed sources shown as dotted lines in which the grey ones and the red ones represent the possible loctation for the missed high luminosity sources with $L_{\rm 5GHz}$ > 26.1 $\rm W\ Hz^{-1}$ and the missed low luminosity sources with $L_{\rm 5GHz}$ < 23.7 $\rm W\ Hz^{-1}$, respectively. \label{fig10}}
	\vskip-10pt
\end{figure}

\section{Conclusions}

We built a largest sample of 126 young radio AGNs with spectroscopy from SDSS DR12 up to now in order to study their accretion, evolution and search NLS1s candidates. Stellar velocity dispersion versus gas velocity dispersion, jet-disk relation and optical variability are also investigated.

% study accretion properties, evolution in young radio AGNs and search NLS1s candidates by analyzing the SDSS spectra of a larger sample as well as stellar versus gas velocity dispersion, disk-jet relation and optical variability.

We find that the black hole masses and Eddington ratios cover a wide range, suggesting that our sources have different levels of accretion activity and not all of young radio AGNs are accreting at high accretion rate. In the distribution of the radio power and linear size, we find that high and low luminosity radio sources 
separately follow the global evolutionary trend with increasing linear size and decreasing accretion rate towards large-scale radio sources. We also discussed the radio properties of sources with low Eddington ratio and low radio luminosity. By comparing the stellar velocity dispersion and line width of [O III], we find that the line width of core of [O III] $\lambda5007$ can be used as a good surrogate of stellar velocity dispersion. The radio power correlates strongly with [O III] $\lambda5007$ core luminosity in our work, suggesting that radio activities and accretion are closely related. We also find one object can be defined as narrow-line Seyfert 1 galaxies (NLS1s), representing a population of young AGNs both with young jets and early accretion activities. The optical variabilities of 15 quasars with multi-epoch spectroscopy were investigated. Our results show that the optical variability in young AGN quasars is similar to radio-quiet quasars, supporting the previous results \citep{od98} that no significant variability in general for young radio AGNs.

%Our results show that young radio sources in low-luminosity and low accreted system, present diversities in radio morphology.

\section*{Acknowledgements}
We thank the anonymous referee for comments and sugges-
tions to improve the manuscript.
We thank Ting Xiao for help with spectral analysis.
This work is supported by the National Science Foundation of China (grants 11873073, U1531245, 11773056, and 11473054, U1831138).
Funding for SDSS-III has been provided by the Alfred P. Sloan Foundation, the Participating Institutions, the National Science Foundation, and the U.S. Department of Energy Office of Science. The SDSS-III web site is http://www.sdss3.org/. This research has made use of the NASA/IPAC Extragalactic Database (NED) which is operated by the Jet Propulsion Laboratory, California Institute of Technology, under contract with the National Aeronautics and Space Administration. We acknowledge the support of the staff of the Lijiang 2.4m telescope. Funding for the telescope has been provided by CAS and the People's Government of
Yunnan Province.

{}

%\begin{landscape}
 \begin{table*}
  \caption{Sample}
  \label{tab:landscape}
  \setlength{\tabcolsep}{0.06in}
  \begin{center}
  \begin{tabular}{lccccclrcrcrcrr}
    \hline \hline
    SDSS name & $z$ & type & refs. & ID & $P_{\rm 5GHz}$ & refs. & LS & refs. & log $M_{\rm BH}$ & method & log $L_{\rm bol}$ & method & log $R_{\rm edd}$ & S/N \\
    (1) & (2) & (3)  & (4)  & (5)  &  (6) & (7) & (8) & (9)  & (10) &(11) &(12) &(13) &(14) & (15) \\
    \hline
    J002225.42+001456.1 & 0.306 & GPS   & S98   & G  & 26.52 & O98&   0.27  & O98 & 8.55 & [O III] & 43.73 & [O III] & $-$2.96 & 7.2 \\
    J002833.42+005510.9 & 0.104 & CSS   & K10   & G     & 24.35 & K10 & 3.35  &  K10 & 7.47 & $\sigma_{\ast}$ & 45.13 &  [O III]& $-$1.48 & 20.9 \\
    J002914.24+345632.2 & 0.517 & CSO   & A12  & G     & 27.13 & H04 & 0.20  &  A12 & 8.57 & [O III] & 43.91 & [O III]& $-$2.80  & 3.0 \\
    J005905.51+000651.6 & 0.719 & CSS & E04  & Q & 27.50 & L97 & . . . &  . . . & 8.32  &  H$\beta $ & 45.73 & H$\beta$ $+$ Mg II &$-$0.72  & 25.7 \\
    J014109.16+135328.3 & 0.621 & CSS  & S89   & Q     & 27.16 & O98 & 6.25  &  O98 & 8.59 & [O III] & 44.67 & [O III] & $-$2.05 & 2.9 \\
    J074417.47+375317.2 & 1.066 & CSS   & S89   & Q     & 27.26 & O98  & 17.87 & O98 & 9.11  & Mg II& 45.99 & Mg II & $-$1.26 & 28.8 \\
    J075303.33+423130.8 & 3.594 & CSO   & T20   & Q     & 28.68 & H07 & . . . & . . . & 8.35  & C IV & 46.86 & C IV & 0.37   & 35.3 \\
    J075448.84+303355.0$^{a}$ & 0.796 & HFP   & S09   & Q     & 26.74 & S09 & . . . &   . . . & 9.20  & Mg II & 45.93 & Mg II & $-$1.41   & 21.3 \\
    J075756.71+395936.1 & 0.066 & CSS   & K10   & G    & 23.50 & K10 & 0.25  &  K10 & 7.69  & $\sigma_{\ast}$ & 43.87 & [O III] & $-$1.97 & 28.0 \\
    J080133.55+141442.8 & 1.196 & CSS   & S89   & Q     & 27.86 & O98 & 33.99 &  O98 & 8.82  & Mg II& 45.63 & Mg II& $-$1.33 & 6.7 \\
    J080413.88+470442.8 & 0.510  & CSS   & F01   & Q     & 26.52 & K10 & 6.18  &  K10 & 8.64  & Mg II & 44.93 & Mg II & $-$1.85 & 5.0 \\
    J080442.23+301237.0 & 1.450  & CSS   & K02   & Q     & 27.72 & K10 & 10.98 &  K10 & 8.97  & Mg II & 45.97 & Mg II + C IV & $-$1.13    & 27.4 \\
    J080447.96+101523.7 & 1.968 & CSS   & S89   & Q     & 28.12 & O98 & 31.02 &  O98 & 8.06  & C IV & 46.45 & C IV & 0.25 & 24.2 \\
    J081253.10+401859.9 & 0.551 & CSS   & F01   & Q     & 26.73 & K10 & 7.72  &  K10 & 8.05 & [O III] & 44.94 & [O III]& $-$1.25 & 4.6 \\
    J081323.75+073405.6 & 0.112 & CSS   & K10   & G     & 24.71 & K10 & 2.81  &  K10 & 8.37  & $\sigma_{\ast}$ & 43.58 &[O III] & $-$2.93    & 20.3 \\
    J082504.56+315957.0 & 0.265 & CSS   & K10   & G     & 24.82 & K10 & 7.75  &  K10 & 9.66  & $\sigma_{\ast}$ &. . . & . . . & . . . & 10.6 \\
    J083139.79+460800.8 & 0.131 & GPS   & S04   & G     & 24.65 & D09 & 0.02  &  D09 & 9.19  & $\sigma_{\ast}$ & 43.29 & [O III] & $-$4.04 & 26.0 \\
    J083411.09+580321.4 & 0.093 & GPS& SW98  & G & 24.13 & B91 & 8.6e-3$^{\ast}$ &  S03 & 8.83  & $\sigma_{\ast}$ & . . . & . . . & . . .& 15.3 \\
    J083637.84+440109.6 & 0.055 & CSS   & S04   & G     & 23.66 & D09 & 1.72  &  D09 & 8.62  & $\sigma_{\ast}$ & 43.44 & [O III]& $-$3.32 & 29.0 \\
    J083825.00+371036.5 & 0.396 & CSS   & K10   & G     & 25.92 & K10 & 1.02  &  K10 & 9.26 & [O III] & 43.76 & [O III]& $-$3.63 & 2.9 \\
    J084856.57+013647.8 & 0.350  & CSS   & K10   & Q     & 25.15 & K10 & 6.16  &  K10 & 8.41 &[O III]  & 44.41 & [O III] & $-$2.14 & 2.9 \\
    J085314.22+021453.8 & 0.459 & CSS   & K10   & G     & 25.42 & K10 & 6.01  &  K10 & . . . & . . . & . . . & . . . &. . . & 2.1 \\
    J085408.44+021316.1 & 0.459   & CSS   & K10   & G     & 25.34 & K10 & 3.43  &  K10 & . . . & . . . & . . . & . . . & . . .& 1.6 \\
    J085601.22+285835.4 & 1.084 & CSS   & K02   & Q     & 27.09 & K10 & 6.53  &  K10 & 8.67  & Mg II & 45.43 & Mg II & $-$1.37 & 10.1 \\
    J090105.25+290146.9 & 0.194 & CSS   & S89   & G     & 25.96 & O98 & 19.33 &  O98 & 7.83  & $\sigma_{\ast}$ & 43.77 & [O III] &  $-$2.20 & 11.8 \\
    J090615.53+463619.0 & 0.085 & GPS   & S04   & G     & 24.49 & D09 & 0.05  &  D09 & 8.91  & $\sigma_{\ast}$ & 43.28 & [O III] & $-$3.77 & 35.0 \\
    J090933.49+425346.5 & 0.670  & CSS   & S89   & G     & 27.23 & O98 & 56.15 &  O98 & . . . & . . . & . . . & . . . &. . . & 7.5 \\
    J091734.79+501638.1 & 0.632 & CSS   & K10   & Q     & 25.82 & K10 & 4.87  &  K10 & . . . & . . . & . . . & . . . &. . . & 6.9 \\
    J092405.30+141021.4 & 0.136 & CSS   & K10   & G     & 24.18 & K10 & 0.74  &  K10 & 9.17  & $\sigma_{\ast}$ & 43.48 & [O III] &  $-$3.83 & 26.4 \\
    J092608.00+074526.6 & 0.442 & CSS   & K10   & Q     & 25.60 & K10 & 7.98  &  K10 & 8.62  & H$\beta $ & 44.81 & H$\beta $ & $-$1.96 & 22.2 \\
    J093430.68+030545.3 & 0.225 & CSS   & K10   & G     & 25.25 & K10 & 1.62  &  K10 & 9.45  & $\sigma_{\ast}$ &  43.88 & [O III] & $-$3.71 & 15.1 \\
    J093609.36+331308.3 & 0.076 & GPS & S04 & G  & 23.84 & D09 & 2.2e-3$^{\ast}$ &  D09 & 8.95 & $\sigma_{\ast}$ & 43.18 & [O III] & $-$3.92 & 33.8 \\
    J094525.90+352103.6 & 0.208 & CSS   & K10   & Q     & 24.72 & K10 & 4.45  &  K10 & 7.70  & $\sigma_{\ast}$ & 44.47 & [O III] & $-$1.37 & 19.2 \\
    J095111.52+345131.9 & 0.358 & HFP   & S09   & G     & 25.43 & O10 & 0.02  &  A12 & 9.27 & [O III] & 43.86 & [O III] & $-$3.56 & 6.1 \\
    J095412.57+420109.1 & 1.787 & CSS   & F01   & Q     & 27.56 & K10 & 16.89 &  K10 & 8.28  & C IV & 45.49 & C IV & $-$0.82 & 2.4 \\
    J100800.04+073016.6 & 0.877 & CSS   & S89   & Q     & 27.88 & O98 & 10.05 &  O98 & 9.84& [O III] & 45.32 & [O III] & $-$2.66 & 2.2 \\
    J100955.51+140154.2 & 0.213 & CSS   & K10   & G     & 25.71 & K10 & 3.29  &  K10 & 9.10  & $\sigma_{\ast}$ & 43.59 & [O III]&  $-$3.65 & 14.1 \\
    J101251.77+403903.4 & 0.506 & CSS   & K02   & G     & 27.01 & K10 & 32.80 &  K10 & . . . & . . . & . . . &  . . . & . . .& 2.3 \\
    J101714.23+390121.1 & 0.211 & CSS   & F01   & G     & 25.80 & K10 & 20.97 &  K10 & . . . & . . . & . . . & . . . & . . . & 6.3 \\
    J102027.20+432056.3 & 1.962 & HFP   & S09   & Q     & 27.80 & S09 & . . . &  . . . & 8.72  & Mg II & 45.95 & Mg II + C IV & $-$0.90 & 13.5 \\
    J102523.78+254158.3 & 0.457 & HFP   & S09   & G     & 25.58 & S09 & . . . &  . . . & 9.28 & [O III] & 44.10 & [O III] &$-$3.32 & 3.1 \\
    J102618.25+454229.3 & 0.152 & CSO   & S04   & G     & 24.55 & D09 & 0.04  &  D09 & 9.08  & $\sigma_{\ast}$ & 43.52 & [O III] & $-$3.70 & 18.1 \\
    J102844.26+384436.8 & 0.362 & CSS   & F01   & Q     & 26.18 & K10 & 16.17 &  K10 & 8.74  & $\sigma_{\ast}$ & 44.10 & [O III] & $-$2.78 & 14.6 \\
    J103507.04+562846.7 & 0.460  & GPS   & S98   & Q     & 27.00 & O98 & 0.23  &  O98 & 9.37 & [O III] & 44.13 & [O III] & $-$3.38 &  3.1 \\
    J103532.58+423019.0 & 2.432 & HFP   & S09   & Q     & 27.71 & S09 & . . . &  . . . & 8.93  & Mg II & 45.86 & Mg II + C IV & $-$1.22 & 9.1 \\
    J103719.33+433515.3 & 0.025 & CSS   & S04   & G     & 22.96 & D09 & 9.4e-3  &  D09 & 8.77  & $\sigma_{\ast}$ & 42.58 & [O III] & $-$4.33 & 54.6 \\
    J104029.94+295757.7 & 0.091 & CSS   & K10   & G     & 24.34 & K10 & 3.67  &  K10 & 8.31  & $\sigma_{\ast}$ & 43.37 & [O III] & $-$3.08 & 29.4 \\
    J104406.33+295900.9 & 2.983 & HFP   & S09   & Q     & 28.14 & S09 & . . . &  . . . & 7.95  & C IV & 46.15 & C IV & 0.06  & 8.9 \\
    J105250.06+335504.9 & 1.405 & HFP   & S09   & Q     & 26.74 & S09 & . . . &  . . . & 9.53  & Mg II & 46.43 & Mg II &  $-$1.23 & 27.9 \\
    J105628.25+501952.0 & 0.820  & CSS   & K10   & Q     & 26.01 & K10 & 8.18  &  K10 & 9.23  & Mg II & 45.76 & Mg II &  $-$1.61 & 22.8 \\
    J105731.17+405646.1 & 0.025 & GPS   & S04   & G & 22.60 & D09 & 3e-4$^{\ast}$ &  D09 & 9.26  & $\sigma_{\ast}$ & 42.47 & [O III] & $-$4.93 & 52.7 \\
    J112027.80+142055.0 & 0.363 & GPS   & S98   & Q     & 26.65 & O98 & 0.40  &  O98 & 9.10& [O III] & 44.11 & [O III] & $-$3.13 & 2.5 \\
    J113138.89+451451.1 & 0.398 & CSS   & F01   & Q     & 26.59 & K10 & 4.88  &  K10 & 8.68  & [O III] & 44.20& [O III]& $-$2.62 & 4.7 \\
    J114311.02+053516.0 & 0.497 & CSS   & K10   & Q     & 25.76 & K10 & 17.03 &  K10 & 7.74  & H$\beta$ & 44.35 & H$\beta$ + Mg II & $-$1.53 & 5.7 \\
    J114339.59+462120.4 & 0.116 & CSS   & F01   & G     & 24.66 & K10 & 17.06 &  K10 & 8.89  & $\sigma_{\ast}$ & . . . & . . . & . . .& 13.0 \\
    J114856.56+525425.2 & 1.632 & HFP   & D00   & Q     & 27.84 & D00 & 6.8e-3  &  A12 & 8.81  & Mg II & 46.58 & Mg II + C IV & $-$0.37 & 33.0 \\
   J115000.08+552821.3 & 0.138 & CSS   & S04   & G     & 24.57 & D09 & 0.10$^{\ast}$  &  D09 & . . . & . . . & . . . & . . . & . . .& 8.9 \\
    J115618.74+312804.7 & 0.417 & CSS   & S89   & Q     & 26.80 & O98 & 4.96  &  O98 & 7.73  & H$\beta$ & 44.68 & H$\beta$ + Mg II & $-$1.19 &  7.6 \\
    J115727.60+431806.3 & 0.230  & CSS   & K10   & Q     & 25.22 & K10 & 4.59  &  K10 & 7.96  & H$\beta$ & 44.72 & H$\beta$ + H$\alpha$ &  $-$1.38 & 29.7 \\
    J115919.97+464545.1 & 0.467 & CSS   & K10   & G     & 26.40 & K10 & 5.00  &  K10 & . . . & . . . & . . . & . . . &. . . & 3.0 \\
    J120321.93+041419.0 & 1.224 & GPS   & S02   & Q     & 27.74 & C10 & 0.60  &  C10 & 8.99  & Mg II & 45.60 & Mg II & $-$1.53 & 10.3 \\
     J120406.82+391218.2 & 0.445 & CSS   & F01   & G     & 26.08 & K10 & 12.06 &  K10 & . . . & . . . & . . . & . . . &. . . & 2.4 \\       
    \hline
  \end{tabular}
  \end{center}
 \end{table*}
 \addtocounter{table}{-1}
%\end{landscape}

\begin{table*}
\setlength{\tabcolsep}{0.06in} \caption{Continued\dots
\label{Sample}}
\begin{center}
\begin{tabular}{lccccclrcrcrcrr}
    \hline \hline
    SDSS name & $z$ & type & refs. & ID & $P_{\rm 5GHz}$ & refs. & LS & refs. & log $M_{\rm BH}$ & method & log $L_{\rm bol}$ & method & log $R_{\rm edd}$ & S/N \\
    (1) & (2) & (3)  & (4)  & (5)  &  (6) & (7) & (8) & (9)  & (10) &(11) &(12) &(13) &(14) & (15) \\
    \hline
    J120624.70+641336.8 & 0.372 & CSS   & S89   & Q     & 26.72 & O98 & 6.98  &  O98 & 8.24  & H $\alpha$ & 44.40 & H $\alpha$ & $-$1.97 & 3.2 \\
    J120902.79+411559.2 & 0.096 & CSS   & S04   & G     & 24.26 & D09 & 0.04  &  D09 & 9.01  & $\sigma_{\ast}$ & . . . & . . . &. . . & 30.0 \\
    J124419.96+405136.8 & 0.249 & CSS   & F01   & Q     & 25.51 & K10 & 3.91  &  K10 & 8.28 & $\sigma_{\ast}$ & 44.16 & [O III] & $-$2.25 & 13.2 \\
    J124449.18+404806.2 & 0.814 & CSS   & F01   & Q     & 27.35 & K10 & 0.30  &   K10 & 8.73  & Mg II & 44.92 & Mg II & $-$1.95 & 6.3 \\
    J124733.31+672316.4 & 0.107 & CSO   & A12  & G     & 24.75 & A12  & 0.02 & A12  & 8.95  & $\sigma_{\ast}$ & 43.17 & [O III] & $-$3.92 & 35.0 \\
    J125226.35+563419.6 & 0.320  & CSS   & S89   & Q     & 26.53 & O98 & 7.78  &  O98 & 8.37  & H$\beta$ & 45.08 & H$\beta$ +H$\alpha$ & $-$1.44 & 14.1 \\
    J125325.72+303635.1 & 1.314 & CSS   & K02   & Q     & 27.31 & K10 & 4.62  &  K10 & 8.30   & Mg II & 45.00 & Mg II &  $-$1.44 & 6.3 \\
    J130941.51+404757.2 & 2.907 & HFP   & S09   & Q     & 27.97 & O10 & 0.01  &   O12 & 8.07  & C IV & 46.30 & C IV & 0.09  & 18.8 \\
    J131057.00+445146.2 & 0.391 & CSS   & K10   & G     & 25.26 & K10 & 3.87  &  K10 & . . . & . . . & . . . & . . . & . . . &  5.2 \\
    J131718.64+392528.1 & 1.563 & CSS   & K02   & Q     & 27.55 & K10 & 0.29  &  K10 & 8.65  & Mg II & 46.06 & Mg II + C IV & $-$0.73 & 18.1 \\
    J131739.20+411545.6 & 0.066 & GPS   & S04   & G     & 24.37 & D09 & 5.1e-3  &  D09 & 8.84  &  $\sigma_{\ast}$ & 42.68 & [O III] & $-$4.30 & 28.1 \\
    J132255.66+391208.0 & 2.987 & HFP   & S09   & Q     & 28.22 & S09 & . . . &  . . . & 8.81  & C IV & 46.90 & C IV & $-$0.05 & 30.1 \\
    J132419.67+041907.0 & 0.263 & CSS   & K10   & G     & 24.99 & K10 & 17.04 &  K10 & . . . & . . . & . . . & . . . & . . . & 6.2 \\
    J132513.37+395553.2 & 0.076 & GPS   & S04   & G     & 23.36 & D09 & 0.01  &  D09 & 8.84  &  $\sigma_{\ast}$ & 42.98 & [O III] & $-$4.00  & 34.2 \\
    J133037.69+250910.9 & 1.055 & CSS   & S89   & Q     & 28.28 & O98 & 0.39  &  O98 & 8.38  & Mg II & 45.79 & Mg II & $-$0.73 & 17.7 \\
    J133108.29+303032.9 & 0.850  & CSS   & S89   & Q     & 28.41 & O98 & 24.51 &  O98 & 8.14  & H$\beta$ & 45.58 & H$\beta$ + Mg II &$-$0.70 & 37.3 \\
    J134035.20+444817.3 & 0.065 & GPS   & S04   & G     & 23.89 & D09 & 4.1e-3  &  D09 & 7.32 & $\sigma_{\ast}$ & 43.49 & [O III] & $-$1.96 & 20.0 \\
    J134536.94+382312.5 & 1.852 & CSS   & F01   & Q     & 27.98 & K10 & 0.93  &  K10 & 9.41  & Mg II & 46.48 & Mg II + C IV & $-$1.07 & 22.6 \\
    J134733.36+121724.2 & 0.120  & GPS   & S98   & Q     & 26.06 & O98 & 0.17  &  O98 & 8.63 & $\sigma_{\ast}$ & 43.84 & [O III] & $-$2.93 & 17.9 \\
    J140028.65+621038.5 & 0.429 & GPS   & S89   & Q     & 27.08 & O98 & 0.39  &  O98 & 8.55 & [O III] & 44.26 & [O III] & $-$2.43 & 3.4 \\
    J140051.58+521606.5 & 0.118 & CSS   & K10 & G & 24.36 & K10 & 0.32$^{\ast}$  &  K09 & 8.87  & $\sigma_{\ast}$ & 43.13 & [O III] & $-$3.88 & 22.0 \\
    J140319.30+350813.3 & 2.291 & CSS   & K02   & Q     & 26.12 & K10 & 10.21 &  K10 & 8.99  & Mg II & 45.63 & Mg II + C IV & $-$1.50 & 8.2 \\
    J140416.35+411748.7 & 0.360  & CSS   & K10   & G     & 25.51 & K10 & 1.01  &  K10 & . . . & . . . &  . . . &  . . . & . . . & 2.6 \\
    J140700.39+282714.6 & 0.077 & GPS   & S98   & Q     & 25.59 & O98 & 0.01  &  O98 & 8.72  &  H$\beta$ & 44.89 &H$\beta$ + H$\alpha$ & $-$1.97 & 47.5 \\
    J140909.50+364208.1 & 2.241 & CSS   & K02   & Q     & 27.04 & K10 & 1.98  &  K10 & 8.54  & Mg II & 45.24 & Mg II + C IV & $-$1.44 & 3.6 \\
    J140942.44+360415.9 & 0.149 & CSS   & K10   & Q     & 24.42 & K10 & 0.07  &  K10 & 7.92  & $\sigma_{\ast}$ & 43.83 & [O III] &  $-$2.22 & 10.2 \\
    J141327.22+550529.2 & 0.282 & CSS   & K10   & G     & 24.92 & K10 & 0.81  &  K10 & . . . & . . . & . . . &  . . . &  . . . & 8.9 \\
    J141414.83+455448.7 & 0.458 & CSO   & A12  & G     & 26.22 & A12 & 0.17  &  A12 & . . . & . . . & . . . &  . . . &. . . &  2.9 \\
    J141558.81+132023.7 & 0.247 & CSO   & A12  & G     & 26.19 & A12 & 0.03  & A12 & . . . & . . . & . . . &  . . . & . . . &  5.3 \\
    J141908.18+062834.7 & 1.436 & CSS   & S89   & Q     & 28.28 & O98 & 12.58 &  O98 & 8.95  & Mg II & 46.40 & Mg II & $-$0.69 & 33.0 \\
    J142051.48+270427.0 & 2.516 & HFP   & S09   & Q     & 27.73 & O10 & . . . &  . . . & 8.62  & C IV & 46.05 & C IV  & $-$0.71 & 4.5 \\
    J142104.24+050844.7 & 0.445 & CSS   & K10   & G     & 25.91 & K10 & 1.68  &  K10 & . . . & . . . & . . . & . . . &. . . & 3.1 \\
    J142123.06+464547.9 & 1.668 & HFP   & S09   & Q     & 27.52 & S09 & . . . &  . . . & 8.94  & Mg II & 46.07 & Mg II + C IV & $-$1.01  & 13.6 \\
    J143009.74+104326.9 & 1.705 & HFP   & D00   & Q     & 28.25 & D00 & . . . &  . . . & 8.69  & Mg II & 46.03 & Mg II + C IV & $-$0.79 & 14.0 \\
    J143521.67+505122.9 & 0.100   & CSS & S04 & G & 24.20 & D09 & 0.28$^{\ast}$  & D09 & 8.32  & $\sigma_{\ast}$ & 42.91 & [O III] &  $-$3.55 & 17.5 \\
    J144516.46+095836.0 & 3.541 & GPS   & S98   & Q     & 29.14 & O98 & 0.15  &  O98 & 9.03  & C IV & 46.59 & C IV  & $-$0.57 & 31.0 \\
     J144712.76+404744.9 & 0.195 & CSS   & F01   & G     & 25.23 & K10 & 26.28 &  K10 & 8.81  & $\sigma_{\ast}$ & 43.36 & [O III] & $-$3.59 & 11.1 \\
    J145958.43+333701.8 & 0.644 & HFP   & S09   & Q     & 26.53 & S09 & . . . &  . . . & 8.98  & H$\beta$ & 46.06 & H$\beta$ + Mg II &  $-$1.06 & 31.6 \\
    J150426.69+285430.5 & 2.285 & CSS   & K02   & Q     & 28.02 & K10 & 0.35  &  K10 & 8.41  & Mg II & 46.14 & Mg II + C IV & $-$0.42 & 23.8 \\
    J151141.26+051809.2 & 0.084 & HFP   & D00   & Q     & 24.97 & D00 & 7.6e-3  &  A12 & 7.88  & $\sigma_{\ast}$ & 43.79 & [O III] & $-$2.23 & 29.6 \\
    J152005.47+201605.4 & 1.572 & CSS   & S89   & Q     & 28.11 & O98 & 8.90  &  O98 & 8.75  & Mg II & 45.49 & Mg II + C IV & $-$1.40 & 9.9 \\
    J152349.34+321350.2 & 0.110  & CSS   & K10   & G     & 24.20 & K10 & 0.40  &  K10 & 7.79  & $\sigma_{\ast}$ & 43.60 & [O III] & $-$2.34& 17.5 \\
    J152837.01+381605.9 & 0.749 & HFP   & S09   & Q     & 26.11 & S09  & . . . &  . . . & 7.83  & Mg II & 44.55 & Mg II & $-$1.42 & 2.2 \\
   J153409.90+301204.0 & 0.929 & CSS   & K10   & Q     & 19.80 & K10 & 4e-3  &  K10 & . . . & . . . & . . . & . . . & . . . & 2.6 \\
    J154349.50+385601.3 & 0.553 & CSS   & K10   & Q     & 25.77 & K10 & 8.10  &  K10 & 8.76 & [O III] & 44.42 & [O III] & $-$2.48 & 2.6 \\
   J154525.48+462244.3 & 0.525 & CSS   & K10   & Q     & 25.79 & K10 & 6.59  &  K10 & 9.76 & [O III] & 44.89  & [O III] & $-$3.01 & 3.1 \\
    J154609.52+002624.6 & 0.558 & GPS   & S02   & G     & 27.02 & L07 & 0.04  &  A12 & 9.13 & [O III] & 44.31 & [O III] & $-$2.96 & 2.9 \\
   J154754.12+351842.2 & 0.620  & HFP   & S09   & Q     & 25.90 & S09 & . . . &  . . . & 7.85  & Mg II & 44.47 & Mg II & $-$1.51  & 1.9 \\
   J155235.38+441905.9 & 0.452 & CSS   & K10   & G     & 25.56 & K10 & 6.93  &  K10 & . . . & . . . & . . . & . . . & . . . & 2.6 \\
   J155927.67+533054.4 & 0.179 & CSS   & K10   & G     & 24.72 & K10 & 5.13  &  K10 & 8.34  & $\sigma_{\ast}$ & 43.73 & [O III] & $-$2.74 & 25.0 \\
   J160239.62+264606.0 & 0.371 & HFP   & S09   & Q     & 25.82 & S09 & . . . &  . . . & 9.66 & [O III] & 44.22 & [O III] & $-$3.58& 3.1 \\
   J160246.39+524358.3 & 0.106 & CSS   & K10   & G     & 24.77 & K10 & 0.38  &  K10 & 8.69  & $\sigma_{\ast}$ & 43.22 & [O III] & $-$3.61 & 26.4 \\
   J160335.16+380642.8 & 0.241 & CSS   & K02   & G     & 27.03 & K10 & 6.09  &  K10 & 9.26 & [O III]  & 43.61 & [O III] &  $-$3.79 & 7.6 \\
   J160913.32+264129.0 & 0.474 & GPS   & S89   & Q     & 27.16 & O98 & 0.30  &  O98 & 9.03  & [O III] & 44.61 & [O III] & $-$2.57 & 3.0 \\
   J161148.52+404020.9 & 0.151 & CSS   & K10   & Q     & 25.01 & K10 & 2.63  &  K10 & 8.39  & $\sigma_{\ast}$ &  43.95 & [O III] & $-$2.58 & 14.0 \\
   J161748.41+380141.8 & 1.609 & HFP   & S09   & Q     & 27.10 & S09 & . . . &  . . . & 9.10  & Mg II & 45.79 & Mg II + C IV & $-$1.45 & 8.2 \\
   J161823.57+363201.7 & 0.733 & CSS   & K02   & G     & 26.82 & K10 & 0.44  &  K10 & . . . & . . . & . . . & . . . & . . . & 16.0 \\
   J162111.27+374604.9 & 1.271 & CSS   & K02   & Q     & 27.27 & K10 & 6.28  &  K10 & 8.68  & Mg II & 45.30 & Mg II & $-$1.53 & 8.4 \\
   J163402.95+390000.5 & 1.083 & CSS   & K02   & Q     & 27.38 & K10 & 6.53  &  K10 & 8.29  & Mg II & 45.29 & Mg II & $-$1.13 & 13.7 \\
  J164311.34+315618.4 & 0.587 & CSS   & K10   & Q     & 25.76 & K10 & 10.58 &  K10 & 8.31  & H$\beta$ & 44.96 & H$\beta$ + Mg II &  $-$1.49 & 9.4 \\
  J164348.60+171549.4 & 0.163 & CSS   & S89   & G     & 25.99 & O98 & 22.40 &  O98 & 8.67  & $\sigma_{\ast}$ & 43.44 & [O III] & $-$3.36 & 25.3 \\
   J165822.18+390625.6 & 0.425 & CSS   & K02   & G     & 27.10 & K10 & 0.97  &  K10 & . . . & . . . & . . . &  . . . & . . . &  2.3 \\
 \hline
\end{tabular}
\end{center}
\end{table*}

\addtocounter{table}{-1}
%\end{landscape}

\begin{table*}
\setlength{\tabcolsep}{0.06in} \caption{Continued\dots
\label{Sample}}
\begin{center}
\begin{tabular}{lccccclrcrcrcrr}
    \hline \hline
    SDSS name & $z$ & type & refs. & ID & $P_{\rm 5GHz}$ & refs. & LS & refs. & log $M_{\rm BH}$ & method & log $L_{\rm bol}$ & method & log $R_{\rm edd}$ & S/N \\
    (1) & (2) & (3)  & (4)  & (5)  &  (6) & (7) & (8) & (9)  & (10) &(11) &(12) &(13) &(14) & (15) \\
    \hline
     J213638.58+004154.2 & 1.941 & GPS   & S98   & Q     & 29.36 & O98 & 0.02  &  O98 & 8.78  & Mg II & 46.76 & Mg II + C IV & $-$0.15 & 32.9 \\
     J225025.34+141952.0 & 0.235 & CSS   & S89   & Q   & 26.30 & O98 & 0.75  &  O98 & 7.75  & H$\beta$ & 44.74 & H$\beta$ + H$\alpha$ &  $-$1.14 & 35.6\\
                     \hline\hline
    object name & $z$ & type & refs. & ID & $P_{\rm 5GHz}$ & refs. & LS & refs. & log $M_{\rm BH}$ & method & log $L_{\rm bol}$ & method & log $R_{\rm edd}$ & S/N \\
    (1) & (2) & (3)  & (4)  & (5)  &  (6) & (7) & (8) & (9)  & (10) &(11) &(12) &(13) &(14) & (15) \\
    HB89 1127$-$145 & 1.184 & GPS   & S98   & Q     & 28.48 & O98 & 0.03  &  O98 & 8.78  & Mg II & 46.48 & Mg II  & $-$0.41 & 32.9 \\
    CGRaBS J1424+2256  & 3.620 & GPS   & S98   & Q     & 28.87 & D00 & . . .  &  . . . & 9.55 & C IV & 47.97 &  C IV & 0.3 & 32.9 \\
    \hline
\end{tabular}
\begin{minipage}{170mm}
Columns are listed as follows: (1) object name; (2) redshift; (3) - (4) type and the reference; (5) source classification: G - galaxy, Q - quasar; (6) - (7) radio power at 5 GHz and the reference; (8) - (9) linear size and the reference; (10) black hole mass; (11) adopted method in estimating the black hole mass; (12) bolometric luminosity; (13) adopted method in estimating the bolometric luminosity; (14) Eddington ratio ($L_{\rm bol}/L_{\rm edd}$); (15) median signal-to-noise ratio of the spectrum.\\
\\
References: A12: \cite{AB12}; B91: \cite{b91}; C10: \cite{C10}; D00: \cite{Dallacasa2000}; D09: \cite{D09}; E04: \cite{E04}; F01: \cite{Fanti2001}; H04: \cite{H04}; H07: \cite{2007ApJ...658..203H}; K02: \cite{Kunert2002}; K10: \cite{2010MNRAS.408.2261K}; L97: \cite{L97};  L07: \cite{L07}; O98: \cite{od98}; O10: \cite{mo10}; O12: \cite{mo12}; S89: \cite{Spencer1989}; S98: \cite{Stanghellini1998}; SW98: \cite{Snellen98}; S02: \cite{Snellen2002}; S03: \cite{2003MNRAS.342..889S}; S04 \cite{Snellen2004}; S09: \cite{S09}; T20: \cite{Peck2000}\\
\\
$\ast$: the upper limit of linear size
%: for [OIII] using equation (6), for H$\beta$ + Mg II using $L_{\rm bol} = 10 \times 555.77 \times (L_{\rm H\beta} + L_{\rm Mg II})/(22+34)$, for Mg II + C IV using $L_{\rm bol} = 10 \times 555.77 \times (L_{\rm Mg II} + L_{\rm C IV})/(34+63)$, for for H$\beta$ + H$\alpha$ using $L_{\rm bol} = 10 \times 555.77 \times (L_{\rm H\beta} + L_{\rm H\alpha})/(22+77)$;
%a: blazar-type source
 \end{minipage}
\end{center}
\end{table*}
%CGRaBS J1424+2256
\begin{table*}
\centering
\caption{Emission line properties} 
%\begin{sidewaystable}[h]
%\begin{landscape} 
  \label{tab:landscape}
  \setlength{\tabcolsep}{0.03in}
  %\Rotatebox{90}{%
  \begin{tabular}{ccccccccccccc} %rrrrrrrrrrrrr
    \hline \hline
    SDSS name & \multicolumn{2}{c}{C IV} & \multicolumn{2}{c}{Mg II} & \multicolumn{2}{c}{H$\beta$}  & \multicolumn{2}{c}{[O III]} &\multicolumn{2}{c}{H$\alpha$}& \multicolumn{2}{c}{[S II]} \\
    (1) & (2) & (3)  & (4)  & (5)  &  (6) & (7) & (8) & (9)  & (10) &(11) &(12) &(13) \\
    \hline
    J002225.42+001456.1 & ...   & ...   & ...   & ...   &  ...   & ...   & 484$\pm$31 & 43$\pm$2 & ...   & ...   & 401$\pm$90 & 13$\pm$4 \\
        J002833.42+005510.9 & ...   & ...   & ...   & ...   & ...   & ...   & 219$\pm$13 & 1472$\pm$22 & ...   & ...   & 452$\pm$7 & 532$\pm$8 \\
        J002914.24+345632.2 & ...   & ...   & ...   & ...   & ...   & ...   & 491$\pm$57 & 21$\pm$2 & ...   & ...   & ...   & ... \\
        J005905.51+000651.6 & ...   & ...   & 3771$\pm$336 & 1477$\pm$46 & 2567$\pm$376 & 884$\pm$60 & 982$\pm$52 & 600$\pm$74 & ...   & ...   & ...   & ... \\
        J014109.16+135328.3 & ...   & ...   & ...   & ...   & ...   & ...   & 494$\pm$0 & 115$\pm$2 & ...   & ...   & ...   & ... \\
        J074417.47+375317.2 & ...   & ...   & 6600$\pm$300 & 975$\pm$19 & ...   & ...   & ...   & ...   & ...   & ...   & ...   & ... \\
        J075303.33+423130.8 & 2944$\pm$55 & 686$\pm$29 & ...   & ...   &  ...   & ...   & ...   & ...   & ...   & ...   & ...   & ... \\
        J075448.84+303355.0 & ...   & ...   & 7554$\pm$655 & 1732$\pm$62 & ...   & ...   & ...   & ...   & ...   & ...   & ...   & ... \\
        J075756.71+395936.1 & ...   & ...   & ...   & ...   & ...   & ...   & 276$\pm$8 & 1860$\pm$17 & ...   & ...   & 351$\pm$9 & 338$\pm$8 \\
        J080133.55+141442.8 & ...   & ...   & 5946$\pm$765 & 324$\pm$9 & ...   & ...   & ...   & ...   & ...   & ...   & ...   & ... \\
        J080413.88+470442.8 & ...   & ...   & 7639$\pm$712 & 518$\pm$35 & ...   & ...   & 597$\pm$5 & 852$\pm$4 & ...   & ...   & ...   & ... \\
        J080442.23+301237.0 & 4396$\pm$385 & 785$\pm$73 & 5533$\pm$333 & 473$\pm$10 & ...   & ...   & ...   & ...   & ...   & ...   & ...   & ... \\
        J080447.96+101523.7 & 2789$\pm$686 & 1146$\pm$56 & ...   & ...   & ...   & ...   & ...   & ...   & ...   & ...   & ...   & ... \\
        J081253.10+401859.9 & ...    & ...   & ...   & ...   &  ...   & ...   & 374$\pm$11 & 331$\pm$14 & ...   & ...   & ...   & ... \\
        J081323.75+073405.6 & ...   & ...   & ...   & ...   &  ...   & ...   & 518$\pm$45 & 265$\pm$15 & ...   & ...   & 558$\pm$12 & 445$\pm$8 \\
        J082504.56+315957.0 & ...   & ...   & ...   & ...   & ...   & ...   & ...   & ...   & ...   & ...   & ...   & ... \\
        J083139.79+460800.8 & ...   & ...   & ...   & ...   & ...   & ...   & 682$\pm$105 & 85$\pm$12 & ...   & ...   & 385$\pm$33 & 72$\pm$9 \\
        J083411.09+580321.4 & ...   & ...   & ...   & ...   & ...   & ...   & ...   & ...   & ...   & ...   & 490$\pm$93 & 63$\pm$6 \\
        J083637.84+440109.6 & ...   & ...   & ...   & ...   & ...   & ...   & 333$\pm$0 & 793$\pm$0 & ...   & ...   & 595$\pm$0 & 355$\pm$0 \\
        J083825.00+371036.5 & ...   & ...   & ...   & ...   & ...   & ...   & 703$\pm$63 & 26$\pm$2 & ...   & ...   & ...   & ... \\
        J084856.57+013647.8 & ...   & ...   & ...   & ...   & ...   & ...   & 450$\pm$11 & 215$\pm$3 & ...   & ...   & 485$\pm$42 & 53$\pm$7 \\
        J085314.22+021453.8 & ...   & ...   & ...   & ...   & ...   & ...   & ...   & ...   & ...   & ...   & ...   & ... \\
        J085408.44+021316.1 & ...   & ...   & ...   & ... & ...   & ...   & ...   & ...   & ...   & ...   & ...   & ... \\
        J085601.22+285835.4 & ...   & ...   & 5684$\pm$529 & 260$\pm$14 & ...   & ...   & ...   & ...   & ...   & ...   & ...   & ... \\
        J090105.25+290146.9 & ...   & ...   & ...   & ...   & ...   & ...   & 698$\pm$94 & 140$\pm$67 & ...   & ...   & 573$\pm$51 & 54$\pm$7 \\
        J090615.53+463619.0 & ...   & ...   & ...   & ...   & ...   & ...   & 541$\pm$88 & 212$\pm$84 & 2322$\pm$55 & 1437$\pm$74 & 518$\pm$17 & 405$\pm$11 \\
        J090933.49+425346.5 & ...   & ...   & ...   & ...   & ...   & ...   & ...   & ...   & ...   & ...   & ...   & ... \\
        J091734.79+501638.1 & ...   & ...   & ...   & ...   & ...  & ...   & ...   & ...   & ...   & ...   & ...   & ... \\
        J092405.30+141021.4 & ...   & ...   & ...   & ...   & ...   & ...   & 771$\pm$54 & 134$\pm$5 & ...   & ...   & 756$\pm$29 & 356$\pm$8 \\
        J092608.00+074526.6 & ...   & ...   & 4475$\pm$719 & 111$\pm$26 & 7062$\pm$320 & 354$\pm$14 & 308$\pm$3 & 496$\pm$4 & ...   & ...   & ...   & ... \\
        J093430.68+030545.3 & ...   & ...   & ...   & ...   & ...   & ...   & 569$\pm$21 & 134$\pm$4 & ...   & ...   & 586$\pm$32 & 120$\pm$4 \\
        J093609.36+331308.3 & ...   & ...   & ...   & ...   & ...   & ...   & 558$\pm$75 & 197$\pm$62 & ...   & ...   & 454$\pm$23 & 205$\pm$10 \\
        J094525.90+352103.6 & ...   & ...   & ...   & ...   & ...   & ...   & 277$\pm$8 & 835$\pm$10 & 3033$\pm$103 & 928$\pm$29 & 327$\pm$23 & 52$\pm$4 \\
        J095111.52+345131.9 & ...   & ...   & ...   & ...   & ...   & ...   & 709$\pm$0 & 44$\pm$1 & ...   & ...   & ...   & ... \\
        J095412.57+420109.1 & 6488$\pm$1069 & 203$\pm$10 & ...   & ...   & ...   & ...   & ...   & ...   & ...   & ...   & ...   & ... \\
        J100800.04+073016.6 & ...   & ...   & 1985$\pm$186 & 43$\pm$4 & ...   & ...   & 956$\pm$0 & 305$\pm$4 & ...   & ...   & ...   & ... \\
        J100955.51+140154.2 & ...   & ...   & ...   & ...   &  ...   & ...   & 618$\pm$59 & 68$\pm$5 & ...   & ...   & 774$\pm$87 & 106$\pm$21 \\
        J101251.77+403903.4 & ...   & ...   & ...   & ...   & ...    & ...   & ...   & ...   & ...   & ...   & ...   & ... \\
        J101714.23+390121.1 & ...   & ...   & ...   & ...   & ...   & ...   & ...   & ...   & ...   & ...   & ...   & ... \\
        J102027.20+432056.3 & 2847$\pm$161 & 383$\pm$43 & 4358$\pm$397 & 186$\pm$9 & ...   & ...   & ...   & ...  & ...   & ...   & ...   & ... \\
        J102523.78+254158.3 & ...   & ...   & ...   & ...   & ...   & ...   & 710$\pm$28 & 48$\pm$2 & ...   & ...   & 555$\pm$119 & 26$\pm$5 \\
        J102618.25+454229.3 & ...   & ...   & ...   & ...   & ...   & ...   & 637$\pm$128 & 117$\pm$7 & ...   & ...   & 773$\pm$149 & 52$\pm$10 \\
        J102844.26+384436.8 & ...   & ...   & ...   & ...   & ...   & ...   & 544$\pm$18 & 84$\pm$2 & ...   & ...   & 662$\pm$34 & 64$\pm$3 \\
        J103507.04+562846.7 & ...   & ...   & ...   & ...   & ...   & ...   & 748$\pm$12 & 53$\pm$1 & ...   & ...   & ... & ... \\
        J103532.58+423019.0 & 5907$\pm$255 & 208$\pm$25 & 6599$\pm$244 & 62$\pm$6 & ...   & ...   & ...   & ...   & ...   & ...   & ...   & ... \\
        J103719.33+433515.3 & ...   & ...   & ...   & ...   & ...   & ...   & 443$\pm$61 & 376$\pm$109 & ...   & ...   & 456$\pm$19 & 741$\pm$26 \\
        J104029.94+295757.7 & ...   & ...   & ...   & ...   & ...   & ...   & 687$\pm$31 & 231$\pm$9 & ...   & ...   & 727$\pm$42 & 378$\pm$19 \\
        J104406.33+295900.9 & 3021$\pm$166 & 210$\pm$32 & ...   & ...   &  ...   & ...   & ...   & ...   & ...   & ...   & ...   & ... \\
        J105250.06+335504.9 & ...   & ...   & 7929$\pm$498 & 1374$\pm$53 & ...   & ...   & ...   & ...   & ...   & ...   & ...   & ... \\
        J105628.25+501952.0 & ...   & ...   & 8762$\pm$641 & 1110$\pm$49 & ...   & ...   & ...   & ...   & ...   & ...   & ...   & ... \\   
    \hline
  \end{tabular}
  %\end{center}
  %\end{landscape} 
  %\end{sidewaystable}
 %  }%
 \end{table*}
 
 \addtocounter{table}{-1}
 %\end{landscape}
 
 \begin{table*}
 \centering
      \setlength{\tabcolsep}{0.02in} \caption{Continued\dots
      \label{Sample}}
      \begin{tabular}{ccccccccccccc}
     \hline \hline
     SDSS name & \multicolumn{2}{c}{C IV} & \multicolumn{2}{c}{Mg II} & \multicolumn{2}{c}{H$\beta$}  & \multicolumn{2}{c}{[O III]} &\multicolumn{2}{c}{H$\alpha$}& \multicolumn{2}{c}{[S II]} \\
     (1) & (2) & (3)  & (4)  & (5)  &  (6) & (7) & (8) & (9)  & (10) &(11) &(12) &(13) \\
     \hline
     J105731.17+405646.1 & ...   & ...   & ...   & ...   & ...   & ...   & 560$\pm$39 & 266$\pm$18 & ...   & ...   & 532$\pm$53 & 264$\pm$23 \\
             J112027.80+142055.0 & ...   & ...   & ...   & ...   &  ...   & ...   & 648$\pm$22 & 86$\pm$3 & ...   & ...   & 993$\pm$128 & 29$\pm$6 \\
             J113138.89+451451.1 & ...   & ...   & ...   & ...   & ...   & ...   & 518$\pm$35 & 64$\pm$8 & 2565$\pm$93 & 225$\pm$12 & 991$\pm$110 & 55$\pm$8 \\
             J114311.02+053516.0 & ...   & ...   & 6971$\pm$1280 & 100$\pm$21 & 3150$\pm$347 & 139$\pm$13 & 767$\pm$9 & 527$\pm$6 & ...   & ...   & ...   & ... \\
             J114339.59+462120.4 & ...   & ...   & ...   & ...   & ...   & ...   & ...   & ...   & ...   & ...   & 450$\pm$22 & 146$\pm$6 \\
             J114856.56+525425.2 & 5036$\pm$152 & 2625$\pm$62 & 3289$\pm$373 & 1128$\pm$49 & ...   & ...   & ...   & ...   & ...   & ...   & ...   & ... \\
             J115000.08+552821.3 & ...   & ...   & ...   & ...   & ... & ...   & ...   & ...   & ...   & ...   & 479$\pm$30   & 28$\pm$2 \\
             J115618.74+312804.7 & ...   & ...   & 4541$\pm$553 & 548$\pm$46 & 3045$\pm$341 & 227$\pm$12 & 528$\pm$23 & 875$\pm$14 & ...   & ...   & ...   & ... \\
             J115727.60+431806.3 & ...   & ...   & ...   & ...   & 3296$\pm$307 & 1592$\pm$74 & 331$\pm$3 & 1051$\pm$8 & 3067$\pm$48 & 4369$\pm$36 & 333$\pm$13 & 66$\pm$3 \\
             J115919.97+464545.1 & ...   & ...   & ...   & ...   & ... & ...   & ...   & ...   & ...   & ...   & ...   & ... \\
             J120321.93+041419.0 & ...   & ...   & 7404$\pm$910 & 283$\pm$19 & ...   & ...   & ...  & ...   & ...   & ...   & ...   & ... \\
             J120406.82+391218.2 & ...   & ...   & ...   & ...   & ...   & ...   & ...   & ...   & ...   & ...   & ...   & ... \\
    J120624.70+641336.8 & ...   & ...   & ...   & ...   & ...   & ...   & 725$\pm$0 & 513$\pm$2 & 7569$\pm$157 & 739$\pm$10 & 993$\pm$23 & 61$\pm$4 \\
       J120902.79+411559.2 & ...   & ...   & ...   & ...   & ...  & ...   & ...   & ...   & ...   & ...   & ...   & ... \\
       J124419.96+405136.8 & ...   & ...   & ...   & ...   & ...   & ...   & 370$\pm$63 & 237$\pm$79 & 2595$\pm$140 & 927$\pm$50 & 605$\pm$112 & 112$\pm$12 \\
       J124449.18+404806.2 & ...   & ...   & 8604$\pm$983 & 162$\pm$14 & ...   & ...   & 763$\pm$28 & 114$\pm$4 & ...   & ...   & ...   & ... \\
       J124733.31+672316.4 & ...   & ...   & ...   & ...   &...   & ...   & 548$\pm$49 & 94$\pm$8 & ...   & ...   & 476$\pm$42 & 62$\pm$4 \\
       J125226.35+563419.6 & ...   & ...   & ...   & ...   & 4431$\pm$670 & 1334$\pm$208 & 525$\pm$4 & 2774$\pm$20 & 3981$\pm$219 & 5025$\pm$147 & 517$\pm$31 & 233$\pm$16 \\
       J125325.72+303635.1 & ...   & ...   & 4955$\pm$548 & 59$\pm$10 & ...   & ...   & ...   & ...   & ...   & ...   & ...   & ... \\
       J130941.51+404757.2 & 3138$\pm$201 & 313$\pm$34 & ...   & ...   &  ...   & ...   & ...   & ...   & ...   & ...   & ...   & ... \\
       J131057.00+445146.2 & ...   & ...   & ...   & ...   &  ...   & ...   & ...   & ...   & ...   & ...   & ...   & ... \\
       J131718.64+392528.1 & 6682$\pm$636 & 981$\pm$100 & 4177$\pm$255 & 293$\pm$11 & ...   &  ...   & ...   & ...   & ...   & ...   & ...   & ... \\
       J131739.20+411545.6 & ...   & ...   & ...   & ...   & ...   & ...   & 514$\pm$78 & 65$\pm$9 & ...   & ...   & 555$\pm$79 & 141$\pm$19 \\
       J132255.66+391208.0 & 4861$\pm$59 & 1165$\pm$76 & ...   & ...   &  ...   & ...   & ...   & ...   & ...   & ...   & ...   & ... \\
       J132419.67+041907.0 & ...   & ...   & ...   & ...   &  ...   & ...   & ...   & ...   & ...   & ...   & 466$\pm$84 & 57$\pm$7 \\
       J132513.37+395553.2 & ...   & ...   & ...   & ...   &  ...   & ...   & 541$\pm$38 & 114$\pm$17 & ...   & ...   & 653$\pm$32 & 291$\pm$16 \\
       J133037.69+250910.9 & ...   & ...   & 3236$\pm$206 & 639$\pm$19 & ...   & ...   & ...   & ...   & ...   & ...   & ...   & ... \\
       J133108.29+303032.9 & ...   & ...   & 2832$\pm$93 & 478$\pm$13 & 2074$\pm$201 & 617$\pm$59 & 567$\pm$26 & 371$\pm$35 & ...   & ...   & ...   & ... \\
       J134035.20+444817.3 & ...   & ...   & ...   & ...   &...   & ...   & 193$\pm$4 & 660$\pm$7 & ...   & ...   & 207$\pm$5 & 246$\pm$5 \\
       J134536.94+382312.5 & 4416$\pm$276 & 1617$\pm$288 & 7325$\pm$377 & 565$\pm$21 & ...  & ...   & ...   & ...   & ...   & ...   & ...   & ... \\
       J134733.36+121724.2 & ...   & ...   & ...   & ...   & 1589$\pm$83 & 413$\pm$37 & 457$\pm$25 & 481$\pm$40 & 2771$\pm$19 & 4645$\pm$110 & 589$\pm$18 & 618$\pm$10 \\
       J140028.65+621038.5 & ...   & ...   & ...   & ...   & ...   & ...   & 484$\pm$18 & 87$\pm$2 & ...   & ...   & 430$\pm$147 & 18$\pm$5 \\
       J140051.58+521606.5 & ...   & ...   & ...   & ...   & ...   & ...   & 669$\pm$66 & 69$\pm$7 & ...   & ...   & 750$\pm$94 & 124$\pm$16 \\
       J140319.30+350813.3 & 11872$\pm$1580 & 109$\pm$5 & 6875$\pm$761 & 78$\pm$13 & ...   & ...   & ...   & ...   & ...   & ...   & ...   &  ... \\
       J140416.35+411748.7 & ...   & ...   & ...   & ...   & ...  & ...   & ...   & ...   & ...   & ...   & ...   & ... \\
       J140700.39+282714.6 & ...   & ...   & ...   & ...   &  7659$\pm$255 & 19313$\pm$253 & 547$\pm$13 & 2794$\pm$51 & 9256$\pm$609 & 75700$\pm$1128 & 433$\pm$30 & 579$\pm$53 \\
       J140909.50+364208.1 & 1821$\pm$150 & 32$\pm$3 & 4788$\pm$858 & 47$\pm$5 & ...   & ...   & ...   & ...   & ...   & ...   & ...   & ... \\
       J140942.44+360415.9 & ...   & ...   & ...   & ...   &...   & ...   & 465$\pm$33 & 298$\pm$10 & ...   & ...   & 489$\pm$47 & 108$\pm$6 \\
       J141327.22+550529.2 & ...   & ...   & ...   & ...   & ...  & ...   & ...   & ...   & ...   & ...   & ...   & ... \\
       J141414.83+455448.7 & ...   & ...   & ...   & ...   & ...   & ...   & ...   & ...   & ...   & ...   & ...   & ... \\
       J141558.81+132023.7 & ...   & ...   & ...   & ...   & ...   & ...   & ...   & ...   & ...   & ...   & ...   & ... \\
       J141908.18+062834.7 & ...   & ...   & 4179$\pm$186 & 1207$\pm$32 & ...   & ...   & ...   & ...   & ...   & ...   & ...   & ... \\
       J142051.48+270427.0 & 7011$\pm$317 & 252$\pm$17 & ...   & ... & ...   & ...   & ...   & ...   & ...   & ...   & ...   & ... \\
       J142104.24+050844.7 & ...   & ...   & ...   & ...   & ...  & ...   & ...   & ...   & ...   & ...   & ...   & ... \\
       J142123.06+464547.9 & 6275$\pm$394 & 883$\pm$132 & 5975$\pm$0 & 228$\pm$0 & ...   & ...   & ...   & ...   & ...   & ...   & ...   & ... \\
       J143009.74+104326.9 & 5868$\pm$363 & 754$\pm$34 & 4481$\pm$465 & 212$\pm$16 & ...   & ...   & ...   & ...   & ...   & ...   & ...   & ... \\
       J143521.67+505122.9 & ...   & ...   & ...   & ...   & ...   & ...   & 469$\pm$44 & 53$\pm$5 & ...   & ...   & 510$\pm$32 & 101$\pm$6 \\
       J144516.46+095836.0 & 7728$\pm$207 & 388$\pm$26 & ...   &  ...   & ...   & ...   & ...   & ...   & ...   & ...   & ...   & ... \\
       J144712.76+404744.9 & ...   & ...   & ...   & ...   & ...   & ...   & 870$\pm$175 & 43$\pm$6 & ...   & ...   & ...   & ... \\
       J145958.43+333701.8 & ...   & ...   & 4596$\pm$304 & 4064$\pm$64 & 4326$\pm$136 & 2461$\pm$34 & 396$\pm$12 & 338$\pm$7 & ...   & ...   & ...   & ... \\
 \hline
 \end{tabular}
  \end{table*}
  \addtocounter{table}{-1}
     %\end{landscape}
     
     \begin{table*}
     \centering
     \setlength{\tabcolsep}{0.03in} \caption{Continued\dots
     \label{Sample}}
      %\Rotatebox{90}{
     \begin{tabular}{ccccccccccccc}
         \hline \hline
         SDSS name & \multicolumn{2}{c}{C IV} & \multicolumn{2}{c}{Mg II} & \multicolumn{2}{c}{H$\beta$}  & \multicolumn{2}{c}{[O III]} &\multicolumn{2}{c}{H$\alpha$}& \multicolumn{2}{c}{[S II]} \\
         (1) & (2) & (3)  & (4)  & (5)  &  (6) & (7) & (8) & (9)  & (10) &(11) &(12) &(13) \\
         \hline
                J150426.69+285430.5 & 3041$\pm$52 & 482$\pm$9 & 3167$\pm$147 & 117$\pm$6 & ...   & ...   & ...   & ...   & ...   & ...   & ...   & ... \\
                J151141.26+051809.2 & ...   & ...   & ...   & ...   &  2483$\pm$112 & 533$\pm$28 & 223$\pm$9 & 898$\pm$23 & 2799$\pm$47 & 4213$\pm$201 & 455$\pm$17 & 289$\pm$10 \\
                J152005.47+201605.4 & 4222$\pm$850 & 220$\pm$14 & 5992$\pm$930 & 121$\pm$7 &  ...   & ...   & ...   & ...   & ...   & ...   & ...   & ... \\
                J152349.34+321350.2 & ...   & ...   & ...   & ...   &  ...   & ...   & 399$\pm$25 & 292$\pm$17 & ...   & ...   & 420$\pm$25 & 70$\pm$4 \\
                J152837.01+381605.9 & ...   & ...   & 3854$\pm$199 & 85$\pm$4 & ...   & ...   & 324$\pm$45 & 17$\pm$2 & ...   & ...   & ...   & ... \\
                J153409.90+301204.0 & ...   & ...   & ...   & ...   & ...   & ...   & ...   & ...   & ...   & ...   & ...   & ... \\
                J154349.50+385601.3 & ...   & ...   & ...   & ...   &  ...   & ...   & 541$\pm$27 & 76$\pm$3 & ...   & ...   & ...   & ... \\
                J154525.48+462244.3 & ...   & ...   & ...   & ...   & ...   & ...   & 918$\pm$0 & 321$\pm$3 & ...   & ...   & 992$\pm$0 & 115$\pm$16 \\
                J154609.52+002624.6 & ...   & ...   & ...   & ...   & ...   & ...   & 659$\pm$40 & 54$\pm$3 & ...   & ...   & ...   & ... \\
                J154754.12+351842.2 & ...   & ...   & 4155$\pm$298 & 113$\pm$5 &...   & ...   & 374$\pm$30 & 22$\pm$1 & ...   & ...   & ...   & ... \\
                J155235.38+441905.9 & ...   & ...   & ...   & ...   & ...   & ...   & ...   & ...   & ...   & ...   & ...   & ... \\
                J155927.67+533054.4 & ...   & ...   & ...   & ...   &...   & ...   & 519$\pm$27 & 149$\pm$4 & ...   & ...   & 550$\pm$33 & 71$\pm$4 \\
                J160239.62+264606.0 & ...   & ...   & ...   & ...   & ...   & ...   & 871$\pm$0 & 111$\pm$4 & ...   & ...   & ...   & ... \\
                J160246.39+524358.3 & ...   & ...   & ...   & ...   &  ...   & ...   & 553$\pm$67 & 111$\pm$8 & 2724$\pm$191 & 690$\pm$111 & 992$\pm$58 & 185$\pm$16 \\
                J160335.16+380642.8 & ...   & ...   & ...   & ...   &...   & ...   & 705$\pm$0 & 54$\pm$2 & ...   & ...   & 935$\pm$96 & 81$\pm$15 \\
                J160913.32+264129.0 & ...   & ...   & ...   & ...   &...   & ...   & 625$\pm$18 & 145$\pm$2 & 3763$\pm$864 & 212$\pm$11 & 993$\pm$160   & 28$\pm$5.1 \\
                J161148.52+404020.9 & ...   & ...   & ...   & ...   &  ...   & ...   & 313$\pm$17 & 394$\pm$14 & ...   & ...   & 609$\pm$21 & 388$\pm$8 \\
               J161748.41+380141.8 & 8309$\pm$2061 & 446$\pm$79 & 7780$\pm$738 & 191$\pm$13 & ...   & ...   & ...   & ...   & ...   & ...   & ...   & ... \\
               J161823.57+363201.7 & ...   & ...   & ...   & ...    & ...   & ...   & ...   & ...   & ...   & ...   & ...   & ... \\
              J162111.27+374604.9 & ...   & ...   & 6320$\pm$844 & 128$\pm$13 & ...   & ...   & ...   & ...   & ...   & ...   & ...   & ... \\
              J163402.95+390000.5 & ...   & ...   & 4028$\pm$621 & 189$\pm$15 & ...   & ...   & ...   & ...    & ...   & ...   & ...   & ... \\
          J164311.34+315618.4 & ...   & ...   & 5187$\pm$709 & 344$\pm$49 &  4187$\pm$281 & 303$\pm$18 & 313$\pm$4 & 790$\pm$8 & ...   & ...   & ...   & ... \\
          J164348.60+171549.4 & ...   & ...   & ...   & ...   &  ...   & ...   & 683$\pm$65 & 81$\pm$17 & ...   & ...   & 482$\pm$56 & 23$\pm$3 \\
         J165822.18+390625.6 & ...   & ...   & ...   & ...   & ...  & ...   & ...   & ...   & ...   & ...   & ...   & ... \\
     J213638.58+004154.2 & 4280$\pm$138 & 3138$\pm$59 & 3325$\pm$345 & 629$\pm$31 & ...   & ...   & ...   & ...   & ...    & ...   & ...   & ... \\
     J225025.34+141952.0 & ...   & ...   & ...   & ...   & 2882$\pm$241 & 1073$\pm$63 & 512$\pm$10 & 1451$\pm$31 & 2990$\pm$88 & 4910$\pm$142 & 473$\pm$35 & 136$\pm$12 \\
     \hline\hline
     SDSS name & \multicolumn{2}{c}{C IV} & \multicolumn{2}{c}{Mg II} & \multicolumn{2}{c}{H$\beta$}  & \multicolumn{2}{c}{[O III]} &\multicolumn{2}{c}{H$\alpha$}& \multicolumn{2}{c}{[S II]} \\
              (1) & (2) & (3)  & (4)  & (5)  &  (6) & (7) & (8) & (9)  & (10) &(11) &(12) &(13) \\
  HB89 1127 $-$ 145 & ...   & ...   & 3677$\pm$370   & 1607$\pm$78   &...  & ... & ... &  ...& ... & ... & ...  & ... \\
  CGRaBS J1424+2256  & 5461$\pm$172   & 9299$\pm$ 101   & ...   & ...   & ... & ... & ... & ... & ... &  ...&...   &  ...\\
  \hline    
     \end{tabular}
     %}
     %\Rotatebox{90}{%
     \begin{minipage}{170mm}
     Columns: (1) object name; (2) - (3) the FWHM and flux for broad C IV; (4) - (5) the FWHM and flux for broad Mg II; (6) - (7) the FWHM and flux for broad H$\beta$; (8) - (9) the FWHM and flux for [O III] line core; (10) - (11) the FWHM and flux for broad H$\alpha$; (12) - (13) the FWHM and flux for [S II] $\lambda 6716$. The FWHM and flux are in units of $\rm km\ s^{-1}$, and $\rm 10^{-17} erg~ s^{-1} cm^{-2}$, respectively.    
      \end{minipage}
      %}
      \end{table*}

 \begin{table*}
   \caption{The optical variability}
   \label{tab:landscape}
   \setlength{\tabcolsep}{0.03in}
   \begin{center}
   \begin{tabular}{lccrrrrrrrrrrr}
     \hline \hline
      SDSS name & type & $z$ & plate & MJD & fiber & S/N & $\Delta$f & $\Delta$$\alpha$ & $\alpha_{\lambda}$ & FWHM (C IV) & f (C IV)  & FWHM (Mg II) & f (Mg II) \\
     (1) & (2) & (3)  & (4)  & (5)  &  (6) & (7) & (8) & (9)  & (10) &(11) &(12) &(13) &(14) \\
     \hline
 %%%%%%%%%%%%%%%%%%%%%%%%%%%%%%%%%%%%%%%%%%%%%%%%%%%%%%%%%%%%%%%%%
 % Table generated by Excel2LaTeX from sheet '工作表1'
 % Table generated by Excel2LaTeX from sheet '工作表1'
 %\begin{table}[htbp]
  % \centering
   %\caption{Add caption}
    % \begin{tabular}{lrrrrrrrrlrrllrr}
     % SDSS name & \multicolumn{1}{l}{type} & \multicolumn{1}{l}{z} & \multicolumn{1}{l}{plate} & \multicolumn{1}{l}{mjd} & \multicolumn{1}{l}{fiber} & \multicolumn{1}{l}{S/N} & \multicolumn{1}{l}{AV } & \multicolumn{1}{l}{CV} & slope & \multicolumn{1}{l}{CIV\_FWHM} & \multicolumn{1}{l}{CIV\_total\_flux} & mgii\_FWHM & mgii\_total\_flux & \multicolumn{1}{l}{hb\_FWHM} & \multicolumn{1}{l}{hb\_total\_flux} \\
   J075303.33+423130.8 & CSO & 3.593 & 434   & 51885 & 71    & 23.6  & 47.0 & BWB & -1.62$\pm$0.06 & 2711$\pm$57 & 532$\pm$48 &. . .     &. . .      \\
          &       &     & 3683  & 55245 & 488   & 29.1  &       &       & -1.95$\pm$0.02 & 2944$\pm$61 & 651$\pm$72  &       &     \\       
        J080442.23+301237.0 & CSS & 1.448 & 860   & 52319 & 495   & 26.1  & 1.2 &       & -1.83$\pm$0.35 & . . .  & . . .   & 5207$\pm$107 & 447$\pm$8   \\
        &       &   & 4451  & 55537 & 546   & 27.4  &   &       & -1.27$\pm$0.16 & . . . & . . . & 5098$\pm$423 & 491$\pm$70 \\     
        J085601.22+285835.4 & CSS & 1.084 & 1589  & 52972 & 37    & 9.3   & 4.7 &       & -1.94$\pm$0.28 &  . . .     & . . .  & 5489$\pm$938 & 300$\pm$65  \\
        &       &  & 1934  & 53357 & 386   & 10.1  &       &       & -1.85$\pm$0.46 &  . . .        &  . . .      & 6014$\pm$1087 & 294$\pm$56   \\      
        J102027.20+432056.3 & HFP & 1.964 & 1218  & 52709 & 607   & 8.4  & 20.0 & BWB & -0.45$\pm$0.03 & 2537$\pm$191 & 568$\pm$77 & 4632$\pm$650 & 165$\pm$10 \\
         &       &  & 3287  & 54941 & 483   & 13.5  &       &       & -0.15$\pm$0.02 & 2886$\pm$144 & 347$\pm$33 & 4245$\pm$336 & 166$\pm$7   \\
        J103532.58+423019.0 & HFP & 2.432 & 1360  & 53033 & 562   & 8.4  & 36.0 &       & -0.90$\pm$0.03 & 6081$\pm$270 & 326$\pm$39 & . . .       &. . .    \\
         &       &  & 4632  & 55644 & 738   & 9.1  &       &       & -0.96$\pm$0.03 & 5926$\pm$238 & 211$\pm$33 &. . .       & . . .   \\
        J104406.33+295900.9 & HFP & 2.981 & 1969  & 53383 & 48    & 7.4  & 50.0 & BWB      & -1.40$\pm$0.06 & 3428$\pm$131 & 526$\pm$64 &. . .       &. . .   \\
        &       &  & 6452  & 56366 & 946   & 8.9  &       &  & -1.09$\pm$0.03 & 3021$\pm$173 & 209$\pm$24 & . . .      &. . .    \\
        J105628.25+501952.0 & CSS & 0.820  & 876   & 52346 & 112   & 22.8  & 19.0 &   & -1.43$\pm$0.03 &  . . .  & . . .      & 7156$\pm$127 & 1074.4$\pm$21.7   \\
         &       &   & 876   & 52669 & 112   & 21.7  &       &       & -1.37$\pm$0.26 & . . . &  . . .     & 7499$\pm$340 & 1127$\pm$93  \\
        J130941.51+404757.2 & HFP & 2.908 & 1460  & 53138 & 267   & 7.6  & 60.0  &  BWB & -1.13$\pm$0.05 & 3273$\pm$194 & 285$\pm$45 &. . .       &. . .   \\
     &       &  & 4706  & 55705 & 768   & 18.8  &       &       & -1.23$\pm$0.01 & 3157$\pm$125 & 326$\pm$28 &. . .       &. . .  \\
       J132255.66+391208.0 & HFP & 2.990  & 1977  & 53475 & 496   & 24.5  & 15.0 &  & -1.18$\pm$0.02 & 4784$\pm$85 & 1112$\pm$84 & . . .      &. . .  \\
       &       &   & 3984  & 55333 & 548   & 30.1  &       &       & -1.32$\pm$0.01 & 4842$\pm$60 & 1112$\pm$75 &. . .       &. . .     \\
        J133108.29+303032.9 & CSS & 0.849 & 2110  & 53467 & 104   & 26.0    & 4.9 &       & -1.64$\pm$0.66 &  . . .       &. . .       & 2865$\pm$344 & 639$\pm$91  \\
        &       &   & 6491  & 56337 & 596   & 37.3  &       &       & -1.84$\pm$0.23 &  . . .       & . . .      & 2944$\pm$109 & 600$\pm$15   \\
        J134536.94+382312.5 & CSS & 1.852 & 2014  & 53460 & 271   & 15.4  & 37.0 & BWB & -0.70$\pm$0.02 & 4997$\pm$290 & 1440$\pm$142 & 6160$\pm$264 & 463$\pm$11  \\
        &       &  & 3987  & 55590 & 958   & 22.6  &       &       & -0.98$\pm$0.01 & 4977$\pm$211 & 1834$\pm$150 & 6452$\pm$241 & 599$\pm$11  \\
        J144516.46+095836.0 & GPS & 3.520  & 1713  & 53827 & 314   & 19.0    & 18.0 &  & -1.85$\pm$0.07 & 7921$\pm$349 & 499$\pm$51 & . . .      & . . .   \\
       &       &  & 5472  & 55976 & 822   & 31.0   &       &       & -2.34$\pm$0.02 & 8076$\pm$251 & 597$\pm$48 &. . .       & . . .  \\
        J150426.69+285430.5 & CSS & 2.282 & 2151  & 54523 & 426   & 12.7  & 2.1 &  & -1.34$\pm$0.02 & 3118$\pm$86 & 598$\pm$55 & . . .      & . . .   \\
         &       &  & 3877  & 55365 & 818   & 23.8  &       &       & -0.89$\pm$0.01 & 3002$\pm$45 & 482$\pm$9 & . . .      & . . .  \\
\hline \hline
      SDSS name & type & $z$ & plate & MJD & fiber & S/N & $\Delta$f & $\Delta$$\alpha$ & $\alpha_{\lambda}$ & FWHM (C IV) & f (C IV)  & FWHM (H$\beta$) & f (H$\beta$) \\
     (1) & (2) & (3)  & (4)  & (5)  &  (6) & (7) & (8) & (9)  & (10) &(11) &(12) &(13) &(14) \\
     J092608.00+074526.6 & CSS & 0.442 & 1195  & 52724 & 453    & 10.8   & 7.0 &      & -0.05$\pm$0.02 &  . . . & . . .       & 7387$\pm$622 & 650$\pm$24   \\
                  &       &  & 2402  & 54176 & 484   & 22.2  &       &       & -0.17$\pm$0.01 &  . . .        &  . . .      & 7560$\pm$453 & 705$\pm$16   \\ 
J115727.60+431806.3 & CSS & 0.230  & 1447  & 53120 & 571   & 15.8  & 41.0 & BWB  & -0.60$\pm$0.44 &  . . .     & . . .    & 3468$\pm$656 & 1368$\pm$432 \\
         &       &   & 6642  & 56396 & 248   & 29.7  &       &       & -1.24$\pm$0.50 & . . . &  . . .     & 3370$\pm$588 & 1751$\pm$429 \\      
     \hline
  \end{tabular}
  
  \begin{minipage}{170mm}
       Columns: (1) object name; (2) source type; (3) redshift; (4) - (6) the plate, MJD, and fiber of spectroscopic observations; (7) median signal-to-noise ratio of the spectrum; (8) the variability in \% (see text for details); (9) color variation: BWB - bluer-when-brighter; (10) the spectral index of power-law continuum; (11) - (12) the FWHM and flux of C IV in units of $\rm km\ s^{-1}$, and $ \rm 10 ^{-17} \rm erg\ s^{-1} cm^{-2}$, respectively; (13) -  (14) same as (11) - (12) but for broad Mg II or broad H$\beta$.
       
      % a: The FWHM and flux measurements for H$\beta$
        \end{minipage}
  \end{center}
 \end{table*}

 \begin{table*}
   \caption{Radio and spectroscopic classification}
   \label{tab:landscape}
   \setlength{\tabcolsep}{0.06in}
   \begin{center}
   \begin{tabular}{lccrrrrrrrrcccc}
     \hline \hline
     SDSS name & type & type & SN & H$\beta$ & [O III] & [O I] & H$\alpha$ & [N II] & [S II] & [S II] & EI & HEG/LEG & log $R_{\rm edd}$ & Type1/Type2 \\
      &  &   &   & $\lambda4861$ &  $\lambda5007$ & $\lambda6300$ & $\lambda6563$ & $\lambda6584$  & $\lambda6717$ &$\lambda6731$ & & & & \\
     (1) & (2) & (3)  & (4)  & (5)  &  (6) & (7) & (8) & (9)  & (10) &(11) &(12) &(13) &(14) & (15) \\
     \hline
         J002833.42+005510.9 & CSS   & G & 21    & 165   & 1472  & 224   & 873   & 1411  & 532   & 443   & 1.06  & HEG   & -1.48  & Type2 \\
         J075756.71+395936.1 & CSS   & G & 28    & 203   & 1860  & 126   & 1038  & 940   & 338   & 292   & 1.35  & HEG   & -1.97    & Type2 \\
         J081323.75+073405.6 & CSS   & G & 20    & 106   & 265   & 192   & 639   & 1071  & 445   & 393   & 0.46  & LEG   & -2.93  & Type2 \\
         J083139.79+460800.8 & GPS   & G & 26    & 24    & 85    & 30    & 117   & 131   & 72    & 63    & 0.72  & LEG   & -4.04    & Type2 \\
         J083637.84+440109.6 & CSS   & G & 29    & 119   & 793   & 139   & 643   & 835   & 355   & 327   & 1.00     & HEG   & -3.32  & Type2 \\
         J090615.53+463619.0 & GPS   & G & 35    & 184   & 212   & 400   & 399   & 283   & 405   & 315   & 0.03  & LEG   & -3.77  & Type1$^{a}$\\
         J092405.30+141021.4 & CSS   & G & 26    & 128   & 134   & 186   & 518   & 736   & 356   & 327   & 0.08  & LEG   & -3.83  & Type2 \\
         J093430.68+030545.3 & CSS   & G & 15    & 43    & 134   & 53    & 239   & 367   & 120   & 124   & 0.64  & LEG   & -3.71  & Type2 \\
         J093609.36+331308.3 & GPS   & G & 34    & 89    & 197   & 128   & 372   & 425   & 205   & 164   & 0.48  & LEG   & -3.92  & Type2 \\
         J100955.51+140154.2 & CSS   & G & 14    & 39    & 68    & 54    & 195   & 251   & 106   & 116   & 0.37  & LEG   & -3.65 & Type2 \\
         J102523.78+254158.3 & HFP   & G & 3     & 11    & 48    & 26    & 51    & 76    & 26    & 26    & 0.67  & LEG   & -3.32 & Type2 \\
         J102618.25+454229.3 & CSO   & G & 18    & 30    & 117   & 63    & 116   & 156   & 52    & 68    & 0.63  & LEG   & -3.70 & Type2 \\
         J102844.26+384436.8 & CSS   & Q   & 15    & 25    & 84    & 38    & 116   & 158   & 64    & 53    & 0.64  & LEG   & -2.78 & Type2 \\
         J103719.33+433515.3 & CSS   & G & 55    & 223   & 376   & 171   & 915   & 1635  & 741   & 585   & 0.33  & LEG   & -4.33 & Type2 \\
         J104029.94+295757.7 & CSS   & G & 29    & 80    & 231   & 127   & 530   & 1197  & 378   & 296   & 0.51  & LEG   & -3.08  & Type2 \\
         J105731.17+405646.1 & GPS   & G & 53    & 94    & 266   & 105   & 535   & 783   & 264   & 197   & 0.66  & LEG   & -4.93 & Type2 \\
         J112027.80+142055.0 & GPS   & Q   & 2     & 19    & 86    & 71    & 77    & 93    & 29    & 36    & 0.67  & LEG   & -3.13  & Type2 \\
         J113138.89+451451.1 & CSS   & Q   & 5     & 11    & 64    & 24    & 74    & 76    & 55    & 47    & 0.89  & LEG   & -2.62 & Type1$^{b}$\\
         J115727.60+431806.3 & CSS   & Q   & 30    & 77    & 1051  & 36    & 320   & 163   & 66    & 57    & 1.69  & HEG   & -1.38  & Type1 \\
         J120624.70+641336.8 & CSS   & Q   & 3     & 51    & 513   & 35    & 290   & 141   & 61    & 70    & 1.53  & HEG   & -1.97    & Type1 \\
         J124419.96+405136.8 & CSS   & Q   & 13    & 87    & 237   & 56    & 152   & 230   & 112   & 90    & 0.48  & LEG   & -2.25 & Type1$^{a}$ \\
         J125226.35+563419.6 & CSS   & Q   & 14    & 315   & 2774  & 114   & 1408  & 517   & 233   & 183   & 1.63  & HEG   & -1.44 & Type1 \\
         J132513.37+395553.2 & GPS   & G & 34    & 90    & 114   & 119   & 408   & 394   & 291   & 232   & 0.25  & LEG   & -4.00    & Type2 \\
         J134035.20+444817.3 & GPS   & G & 20    & 219   & 660   & 58    & 1365  & 682   & 246   & 195   & 1.20   & HEG   & -1.96    & Type2 \\
         J134733.36+121724.2 & GPS   & Q   & 18    & 163   & 481   & 654   & 1201  & 1193  & 618   & 461   & 0.57  & LEG   & -2.93  & Type1$^{a}$\\
         J140028.65+621038.5 & GPS   & Q   & 3     & 9     & 87    & 33    & 131   & 71    & 18    & 21    & 1.44  & HEG   & -2.43  & Type2 \\
         J140700.39+282714.6 & GPS   & Q   & 47    & 543   & 2794  & 979   & 2139  & 2045  & 579   & 519   & 0.93  & LEG   & -1.97    & Type1 \\
         J140942.44+360415.9 & CSS   & Q   & 10    & 79    & 298   & 76    & 270   & 246   & 108   & 80    & 0.83  & LEG   & -2.22  & Type2 \\
         J143521.67+505122.9 & CSS   & G & 18    & 23    & 53    & 24    & 161   & 221   & 101   & 73    & 0.58  & LEG   & -3.55  & Type2 \\
         J151141.26+051809.2 & HFP   & Q   & 30    & 118   & 898   & 220   & 487   & 410   & 289   & 250   & 1.01  & HEG   & -2.23  & Type1$^{a}$ \\
         J152349.34+321350.2 & CSS   & G & 17    & 25    & 292   & 34    & 136   & 309   & 70    & 78    & 1.13  & HEG   & -2.34  & Type2 \\
         J154525.48+462244.3 & CSS   & Q   & 3     & 54    & 321   & 96    & 328   & 175   & 115   & 80    & 1.12  & HEG   & -3.01    & Type2 \\
         J155927.67+533054.4 & CSS   & G & 25    & 36    & 149   & 31    & 136   & 214   & 71    & 64    & 0.77  & LEG   & -2.74  & Type2 \\
         J160246.39+524358.3 & CSS   & G & 26    & 37    & 111   & 48    & 199   & 401   & 185   & 125   & 0.51  & LEG   & -3.61 & Type1$^{a}$ \\
         J160335.16+380642.8 & CSS   & G & 8     & 11    & 54    & 32    & 92    & 84    & 81    & 44    & 0.80   & LEG   & -3.79  & Type2 \\
         J160913.32+264129.0 & GPS   & Q   & 3     & 33    & 145   & 130   & 138   & 109   & 28    & 83    & 0.72  & LEG   & -2.57  & Type1$^{b}$ \\
         J161148.52+404020.9 & CSS   & Q   & 14    & 94    & 394   & 175   & 567   & 691   & 388   & 374   & 0.72  & LEG   & -2.58  & Type2 \\
         J164348.60+171549.4 & CSS   & G & 25    & 13    & 81    & 39    & 69    & 249   & 23    & 21    & 0.76  & LEG   & -3.36  & Type2 \\
         J225025.34+141952.0 & CSS   & Q   & 36    & 201   & 1451  & 37    & 718   & 411   & 136   & 109   & 1.52  & HEG   & -1.14 & Type1 \\       
     \hline
\end{tabular}
\begin{minipage}{170mm}
Columns are listed as follows: (1) object name; (2) radio type: CSS/GPS/HFP/CSO; (3) source SDSS calssification: G - galaxy, Q - quasar; (4) median signal-to-noise ratio of the spectrum; (5) - (11) The flux of each narrow emission line in units of $ \rm 10 ^{-17} \rm erg\ s^{-1} cm^{-2}$ with typical uncertainty of $<$ 10\% ; (12) the EI index; (13) HEG and LEG classification based on EI index (separation:0.95); (14) the estimated Eddington ratio ($L_{\rm bol}/L_{\rm edd}$); (15) optical classification based on whether the exsistence of the broad lines. ($^{a}$: the source have broad lines and $\sigma_{\ast}$ measurements; $^{b}$: the source have broad lines measurements but $D_{4000} > 1$ and SN $<$ 10)\\
\\
 \end{minipage}
\end{center}
\end{table*}

\bsp	% typesetting comment
\label{lastpage}
\end{document}